%% file: 0_main_08232026.tex
\documentclass[11pt,letter]{article}
\usepackage{graphicx} 
\usepackage{amsfonts,amsmath,amsthm,amssymb,mathtools}
 \usepackage{multirow}
\usepackage{booktabs}
\usepackage[authoryear]{natbib}
\usepackage{xcolor}
\usepackage{algorithm}
\usepackage{algpseudocode}
\usepackage{blkarray}
\usepackage{comment}
\usepackage{tikz}
\usepackage{rotating}
\usepackage{subcaption}
\usepackage{url}
\usepackage{enumitem}
\newtheorem{lemma}{Lemma}
\newtheorem{remark}{Remark}
\newtheorem{definition}{Definition}

\newtheorem{theorem}{Theorem}
\newtheorem{proposition}{Proposition}
\usepackage{amsmath,amsfonts,amssymb,amsthm}
\usepackage[edges]{forest} 
\usepackage[colorinlistoftodos]{todonotes} 
\usepackage{hyperref}
\usepackage{tikz}
\usepackage{subcaption}
\usepackage{adjustbox}
\usepackage{geometry, diagbox,tabularx}
\usepackage{bm}
\newtheorem{example}{Example}
\newenvironment{excont}[1]
  {\par\medskip\noindent\textbf{Example~\ref{#1} (continued).}\ \itshape}
  {\par\medskip}
\newcommand{\D}{\mathcal{D}}
\newcommand{\B}{\mathcal{B}}
\newcommand{\conv}{\mathrm{conv}}
\newcommand{\Proj}{\mathrm{proj}}
\newcommand{\fb}{\mathbf{f}}
\newcommand{\ff}{\mathbf{f}^{\textrm{feas}}}
\newcommand{\fp}{\mathbf{f}^{+}}
\newcommand{\Ppf}{\mathcal{P}_{p,f}}
\newcommand{\x}{\mathbf{x}}
\newcommand{\y}{\mathbf{y}}
\newcommand{\ub}{\mathbf{u}}
\newcommand{\rb}{\mathbf{r}}
\newcommand{\vb}{\mathbf{v}}
\newcommand{\pD}{\overline{\mathbf{p}}^D}
\newcommand{\pcalD}{\overline{\mathbf{p}}_\D}

\newcommand{\Prum}{\mathcal{P}^{\textrm{RUM}}}
\usetikzlibrary{arrows.meta,positioning}

\usepackage{xcolor}
\newcommand{\mx}[1]{\textcolor{blue}{#1}}
\newcommand{\ks}[1]{\textcolor{purple}{#1}}

\newcommand{\blambda}{\boldsymbol{\lambda}}
\newcommand{\btheta}{\boldsymbol{\theta}}

\usepackage{xspace}
\newcommand{\DRUM}{\text{D-RUM}\xspace}

\title{Alternative Computational Formulations for Testing Discrete-Choice Random Utility Models: Half-Space and Network-Flow Formulations
\footnote{ 
We thank Mark Dean, Dilip Ravindran and Jörg Stoye for providing the dataset used in the paper. We thank Zhuo Chen, Junnan He, Mark Dean, Donald Green, Simon Lee, Huikang Liu, Bernard Salani\'{e}, Andres Santos, Jörg Stoye, Pietro Tebaldi, Clayton Thomas, Maxim Ventura, Yinyu Ye and Yuichi Kitamura for comments and feedback.  We thank Arya Gadage for excellent research assistance. This paper was written with the assistance of ChatGPT and Claude Code, which were used to suggest mathematical arguments, assist with drafting and revising the manuscript, and help implement the code. Solvers implementing our methods are publicly available on GitHub at \href{https://github.com/mingxiz/d-rum-testing-gpu}{\texttt{d-rum-testing-gpu}} and \href{https://github.com/hchang1508/consumption-budget-rum-testing-gpu-cpu}{\texttt{consumption-budget-rum-testing-gpu-cpu}}. }}
\date{\today}

\author{
Haoge Chang\thanks{Department of Economics, Columbia University, \href{mailto:hc3516@columbia.edu}{\nolinkurl{hc3516@columbia.edu}}}
\and
Kota Saito\thanks{Division of Humanities and Social Science, California Institute of
Technology, \href{mailto:saito@caltech.edu}{\nolinkurl{saito@caltech.edu}}}
\and
Mingxi Zhu\thanks{Scheller College of Business, Georgia Institute of Technology, \href{mailto:mingxi.zhu@scheller.gatech.edu}{\nolinkurl{mingxi.zhu@scheller.gatech.edu
}}}
}
\begin{document}
\maketitle

\begin{abstract}
This paper develops computational algorithms for testing discrete-choice random utility models (RUMs). The set of RUM-consistent choice probabilities forms a polytope, and existing computational methods rely on its vertex representation. We exploit alternative half-space (Block--Marschak inequalities) and network-flow representations to develop algorithms based on the alternating direction method of multipliers and column generation. We also develop a network-flow formulation for random utility models in the consumption-budget setting. Our algorithms are better suited to GPU implementation than existing vertex-based methods. We establish convergence guarantees and provide optimality certificates for all proposed algorithms. In numerical experiments, our algorithms achieve $100$--$600\times$ speedups over existing methods.
\end{abstract}
\newpage
\tableofcontents
\newpage
\section{Introduction}
\input{paper_main/A_Introduction}

\subsection{Road Map}
\input{paper_main/A_road_map}
\subsection{Related Literature}\label{section:literature}
\input{paper_main/A_Related_Literature}

\section{Setup, Projection Problem and the Hardness Result}\label{section:computational_problem}
\input{paper_main/A_Computation_Problem}

\section{Existing Approaches}\label{section:existing_approach}
\input{paper_main/A_Existing_Approach}

\section{BM and Flow Representations, Projection Approach, and Overview of Our Algorithms  }\label{section:overview}
This section provides an overview of our approach. We begin by reviewing existing characterizations of the \DRUM polytope in Sections \ref{section:representations} and \ref{section:flow_rep}. We then provide an overview of our proposed methods in Sections \ref{section:projection_approach} and \ref{section:overview_algo}.

\subsection{Block--Marschak Representation of the \DRUM Polytope}\label{section:representations}
\input{paper_main/A_RUM_Characterization}

\subsection{Lifting-Based Approaches via the BM and Flow Representations}\label{section:projection_approach}
\input{paper_main/A_Extended_Formulation}

\subsection{Notation}\label{section:notations_and_roadmap}
\input{paper_main/A_Notations_Roadmap}

\section{Block--Marschak Formulation via ADMM}\label{section:bm_admm}
\input{paper_main/A_BM_08232026.tex}

\section{Flow Formulation: Column Generation and ADMM}\label{section:flow}
\input{paper_main/A_Flow_Formulation}

\subsection{Flow-Based Column Generation (Flow-CG)}\label{section:BellmanFord}
\input{paper_main/A_Bellman_Ford}

\section{Random Utility Models in the Consumption-Budget Setting}\label{section:consumption_budget}
\input{paper_main/A_BudgetSet}

\section{Numerical Evidence}\label{section:numerical_evidence}
\input{paper_main/A_Numerical_Final}

\section{Application: Choice Overload}\label{section:applications}
\input{paper_main/A_Application}

\appendix
\newpage
\clearpage

\section*{Supplemental Material}
\addcontentsline{toc}{section}{Supplemental Material}

\setcounter{section}{0}
\renewcommand{\thesection}{S.\arabic{section}}
\renewcommand{\theHsection}{S.\arabic{section}}
\section{NP-Hardness of the \DRUM Projection Problem}\label{section:NP_statement}
\input{paper_appendix/Z_NP_Hardness}
\clearpage

\section{Flow-ADMM}\label{section:ADMM}
\input{paper_main/A_ADMM_algorithm_flow_mingxi_08232026}

\clearpage

\section{Optimality Certificate and Convergence Guarantee}\label{section:certificate_guarantee}
\input{paper_main/A_FW_Guarantee}

\input{paper_main/A_ADMM_certificate_flow_08232026}
\clearpage

\section{Additional Applications}\label{sec:additional_applications}
\input{paper_appendix/Z_Additional_Applications}
\clearpage

\section{Indexing Choice Probabilities or Arc Flows for a Given Alternative $x$}\label{section:index}

\input{paper_appendix/Z_Index}


\newpage

\section{Evaluating the Simplex Proximal Operator}\label{section:proximal_operator_simplex}
\input{paper_appendix/Z_proximal}

\newpage
\section{Additional Simulation Results From Section \ref{section:compare_with_in}}
\label{sec:supp_simulations}
\input{paper_main/A_Additional_Simulations}

\newpage

\section{Statistical Inference}\label{sec:statistical}

\input{paper_main/A_Statistical_Application}

\section{Sample Prompt}\label{sec:sample_prompt}
\input{paper_main/A_Sample_Prompt}
\clearpage

\section*{Online Appendix}
\addcontentsline{toc}{section}{Online Appendix}
\setcounter{section}{0}
\renewcommand{\thesection}{A.\arabic{section}}
\renewcommand{\theHsection}{A.\arabic{section}}
\section{Additional Results for Flow-ADMM}
\input{paper_appendix/Z_Appendix_hypercube_Laplacian}
\input{paper_appendix/Z_Appendix_ADMM_Efficient_Solve}

\section{Convergence of Relaxed ADPMM with an Augmented Lagrangian Weighted by a Positive-Definite Matrix}\label{section:admm_convergence}
\input{paper_appendix/Z_ADMM_Convergence}
\section{Additional Algorithms}

\subsection{Integer Programming and Heuristic Methods for the Pricing Problem}\label{section:pricing_ip_heuristic}
\input{paper_appendix/Z_Appendix_column_generation}
\subsection{BM-ADMM with Over-Relaxation}\label{section:bm_with_relaxation}
\input{paper_appendix/Z_Appendix_BM_ADMM_Complete}

\subsection{Flow-ADMM with Over-Relaxation}\label{section:flow_with_relaxation}
\input{paper_appendix/Z_Appendix_Flow_ADMM_Complete}

\subsection{ADMM Optimality Certificates}
\input{paper_appendix/Z_appendix_ADMM_certificate}


\newpage
\section{Extensions}\label{section:Extension}
\input{paper_main/A_Add_Delete_Ranking}

\input{paper_appendix/Z_Proof_Add_Delete_Ranking}
\input{paper_main/A_Simplification_Correct}

\newpage 

\newpage
\section{Proofs}\label{section:proofs}
\subsection{Proof of Proposition \ref{prop:NPhard} }\label{section:NP_proof}

\input{paper_appendix/Z_Hardness_Proof}
\subsection{Proof of Lemma \ref{lemma:mobius_eigen}}\label{section:BM}
\input{paper_appendix/Z_BM_Lemma}
\newpage
\subsection{Proof of Proposition \ref{prop:convergence_of_FW}} 
\label{proof:IP_certificate}
\input{paper_appendix/Z_Proof_Bellman_Ford}

\newpage
\subsection{Proof of Proposition \ref{prop:repair_flow}}\label{proof:push_back}
\input{paper_appendix/Z_Flow_Projection_Error_Bound}

\subsection{Proof of Lemma \ref{lemma:laplacian}}
\input{paper_appendix/Z_lemma_laplacian_proof}

\subsection{Proof of Proposition \ref{prop:admm_convergence}}\label{proof:admm_convergence}
\input{paper_appendix/Z_Proof_ADMM_Convergence}

\subsection{Proof of Proposition \ref{prop:admm_certificate_overrelaxation}}\label{proof:admm_certificate_proof}
\input{paper_appendix/Z_proof_admm_certificate}

\subsection{Proof of Proposition \ref{prop:delete}}\label{proof:delete}
\input{paper_appendix/Z_Appendix_Delete_Rankings_Proof.tex}

\subsection{Proof of Proposition \ref{theorem:simplification}}
\input{paper_appendix/Z_Proof_Simplication}

\subsection{Proof of Theorem \ref{thm:network}}\label{proof:KS_network}
\input{paper_appendix/Z_KS_Network}


\newpage

\newpage



\bibliography{ref, ref_theory}
\bibliographystyle{plainnat}

\end{document}

%% file: paper_main/A_Introduction.tex
The random utility model (RUM) is a foundational framework for studying stochastic choices in economics \citep{block1960random, McFadden73}. A key question in this literature is whether observed choice probabilities are RUM-rationalizable, that is, whether they are consistent with aggregate choices generated by a population of agents with heterogeneous strict preference rankings. This question is important for at least two reasons. First, RUM underlies a wide range of empirical analyses, so its rejection has direct implications for applied work built upon it. Second, RUM provides a natural benchmark for testing stochastic rationality \citep{regenwetter2011transitivity,mccausland2020testing}, and its rejection can provide evidence that preferences alone are insufficient to explain choice behavior \citep{aguiar2023random, dean2022better}.

In this paper, we focus on the RUM in a discrete-choice setting, in which agents choose from a finite set of alternatives. We refer to this class of models as D-RUM. It is well known that the set of \DRUM-rationalizable choice
probabilities forms a polytope and that determining
rationalizability is equivalent to checking membership
in this polytope \citep{falmagne1978representation, federicobook}. A common approach to checking membership is to project observed choice probabilities onto the polytope, with the resulting distance serving as a measure of stochastic rationality and/or as an input into statistical testing procedures \citep{kitamura2018nonparametric}. We shall hereafter denote the vector of observed choice probabilities as $\widehat{p}$ and the \DRUM polytope as $\mathcal{P}_{\mathrm{RUM}}$.  The projection problem takes the form:\footnote{A formal definition is given in Section~\ref{section:computational_problem}.}
\begin{equation}\label{def:projection}
    \min_{p}\ \|\widehat{p}-p\|_2^2
    \quad \text{s.t.} \quad
    p \in \mathcal{P}_{\mathrm{RUM}}.
\end{equation}

This problem can be computationally challenging. Intuitively, the number of strict preference rankings grows factorially with the number of alternatives: with $n$ alternatives, there are $n!$ possible rankings.\footnote{More fundamentally, deciding membership in the \DRUM polytope is NP-hard.} Existing methods can be slow even for moderately large choice problems.
This computational burden is further compounded by the need to solve
the projection problem repeatedly in many statistical procedures,
such as the bootstrap.
As we show in our application section, with existing methods, applying a
commonly used bootstrap procedure to the 25-alternative pilot dataset
of \citet{dean2022better} takes 57 hours.\footnote{The final version of \cite{dean2022better} uses a different, smaller 
dataset, in large part because the 25-alternative dataset proved to be 
computationally challenging.}


The \DRUM polytope is often described by the convex hull of choice vectors generated by strict preference rankings. This is referred to as the vertex representation ($V$-representation) of the polytope. 
All existing computational approaches \citep{kitamura2018nonparametric,smeulders2021nonparametric} rely on the $V$-representation of the \DRUM polytope.\footnote{See Section \ref{section:existing_approach} for a detailed discussion.} It is well known that the \DRUM polytope also admits a half-space representation ($H$-representation)  based on the Block--Marschak (BM) inequalities \citep{block1960random,falmagne1978representation} and a network-flow representation \citep{fiorini2004short}. Our paper develops substantially faster algorithms, together with theoretical guarantees, by exploiting the alternative formulations.\footnote{The structures we identify in the two representations have applications beyond the quadratic optimization problem studied here. In particular, our flow-based column-generation algorithm extends readily to general differentiable convex objectives. We also expect the algebraic structures identified in this paper can be used to develop efficient algorithms for related linear programming problems.}  

As a headline number, our algorithm reduces the computation time for the application mentioned above from 57 hours to about 25 minutes, while preserving the same computational and statistical guarantees. This gain is a result of three complementary ingredients: alternative representations of the \DRUM polytope motivated by economic theory, optimization algorithms tailored to their special structures, and modern but accessible computing hardware, namely graphics processing units (GPUs). We now give a quick overview.

Developing efficient algorithms based on the two representations poses two difficulties.
First, the BM representation and flow representation are available only under relatively restrictive data conditions: either in full-menu settings, where choice frequencies are observed across all possible choice sets \citep{block1960random, falmagne1978representation}, or under specific structural assumptions on the menu collection \citep{suck2002binary,mcfadden2005revealed,kono2026}. Available datasets are usually incomplete and often do not have the particular structure assumed in the theory literature. Second, with \(n\) alternatives, these representations involve \(n2^{n-1}\) variables. Although smaller than the \(n!\) rankings in the \(V\)-representation, this number remains exponential in \(n\) and appears impractical for realistic applications.

The first difficulty can be resolved via a simple coordinate projection (lifting) argument. Let $d$ be the dimension of the ambient Euclidean space of $\mathcal{P}_{\mathrm{RUM}}$. Let ${\cal Q} \subset \mathbb{R}^{d+k}$ be a higher-dimensional \DRUM polytope on a richer domain (such as all nonempty subsets of the set of alternatives) that is characterized by the BM or flow-based representation. Then, it can be easily shown that $\mathcal{P}_{\mathrm{RUM}} = \{p \in \mathbb{R}^d : \exists\, \tilde{p} \in \mathbb{R}^k \text{ with } (p,\tilde{p}) \in {\cal Q}\}$, where $\tilde{p}$ denotes choice probabilities for unobserved menus.\footnote{Our approach is motivated by the theory of extended formulations \citep{yannakakis1988expressing,conforti2010extended}, in which a polytope is represented as the projection of a higher-dimensional polyhedron. \cite{davis2018extended} made a similar observation in the context of the flow representation.}
 Moreover, it can be shown that the optimization problem (\ref{def:projection}) over $\mathcal{P}_{\mathrm{RUM}} $ is equivalent to an optimization problem over $\cal Q$:  
\begin{equation}\label{def:projection2}
   \min_{(p,\tilde{p})}\ \|\widehat{p}-p\|_2^2
    \quad \text{s.t.} \quad
    (p,\tilde{p}) \in {\cal Q}.
\end{equation}
In this way, the optimization problem (\ref{def:projection}) over a low-dimensional set $\mathcal{P}_{\mathrm{RUM}}$  can be reformulated as the  optimization problem (\ref{def:projection2}) over the higher-dimensional polytope ${\cal Q}$,  which  is easier to solve thanks to the BM or flow-based formulation.


To address the second difficulty, we show that the computational burden of exponential-size formulations can be reduced by exploiting their algebraic and combinatorial structures. In the BM formulation with $n$ alternatives, the coefficient matrix $K_n$ of the BM polynomials, defined in Section~\ref{section:overview}, is a M\"obius matrix with a Kronecker product structure. Specifically, $K_n$ is block diagonal with $n$ identical blocks, each of the form
\begin{equation*}
    M^{\otimes(n-1)}
    =
    \underbrace{M \otimes M \otimes \cdots \otimes M}_{n-1\text{ factors}},
\end{equation*}
where $M$ is a $2\times2$ matrix. We show that this structure allows us to solve the linear systems arising in our algorithms in time\textit{ nearly linear} in their dimension.\footnote{The underlying idea is similar to that of the fast Fourier transform. We provide details in Section~\ref{section:bm_efficient}.}
This implies that we can avoid the quadratic cost of dense matrix--vector multiplication and the cubic cost of conventional dense factorization. This structure also allows implementations that avoid explicitly storing the Kronecker product matrices, substantially reducing memory requirements. For the flow formulation, we show that similar algebraic structures arise through the eigenstructure of the hypercube Laplacian. 

Building on these observations, we develop alternating direction method of multipliers (ADMM) algorithms based on both the BM and flow formulations. ADMM is an operator-splitting method that decomposes a complex constrained optimization problem into simpler subproblems that are solved iteratively.  We show that our ADMM algorithms exploit the algebraic structure above so that their linear-system steps can be implemented in a matrix-free manner with near-linear complexity. We refer to the resulting algorithms as BM-ADMM and Flow-ADMM.

We also develop a flow-based column-generation algorithm (Flow-CG). Flow-CG follows the same structure as the existing column-generation algorithm of \citet{smeulders2021nonparametric}. The algorithm alternates between a master problem, which solves the projection problem with a restricted set of preference types, and a pricing problem, which solves a linear optimization problem over the polytope to identify a new preference type. In the existing approach, the pricing problem is formulated as an integer program and constitutes the main computational bottleneck. We observe that a linear optimization problem over the \DRUM polytope can be reformulated as a linear optimization problem over the flow polytope of a power-set digraph. This reduces the pricing problem to a shortest-path problem on the graph. We exploit the graph's layered, acyclic structure to develop a tailored implementation of the Bellman--Ford algorithm, which efficiently computes a shortest path from the source node (the empty set) to the sink node (the full set of alternatives) on the power-set digraph.

In Section~\ref{section:consumption_budget}, we extend the flow representation result in the discrete-choice setting to the consumption-budget setting of \citet{kitamura2018nonparametric}. The key insight is that the KS model also has a network-flow representation but possibly involves \textit{multiple arcs} between the same pair of nodes. Practically, this implies that the Flow-CG algorithm can be adapted to the consumption-budget setting.\footnote{We provide a supplemental note that further clarifies the relationship between the consumption-budget setting and the discrete-choice setting \citep{chang2026random}.}

Finally, and importantly, Flow-CG, Flow-ADMM, and BM-ADMM are particularly well suited to GPU implementation because they involve large numbers of simple arithmetic operations which can be efficiently parallelized.  In contrast, existing approaches based on the $V$-representation are less amenable to such parallelization. All computations using our algorithms in this paper were performed on a workstation equipped with a single consumer-grade GPU, which is easily accessible. Our algorithms can also be deployed on data-center-grade GPUs, potentially delivering further performance gains.\footnote{For reference, the cheapest options for renting a single data-center-grade GPU, such as an NVIDIA A100, \href{https://www.devzero.io/instances/gpu/a100}{cost} between \$0.68 and \$3.60 per hour.}

Our numerical experiments highlight the complementary strengths of the new  algorithms. Flow-CG performs best when the underlying menu collection is sparse, such as with binary menus. BM-ADMM performs better when the menu collection is dense, for example, when at least \(1\%\) of all possible menus are observed. We show that the proposed methods deliver order-of-magnitude speedups over existing approaches. Flow-CG is often at least 100 times faster than the best existing method on binary menus. For dense menu collections, existing methods quickly become computationally infeasible.

We apply our algorithms to three datasets and show that they
substantially reduce the time required to test RUM. The first
is a 25-alternative pilot dataset from \citet{dean2022better},
who use tests of RUM to study choice overload. The best existing methods can take about 8 hours and 57 hours, depending on menu configurations, whereas our method takes 12 minutes and 25 minutes, respectively.\footnote{The speedup is reduced due to the fact that many bootstrap draws do not need to be solved to high accuracy.} We examine
how the test results change when fewer alternatives are
included. Specifically, we compare full-sample $p$-values
with those obtained from randomly sampled subsets, each
containing 12 nondefault alternatives and the default option.
The full-sample $p$-values lie at or below the first quartile
of the corresponding subsample distributions, suggesting
that including more alternatives can strengthen the evidence
against RUM in this application. The second dataset consists of choices generated by two
LLM-based AI systems, DeepSeek V4 Flash and GPT-6 Luna. Existing methods do not converge within 24 hours whereas our method takes about 2 minutes.
We find that both systems exhibit choice behavior that is
not stochastically rationalizable. Using distance to the
RUM polytope as a measure of departures from stochastic
rationality, we find smaller deviations for DeepSeek V4 Flash
than for GPT-6 Luna, and this difference is statistically
significant. Finally, we revisit the 25-period, three-good dataset studied by \cite{kitamura2018nonparametric} and \cite{smeulders2021nonparametric}. Our algorithm reduces computation time from 9.6 hours to 49 seconds, a roughly 700-fold speedup.

To further support our algorithms, we establish several theoretical results for the proposed algorithms and develop several extensions. First, we establish convergence guarantees and derive optimality certificates for our algorithms. In particular, we introduce a \textit{pushback} procedure that restores feasibility to intermediate ADMM iterates and can be applied to bootstrap samples for early rejection. This allows us to construct valid upper and lower bounds on the optimal objective value, yielding a principled stopping criterion. To our knowledge, the pushback procedure and its associated guarantees are novel.
Second, we show how the flow and BM representations can be modified to impose preference restrictions or expand the set of admissible choice types. The preference restrictions that we consider require one alternative to be ranked above another in every admissible ranking. We also explain how our algorithms can be adapted to these modifications.
Third, we introduce a graph-theoretic concept, which we call the \textit{co-occurrence graph}, and show how it can substantially simplify computations for several existing datasets.

We do, however, want to point out a limitation of our current approach. Because our algorithms require storing exponentially many variables, the largest number of alternatives they can handle is limited by GPU memory.\footnote{On a desktop with 32 GB of GPU memory, our implementations handle instances with up to 24 alternatives for BM-ADMM and Flow-ADMM and up to 35 alternatives for Flow-CG.} Existing methods based on integer programming do not face this memory constraint, but can be prohibitively slow: as we show numerically, the existing column-generation method can take more than seven days for binary-menu instances when $n=35$ and becomes infeasible for only slightly denser menu collections. Despite this memory limitation, our algorithms offer substantial computational gains for economically relevant problem sizes.\footnote{ We are aware of a hybrid approach that could relax the memory constraint at the cost of longer running times, but we leave this extension for future work.}

%% file: paper_main/A_road_map.tex
Section \ref{section:literature} discusses related literature.
Section \ref{section:computational_problem} formally introduces the computational problem. Section \ref{section:existing_approach} reviews existing approaches. 
Section \ref{section:overview} reviews the BM and flow representations and provides an overview of our approach. Section \ref{section:bm_admm} introduces the BM-ADMM algorithm based on the BM formulation and discusses the key algebraic properties of the BM coefficient matrix for efficient computation.
Section \ref{section:flow} introduces the flow-based column-generation (Flow-CG) and Flow-ADMM algorithms. Section~\ref{section:consumption_budget} presents the network-representation result in the consumption-budget setting.
Section \ref{section:numerical_evidence} presents numerical results, and Section \ref{section:applications} presents the choice-overload application.
Section \ref{section:certificate_guarantee} establishes global convergence guarantees and provides optimality certificates for our algorithms.
Section \ref{sec:additional_applications} presents two additional applications: testing the stochastic rationality of LLM-based AI systems, and testing RUM with the consumption-budget data.
Section \ref{section:Extension} presents results on incorporating additional restrictions on preferences and augmenting the set of admissible choice types, as well as results on the co-occurrence graph.
Proofs are collected in Section \ref{section:proofs}.
The extensions and proofs are available in the arXiv version of the paper.

%% file: paper_main/A_Related_Literature.tex
This paper contributes to the literature on the empirical testing of the RUM. Statistical formulations and inferential methods have been developed by \citet{regenwetter2011transitivity}, \citet{regenwetter2014qtest}, \citet{hoderlein2014revealed}, \citet{kitamura2018nonparametric}, and \citet{deb2023revealed}. The computational aspects of RUM testing have been studied by \citet{smeulders2015testing}, \citet{jagabathula2019limit}, and \citet{smeulders2021nonparametric}. Applications in economics include \citet{dean2022better, aguiar2023random, kashaev2024entangled}. As mentioned, all existing methods use the $V$-representation. Our contribution is to exploit alternative $H$- and flow representations and develop substantially faster algorithms.

Our work builds on \citet{kitamura2018nonparametric}, who develop a general nonparametric test of RUM rationalizability in the budget-set setting and solve the associated projection problem by exhaustively enumerating rational choice types. \citet{smeulders2021nonparametric} replace exhaustive enumeration with column generation, adding types iteratively by solving a pricing problem. Both approaches rely on the $V$-representation. \citet{davis2018extended}, \citet{turansick2025alternative}, and \citet{kono2026} advocate extended formulations based on flow and BM representations for computational problems
associated with the \DRUM model. They are motivated by
the smaller size of these formulations relative to a
fully enumerated vertex formulation. None of the papers develops algorithms tailored to exploit the structure or provides numerical evidence on performance. We develop algorithms that exploit the structure of the $H$- and flow representations and are well suited to modern hardware architectures. Our numerical experiments illustrate the value of these specialized algorithms: directly solving the projection problem with BM and flow formulations and general-purpose solvers does {\it not} yield comparable computational gains over existing column-generation methods, while our algorithms achieve substantial improvements in computational efficiency. \cite{shiraikoida} study the $H$-representation of the random utility polytope in the consumption-budget setting of \cite{kitamura2018nonparametric}. In contrast, we develop a network-flow representation with a focus on its computational properties.

Our algorithms build on the $H$-representation and the flow representation of the \DRUM polytope. Under the assumption that choice frequencies are observed for every
alternative from every choice set, \citet{falmagne1978representation}
obtained the first such characterization using the nonnegativity of
the Block--Marschak polynomials. \citet{block1960random} had
previously shown that this condition is necessary. \citet{fiorini2004short} later provided an alternative proof using network-flow theory. The assumption that all menus are observed has since been relaxed by \citet{mcfadden2005revealed}, \citet{suck2002binary}, and \citet{kono2026}. \citet{mcfadden2005revealed} and \citet{suck2002binary} show that the characterization remains valid for incomplete datasets provided the menu collection has a nested (upper-set) structure. More recently, \citet{kono2026} extend this line of work using the flow representation to obtain a characterization that accommodates a broader class of missing-data patterns. 


We use Frank--Wolfe (conditional gradient) methods, including column generation as a special case \citep{frank1956algorithm,holloway1974extension,braun2025conditional}, and the alternating direction method of multipliers (ADMM) \citep{gabay1976dual,glowinski1975approximation,eckstein1992douglas,neal2011distributed}. We build on work on GPU acceleration for linear and quadratic programming \citep{lu2025overview,huang2025restarted}, general-purpose ADMM \citep{schubiger2020gpuadmm}, and economic applications \citep{aldrich2011tapping,aldrich2014gpu}. Our contribution is to exploit the algebraic and network structure of the \DRUM projection problem to simplify computations and accelerate them on GPUs.

With binary menus, the pricing subproblem reduces to the classical linear ordering problem \citep{marti2011linear}. Computational studies of linear ordering often consider individual instances with hundreds of alternatives. In \DRUM projection, the challenge is to solve many moderately sized pricing instances exactly and quickly.\footnote{With column generation, each projection requires a sequence of pricing steps, and bootstrap inference requires the projection problem itself to be solved repeatedly. Consequently, a running time that is acceptable for a single instance can become prohibitive in our setting.} Flow-CG exploits the structure of the power-set digraph to solve these instances on GPUs. In our experiments, it solved the full projection problem at least 100 times as fast as the benchmark methods on challenging instances with $n\approx30$.

The distance from observed choice probabilities to the \DRUM polytope provides a natural measure of departure from stochastic rationality. In this respect, our paper relates to the literature on measures of rationality, including \citet{apesteguia2015measure}, \citet{ok2023measuring}, and \citet{carpentiere2023rational} in the discrete-choice setting for stochastic choice and \citet{echenique2011money} and \citet{dean2016measuring} in the consumption-budget setting for deterministic choice. 

We note that in this paper we use the numerical delta method of \citet{hong2018numerical} for bootstrap inference.\footnote{We thank Andres Santos for pointing us to this paper.} This procedure differs from the cone-tightening bootstrap of \citet{kitamura2018nonparametric}. Under the regularity conditions of \citet{hong2018numerical}, the convexity of the distance function yields a uniform asymptotic validity guarantee analogous to that of the cone-tightening approach. We choose the numerical delta method for ease of implementation: it requires repeatedly solving the same projection problem and can therefore be implemented directly using any of our proposed algorithms. See Section~\ref{sec:statistical} of the appendix for implementation details.





%% file: paper_main/A_Computation_Problem.tex
\subsection{Setup: Discrete-choice Random Utility Models}
Let \(X\) denote a finite set of alternatives. Without loss of generality, we identify
\(X\) with the set of integers \(\{1,2,\ldots,|X|\}\). Let \(2^X\) denote the power set of \(X\). Unless noted otherwise, we assume that $2^X$ excludes the empty set. For each menu
\(D\in 2^X\), let $|D|$ denote its cardinality and \(\Delta^D\) denote the unit simplex over \(D\):
\begin{equation}\label{eq:simplex}
\Delta^D
=\left\{p(\cdot,D)\in[0,1]^{|D|}:
\sum_{x\in D}p(x,D)=1\right\}.
\end{equation}

Let
\(\mathcal D \subseteq 2^X\) be an arbitrary collection of menus.  We define the set of \textbf{stochastic choice} functions on \(\mathcal D\) as
\begin{equation}\label{defn:sc}
    \mathcal P^{\mathcal D} \;=\; \prod_{D\in\mathcal D}\Delta^D,
\end{equation}
where \(\prod\) denotes the Cartesian product. For notational simplicity, we assume that choice probabilities are observed for every element in each choice set \(D \in \mathcal{D}\). All results and discussions in this paper extend straightforwardly to settings with partially observed choice probabilities. We index the coordinates of $\mathcal{P}^{\mathcal D}$ by the set $\mathcal{J}=\{(x,D): x\in D,\; D\in\mathcal{D}\}$. The ambient Euclidean space of $\mathcal{P}^{\mathcal D}$ is therefore $\mathbb{R}^{|\mathcal{J}|}$.

Given the finite set $X$, we consider the set $\Pi$ of all strict preferences on $X$. In particular, each element of $\Pi$ is identified with a one-to-one function $\pi:X\to\{1,...,|X|\}$. Given a $\pi\in \Pi$ and $x,y\in X$, we say $x$ \emph{is preferred over} $y$ if $\pi(x)>\pi(y)$. We write $\pi(x)\geq \pi(D)$ if $\pi(x)>\pi(y)$ for all $y\in D\backslash\{x\}$.

A \textbf{discrete-choice random utility model (\DRUM)} is defined as a probability distribution $v$ on $\Pi$ where $v:\Pi\to [0,1]$, $\sum_{\pi\in \Pi}v(\pi)=1$. Denote the set of such probability distributions as $\Delta(\Pi)$. Given $X$ and a set $\D$ of choice sets, a stochastic choice function $p\in \mathcal P^\D$ is \DRUM-\textbf{rationalizable} if there exists a $v\in \Delta(\Pi)$ such that for all $D\in \D$ and all $x\in D$, we have
\begin{equation*}
    p(x,D)= \sum_{\pi\in\Pi}v(\pi)1\{\pi(x)\geq \pi(D)\}.
\end{equation*}
We denote by $p^\pi$ the stochastic choice function induced by a deterministic ranking $\pi$, suppressing its dependence on $\mathcal{D}$ unless noted otherwise.

The set of rationalizable stochastic choice functions is a polytope given by the convex hull of the stochastic choice functions induced by all deterministic rankings. We denote the polytope as
\begin{equation}\label{eqn:v_representation}
\mathcal{P}_{\mathrm{RUM}}^{\D}
=
\mathrm{conv}\Big\{
p^{\pi} : \pi\in\Pi
\Big\},
\end{equation}
where for each $\pi\in \Pi$, $p^{\pi}(x,D)=1\{\pi(x)\ge \pi(D)\},\forall x\in D,\ D\in\D$. 
\subsection{\DRUM Projection Problem}
In this paper, we are concerned with the following quadratic projection problem. Given a collection of menus $\D$ and a vector 
$\widehat{p}\in \mathcal{P}^\D$, compute
\begin{equation}\label{eqn:DC-RUM_Proj}
    \min_{p\in \mathcal{P}^{\D}_{\mathrm{RUM}} } \|\widehat{p}-p\|_{\Omega}^2,  \qquad \text{where} \qquad\|\widehat{p}-p\|_{\Omega}^2= (\widehat{p}-p)^\top \Omega (\widehat{p}-p)
\end{equation}
and $\Omega$ is a positive semidefinite matrix of appropriate dimension. That is, we seek the projection of $\widehat p$ onto the polytope 
$\mathcal{P}^{\D}_{\mathrm{RUM}}$ under the quadratic seminorm induced by $\Omega$. We refer to this optimization problem as the \textbf{\DRUM Projection} problem. This problem is a convex quadratic program with linear constraints. 

Problem \eqref{eqn:DC-RUM_Proj} arises in econometrics as a key subproblem in testing the discrete-choice random utility model via a quasi-likelihood ratio test \citep{kitamura2018nonparametric,smeulders2021nonparametric,dean2022better}. In these applications, $\widehat{p}$ denotes an observed stochastic choice function, and $\Omega$ could be the identity matrix or an estimate of the variance--covariance matrix.

It is well known that the \DRUM Projection problem is NP-hard; we include a self-contained proof in Section~\ref{section:NP_statement}.

%% file: paper_main/A_Existing_Approach.tex
The standard approach to solving the \DRUM projection problem is the column generation (CG) method based on the $V$-representation \citep{smeulders2021nonparametric} (Algorithm~\ref{alg:CG}).\footnote{One can also enumerate all $n!$ deterministic preference rankings and solve the resulting optimization problem with a commercial solver \citep{kitamura2018nonparametric,dean2022better}. The factorial growth limits this approach to small problems. For example, with $n=13$ and only binary menus, storing all $13!$ rankings at one bit per pairwise comparison takes about $60.7$ GB.}
The algorithm
begins with a restricted set of rankings and iteratively adds new columns.
Each iteration requires solving a restricted master problem
(Algorithm~\ref{alg:master}), which is the same quadratic program as
\eqref{eqn:DC-RUM_Proj} but defined over a restricted subset of rankings, and
a \emph{pricing problem} (Algorithm~\ref{alg:pricing}), which corresponds to
linear optimization over the \DRUM polytope. It should be noted that the CG algorithm in this setting can be viewed as an instance of the fully corrective Frank--Wolfe
algorithm \citep{holloway1974extension,braun2025conditional}, from which its convergence and convergence-rate guarantees follow.

The pricing problem can be computationally challenging. It is NP-hard in general and 
can be formulated and solved as an integer programming (IP) problem. To alleviate 
this computational burden, \cite{smeulders2021nonparametric} propose solving the 
pricing problem heuristically. While this heuristic can accelerate the algorithm in practice,
the final verification step requires solving the pricing problem
to a prescribed optimality tolerance. If this verification relies only on a heuristic pricing solution
without a certified error bound, the algorithm may terminate
before achieving the prescribed accuracy. A practical strategy is therefore to use heuristics 
to generate candidate rankings and occasionally solve the pricing problem exactly 
to certify optimality. However, in practice this approach may still require frequent 
calls to the exact pricing routine, which can substantially increase total running time.\footnote{Due to space constraints, the integer programming formulation of the pricing problem 
(Algorithm~\ref{alg:IP_pricing}) and the heuristic pricing algorithm 
(Algorithm~\ref{alg:pricing_heuristic_random}) are deferred to Section~\ref{section:pricing_ip_heuristic}
of the online appendix. The heuristic is based on an adaptation of the random best-insertion 
procedure of \cite{marti2011linear}.} We refer to the column generation method that 
solves the pricing problem via integer programming as \textbf{IP-CG}, and the version using the 
heuristic as \textbf{IP-Heuristic-CG}.

\begin{algorithm}[htbp]
\caption{Column Generation for the \DRUM Projection Problem \eqref{eqn:DC-RUM_Proj}}
\label{alg:CG}
\begin{algorithmic}[1]
\State \textbf{Input:} choice probabilities $\widehat{p}$ and an initial restricted set of rankings $\emptyset \not=\Pi^r \subseteq \Pi$
\While{termination criterion is not met}
    \State Solve the restricted master problem: $\left(v_{\mathrm{opt}},p\right)
        \gets
        \texttt{DRUM\_RM}\left(\widehat{p},\hspace{2pt}\Pi^r\right)$
    \State Solve the pricing problem
        $\left(c,\pi\right)
\gets\texttt{DRUM\_Pricing}\left(\widehat{p},\hspace{2pt} p\right)$
    \State Update $\Pi^r \gets \Pi^r \cup \{\pi\}$
\EndWhile
\State \textbf{Return:} $v_{\mathrm{opt}}$ and $p$
\end{algorithmic}
\end{algorithm}
\begin{algorithm}[htbp]
\caption{\DRUM Projection Restricted Master Problem $\left(\texttt{DRUM\_RM}\right)$}
\label{alg:master}
\begin{algorithmic}[1]
\State \textbf{Input:} choice probabilities $\widehat p$ and a restricted set of rankings $\Pi^r\subset \Pi$
\State Solve $\min_{p\in \mathbb{R}^{|\mathcal{J}|},\lambda \ge 0} 
\;\|p-\widehat{p}\|_{\Omega}^2
\quad 
\text{s.t.} \quad 
p=\sum_{\pi\in\Pi^r}\lambda_\pi p^\pi,
\quad 
\sum_{\pi\in\Pi^r}\lambda_\pi =1$
\State \textbf{Output:} optimal value $v_{\mathrm{opt}}$ and solution $p$.
\end{algorithmic}
\end{algorithm}

\begin{algorithm}[htbp]
\caption{\DRUM Projection Pricing Problem $\left(\texttt{DRUM\_Pricing}\right)$}
\label{alg:pricing}
\begin{algorithmic}[1]
\State \textbf{Input:} choice probabilities $\widehat p$, current solution $p$
\State Exactly or heuristically solve
\begin{equation}\label{eqn:pricing}
\max_{\pi\in \Pi} \;
\left\langle
\widehat p-p,\, \Omega \left(p^\pi-p
\right)\right\rangle
\end{equation}
\State \textbf{Output:} computed objective value $c$ and ranking $\pi$
\end{algorithmic}
\end{algorithm}

%% file: paper_main/A_RUM_Characterization.tex
Besides the $V$-representation of the \DRUM polytope given in \eqref{eqn:v_representation}, two other representations of the \DRUM polytope are known in settings where choice probabilities are observed for all possible alternative--choice-set pairs. The half-space
($H$-) representation of the \DRUM polytope
\citep{block1960random,falmagne1978representation} consists of the
\textit{Block--Marschak (BM) inequalities}
\begin{equation}\label{eqn:BM_inequality}
  \sum_{\{E \subseteq X:\, D \subseteq E\}}
  (-1)^{|E \setminus D|} p(x,E)
  \geq 0,
  \qquad
  \forall\, D \in 2^X,\ \forall\, x \in D,
\end{equation}
together with the requirement that probabilities sum to one in each menu. Each Block--Marschak inequality is associated with a distinct
alternative--choice-set pair $(x,D)$. When there are $n$ alternatives, the
half-space representation consists of $n2^{n-1}$ distinct Block--Marschak
inequalities. We write these inequalities in matrix form as
\begin{equation}\label{def:bm_inequalities_matrix_form}
   K_n p \geq 0,
\end{equation}
where $K_n\in \mathbb{R}^{n2^{n-1}\times n2^{n-1}}$ and
$p\in \mathbb{R}^{n2^{n-1}}$. The ordering of the rows of $K_n$ and the
entries of $p$ is immaterial for now; we specify a particular ordering in
Section \ref{section:bm_efficient}, where it becomes important for exploiting
the structure of the matrix.

\subsection{Flow Representation of the \DRUM Polytope}\label{section:flow_rep} 

\citet{fiorini2004short} provides an alternative characterization
based on network flows. Specifically, it constructs a linear bijection between
choice probabilities in the \DRUM polytope and flows in the unit-flow polytope
of a digraph. Because we use this flow representation extensively below, we describe it here.

\begin{figure}[htbp]
\centering
\begin{tikzpicture}[
    x=0.78cm, y=0.78cm,
    every node/.style={font=\scriptsize},
    setnode/.style={circle,draw=black!55,fill=white,inner sep=0.6pt,minimum size=5mm},
    edge/.style={draw=gray!45,line width=0.35pt}
]

\node[setnode] (e)   at (0,0) {$\emptyset$};

\node[setnode] (1)   at (1.5, 1.35) {$1$};
\node[setnode] (2)   at (1.5, 0)    {$2$};
\node[setnode] (3)   at (1.5,-1.35) {$3$};

\node[setnode] (12)  at (3, 1.35) {$12$};
\node[setnode] (13)  at (3, 0)    {$13$};
\node[setnode] (23)  at (3,-1.35) {$23$};

\node[setnode,font=\tiny] (123) at (4.5,0) {$123$};

\draw[edge] (e)--(1);   \draw[edge] (e)--(2);   \draw[edge] (e)--(3);
\draw[edge] (1)--(12);  \draw[edge] (1)--(13);
\draw[edge] (2)--(12);  \draw[edge] (2)--(23);
\draw[edge] (3)--(13);  \draw[edge] (3)--(23);
\draw[edge] (12)--(123); \draw[edge] (13)--(123); \draw[edge] (23)--(123);

\end{tikzpicture}
\caption{Power-set digraph for $X=\{1,2,3\}$. All arcs run from left to right and add one element. Node labels list the elements of each set, so $12$ denotes $\{1,2\}$. Paths from $\emptyset$ to $X$ correspond to rankings.}\label{example:powerset_diagraph}
\end{figure}
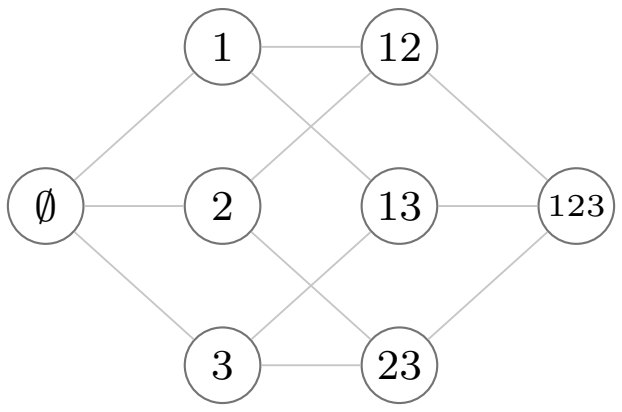

Let \(X\) be the set of $n$ alternatives, and consider a directed graph
\(G=(N,E)\), where the node set is the power set \(N=2^X\) (including the empty set) and the arc set is
\[
E=\{(D\setminus\{x\},D): x\in D,\; D\in 2^X\}.
\]
With \(n\) alternatives, the graph has \(2^n\) nodes and \(n2^{n-1}\) arcs.
Each arc corresponds to adding a single alternative to a subset. We refer to
this graph as the \emph{power-set digraph}.\footnote{Its underlying undirected
graph is the \emph{hypercube graph}. The \emph{\(n\)-dimensional hypercube
graph} (or \(n\)-cube), denoted \(Q_n\), is the graph whose vertices are all
binary vectors in \(\{0,1\}^n\), with an edge between two vertices if and only
if they differ in exactly one coordinate. This structure will have important implications for the Flow-ADMM algorithm
developed later.}
Figure~\ref{example:powerset_diagraph} illustrates the power-set digraph for
the case of three alternatives. We note that the power-set digraph has a natural layer structure, where the $k$th layer, $k\geq 0$, corresponds to all sets of cardinality $k$.

For each arc \(\left(D\setminus\{x\},D\right)\), let
\(f(D\setminus\{x\},D)\) denote the flow on that arc. A unit
\(\emptyset\)-\(X\) flow is defined as a nonnegative flow that sends one unit
of flow from the source node \(\emptyset\) to the sink node \(X\) and satisfies
flow conservation (i.e., total inflows equal total outflows) at all
intermediate nodes. The set of feasible flows forms the
\(\emptyset\)-\(X\) unit-flow polytope
\begin{equation}\label{eqn:flow_polytope}
\mathcal{F}_X=\{\, f\in\mathbb{R}^E: Af=b,\; f\ge0 \,\},
\end{equation}
where \(A\in\mathbb{R}^{2^n\times n2^{n-1}}\) is the node--arc incidence
matrix of \(G\), under the convention that each arc contributes \(-1\) at
its tail node and \(+1\) at its head node. The vector \(b\) has entry \(-1\)
at \(\emptyset\), entry \(1\) at \(X\), and zeros elsewhere. We provide an example with three alternatives, writing out the constraint $Af=b$ explicitly in Online Appendix \ref{section:power_set_digraph}.

\cite{fiorini2004short} establishes a bijective correspondence between unit flows and choice probabilities in the \DRUM polytope. Specifically, for
each flow \(f\in\mathcal{F}_X\), the corresponding choice probability vector
\(p\in\mathcal{P}_{\mathrm{RUM}}^{2^X}\) is given by
\begin{equation}\label{eqn:flow_to_p}
    p(x,D)
    =
    \sum_{\{D'\subseteq X:\,D\subseteq D'\}}
    f\left(D'\setminus\{x\},D'\right),
    \qquad
    \forall\, x\in D,\ D\in 2^X.
\end{equation}
Conversely, each choice probability vector \(p\) in the RUM polytope is
mapped to a flow \(f\in\mathcal{F}_X\) according to
\begin{equation}\label{eqn:p_to_flow}
    f\left(D\setminus\{x\},D\right)
    =
    \sum_{\{E\subseteq X:\,D\subseteq E\}}
    (-1)^{|E\setminus D|}p(x,E),
    \qquad
    \forall\, x\in D,\ D\in 2^X.
\end{equation}
Comparing \eqref{eqn:BM_inequality} and \eqref{eqn:p_to_flow}, we see that
the flow variables are exactly the values of the Block--Marschak
polynomials. We write the linear mapping in \eqref{eqn:flow_to_p} in matrix
form as
\begin{equation}\label{eqn:flow_to_p_matrix_form}
    p = Z_n f,
\end{equation}
where the ordering of the entries of \(p\) and \(f\) will be specified when
needed. Since \(K_n\) maps choice probabilities to their Block--Marschak
polynomials, it follows that 
\begin{equation}\label{eq:inverse}
Z_n = K_n^{-1}.
\end{equation}

%% file: paper_main/A_Extended_Formulation.tex
As noted in the introduction,  a central difficulty is that both representations are available only
under relatively restrictive menu configurations, while deriving
representations tailored to a particular collection of observed menus is often
challenging. For example, it is well known that the $H$-representation
of the linear ordering polytope---which corresponds to the \DRUM polytope when the observed menus consist of all binary choice sets---is difficult to characterize
\citep{fishburn1992induced,marti2011linear}.

To exploit these representations, we use the notion of (coordinate) \textit{projection}. Let
${\cal P}\subset\mathbb{R}^d$ be a low-dimensional polytope of interest, and suppose
there exists a higher-dimensional polytope ${\cal  Q}\subset\mathbb{R}^{d+k}$ such
that ${\cal  P}$ is the coordinate projection of ${\cal  Q}$. That is,
$p\in\mathcal P$ if and only if there exists $\tilde p\in\mathbb R^k$
such that $(p,\tilde p)\in\mathcal Q$. For any real-valued function $g$ on
$\mathcal P$, it follows that
$\inf\{g(p):p\in\mathcal P\}=\inf\{g(p):(p,\tilde p)\in\mathcal Q\}$.%
\footnote{For a formal proof, see Section 3.15 of \cite{ip}.}
In other words, optimization over the low-dimensional polytope ${\cal P}$ can be carried out equivalently over its higher-dimensional formulation ${\cal Q}$.

As mentioned in the introduction, we let ${\cal P}={\cal P}_{\mathrm{RUM}}^{\D}$ and ${\cal Q}={\cal P}_{\mathrm{RUM}}^{2^{X}}$ since the BM and flow representations are available on $2^X$ but not on $\D$ in general. Also by letting  $g(p)=\|p-\hat{p}\|^2_{\Omega}$, the original \DRUM projection  problem \eqref{eqn:DC-RUM_Proj} becomes equivalent to the following:
\begin{equation}\label{eqn:proj3}
   \min_{(p,\tilde{p})}\ \|\widehat{p}-p\|_{\Omega}^2
    \quad \text{s.t.} \quad
    (p,\tilde{p}) \in {\cal P}_{\mathrm{RUM}}^{2^{X}},
\end{equation}
where the set ${\cal P}_{\mathrm{RUM}}^{2^{X}}$ is characterized by the BM and flow representations as explained in the previous sections.

For choice models in general (not restricted to \DRUM), this result has the
trivial yet useful implication that even when tractable empirical restrictions
can only be derived under rich, and sometimes full, menu variation, these
restrictions can be carried over to incomplete-menu settings via a projection
argument. This is particularly useful for statistical testing, where observed
data are frequently available only for a limited collection of menus.


\subsection{Overview of Our Algorithms}\label{section:overview_algo}

Even with the projection approach, there is another  challenge:  the BM and flow-based formulations still involve an exponential number of variables---$n2^{n-1}$ when there are $n$ alternatives. Although this is much smaller than $n!$, which is the number of variables in $V$-representations, our approach  may still appear computationally impractical for problem sizes of economic interest. 

The remainder of the paper develops algorithms (BM-ADMM, Flow-CG and Flow-ADMM) that exploit the algebraic structure of the BM and flow formulations to address this challenge. With the projection approach, we solve (\ref{eqn:proj3}) by exploiting  the BM formulation for BM-ADMM or by using the flow representation  for Flow-CG and Flow-ADMM.

For BM-ADMM, we exploit Kronecker-product
structures in the BM transformation. In particular, in Lemma \ref{lemma:mobius_eigen} of Section \ref{section:bm_admm}, we obtain the spectral properties of the matrix $K_n^\top K_n$, where $K_n$ is the coefficient matrix of the BM inequalities defined by \eqref{def:bm_inequalities_matrix_form}. 
This result allows us to solve the relevant linear systems efficiently. For Flow-ADMM, we achieve similar computational gains by exploiting the spectral decomposition of the hypercube graph Laplacian.

For Flow-CG, in Section \ref{section:BellmanFord}, we show that the underlying directed-graph structure allows the pricing problem to be solved efficiently. In particular, we show that solving the pricing problem leads to solving 
the well-known {\it shortest-path problem} on a power-set digraph and it can be readily solved using the Bellman--Ford algorithm.
In Section~\ref{section:consumption_budget}, we provide a network-flow representation of RUM in the consumption-budget setting of \citet{kitamura2018nonparametric}. This result allows us to adapt the Flow-CG algorithm to testing RUM in the consumption-budget setting as well.

We will observe that these structural properties of the three algorithms can also be naturally exploited on modern parallel hardware, particularly GPUs.

%% file: paper_main/A_Notations_Roadmap.tex
Finally, we introduce notation that we will use to explain our algorithms in the rest of the paper.

We use $\mathbb{R}^d$ to denote the $d$-dimensional Euclidean space. For a positive integer $n$, we write $[n]=\{1,\ldots,n\}$. For any set $C$, its indicator function is defined by $\iota_C(x)=0$ if $x\in C$ and $\iota_C(x)=+\infty$ otherwise. We denote the $k\times k$ identity matrix by $I_k$. The distance from a point $x$ to a set $C$ is defined as $\operatorname{dist}(x,C)=\inf_{y\in C}\|x-y\|_2$.

For matrices $A=(a_{ij})\in\mathbb{R}^{m\times n}$ and $B\in\mathbb{R}^{p\times q}$, their Kronecker product is denoted by $A\otimes B\in\mathbb{R}^{mp\times nq}$ and defined by $
A\otimes B=\bigl[a_{ij}B\bigr]_{i\in [m],j\in [n]}$. We write either $\langle a,b\rangle$ or $a^\top b$ for the standard Euclidean inner product of vectors $a$ and $b$.

In the optimization problems below, we use boldface symbols to denote decision variables and regular font to denote coefficients and input data.

%% file: paper_main/A_BM_08232026.tex
\subsection{BM-ADMM}\label{sec:bm-admm}

Based on the Block--Marschak representation, the
\textit{Block--Marschak formulation} (BM formulation hereafter)
for the \DRUM projection problem \eqref{eqn:proj3}
with $n$ alternatives is
\begin{equation}\label{eqn:bm_formulation}
\min_{\mathbf{p}} \quad 
\|\mathbf{p}_{\mathcal{D}} - \widehat{p}\|_\Omega^2
\qquad
\text{s.t.} \quad 
K_n\mathbf{p} \ge 0, 
\quad 
\mathbf{p}^D \in \Delta^D \ \text{for all } D \in 2^X,
\end{equation}
where  $\mathbf{p}\in \mathbb{R}^{n2^{n-1}}$ is the decision variable, $\widehat{p}$ denotes the observed choice probabilities, $\mathbf{p}_{\mathcal{D}}$ denotes the restriction of $\mathbf{p}$ to the observed choice set collection $\mathcal{D}$, and $\mathbf{p}^D$ denotes the choice probability vector on a choice set $D$. $\Delta^D$ is the probability simplex associated with the choice set $D$, as defined in \eqref{eq:simplex}, and $K_n$ is the coefficient matrix of the Block--Marschak inequalities defined in \eqref{def:bm_inequalities_matrix_form}.

This section develops an alternating direction method of multipliers (ADMM) approach to the projection problem \eqref{eqn:DC-RUM_Proj}. Our main insight is that ADMM can exploit the Kronecker-product
structure of the BM formulation through computations similar
to those used in the fast Fourier transform, as explained in
Section~\ref{section:bm_efficient}. Consequently, despite the formulation's exponential size, the resulting algorithm remains computationally tractable and efficient for moderately large instances. This particular structural compatibility motivates our choice of ADMM over other widely used optimization methods, such as interior-point methods.\footnote{Convex optimization algorithms are often classified as first- or second-order methods. ADMM is a first-order method, as is column generation when interpreted as a Frank--Wolfe method. Although second-order methods such as interior-point methods typically offer stronger convergence guarantees, their computational demands make them poorly suited to the exponentially large formulations considered here. Applied generically to the BM or flow formulation, these methods require solving large linear systems whose coefficient matrices vary across iterations. Direct factorization is prohibitively expensive in both time and memory, while the changing coefficients make it difficult to exploit the special structure used by our algorithms. For example, when these linear systems are treated as dense,
a direct solve costs $O\left(n^3 2^{3(n-1)}\right)$ operations,
while an iterative method such as conjugate gradients costs
$O\left(Kn^2 2^{2(n-1)}\right)$ operations over $K$ iterations. In contrast, we show that the linear systems arising in our ADMM algorithms can be solved in $O\left(n^2 2^{n-1}\right)$ operations for either the BM or the flow formulation by exploiting the structure of each formulation. We have not found an analogous way to exploit these structures within an interior-point method.} 

We give a quick overview of the ADMM algorithm \citep{neal2011distributed}. ADMM is an operator-splitting method. It splits constraint sets and objective functions into subproblems that are easier to solve than the original problem. Specifically, the algorithm solves problems of the form 
\begin{equation}\label{eqn:admm_problem}
\begin{aligned}
\min_{\x\in\mathbb{R}^n,\y\in\mathbb{R}^m}\quad & f(\x)+g(\y) \qquad \text{s.t.}\quad & A\x+B\y=b
\end{aligned}
\end{equation}
where $A\in \mathbb{R}^{p\times n}$, $B\in \mathbb{R}^{p\times m}$, and $b\in \mathbb{R}^p$, and $f(\cdot)$ and $g(\cdot)$ are  closed, proper, and convex functions. The algorithm uses the augmented Lagrangian
\begin{equation*}
\mathcal L_\rho(\x,\y,\blambda)
=
f(\x)+g(\y)+\blambda^\top(A\x+B\y-b)
+\frac{\rho}{2}\|A\x+B\y-b\|_2^2 ,
\end{equation*}
where \(\rho>0\) is the penalty parameter and $\blambda\in \mathbb{R}^p$ is the dual variable. The first three terms constitute the standard Lagrangian, while the augmented quadratic term penalizes violations of primal feasibility constraints.

ADMM consists of iterations that alternately minimize the augmented Lagrangian with respect to $\x$ and $\y$, followed by an update of $\blambda$. For example, at iteration $k+1$, we solve
\begin{align}
\x^{k+1}
&=
\arg\min_{\x\in\mathbb{R}^n}\;
\mathcal{L}_\rho(\x,\y^k,\blambda^k), \label{eqn:x_step}\\
\y^{k+1}
&=
\arg\min_{\y\in\mathbb{R}^m}\;
\mathcal{L}_\rho(\x^{k+1},\y,\blambda^k),\label{eqn:y_step}\\
\blambda^{k+1}
&=
\blambda^k+\rho\big(A\x^{k+1}+B\y^{k+1}-b\big), \nonumber 
\end{align}
where $(\x^{k},\y^{k},\blambda^k)$ are the outputs at iteration $k$.
Each iteration first minimizes over \(\x\), with \(\y\) and \(\blambda\) fixed, then minimizes over \(\y\) with the updated \(\x\) and current \(\blambda\), and finally updates the dual variable using the current primal residual. A problem is suitable for ADMM if the two minimization problems in
\eqref{eqn:x_step} and \eqref{eqn:y_step} can be solved substantially more efficiently than the corresponding joint minimization over
$(\x,\y)$.

To transform the BM formulation \eqref{eqn:bm_formulation} into the form 
\eqref{eqn:admm_problem}, we introduce auxiliary variables and consider the 
equivalent reformulation
\begin{align}\label{eqn:bm_split}
\min_{\mathbf{p}, \overline{\mathbf{p}}, \mathbf{s}} \quad 
&\|\pcalD - \widehat{p}\|_{\Omega}^2
+ \iota_{\mathbb{R}_+^{n2^{n-1}}}(\mathbf{s})  + \sum_{D\in 2^X}\iota_{\Delta^D}\left(\pD\right)
\quad \text{s.t.} \quad 
\begin{cases}
\mathbf{p} - \overline{
\mathbf{p}} = 0, \\
K_n\mathbf{p} - \mathbf{s} = 0.
\end{cases}
\end{align}
Here, $\mathbf{p}$, $\overline{\mathbf{p}}$, and $\mathbf{s}$ are decision variables, 
each of dimension $n2^{n-1}$, where $n$ denotes the number of alternatives. 
The set $\mathbb{R}_+^{n2^{n-1}}$ denotes the nonnegative orthant of 
$\mathbb{R}^{n2^{n-1}}$, and $\iota_{\mathbb{R}_+^{n2^{n-1}}}$ denotes its 
indicator function as defined in Section \ref{section:notations_and_roadmap}. $\iota_{\Delta^D}$ is the indicator function of the probability simplex $\Delta^D$. Other decompositions are also possible, for example, by replacing 
$\pcalD$ with $\mathbf{p}_{\mathcal D}$ in the objective 
or by including the distance term with both variables. 

The corresponding augmented Lagrangian is
\begin{align*}
\mathcal{L}_\rho(\mathbf{p},\overline{\mathbf{p}},\mathbf{s},\mathbf{u},\mathbf{v})=  &\|\pcalD-\widehat{p}\|_{\Omega}^2
  + \iota_{\mathbb{R}_+^{n2^{n-1}}}(\mathbf{s})
  + \sum_{D\in 2^X}\iota_{\Delta^D}(\pD)
  \\
   &+ \langle \mathbf{u},\mathbf{p}-\overline{\mathbf{p}}\rangle
  + \langle \mathbf{v},K_n\mathbf{p}-\mathbf{s}\rangle 
  + \frac{\rho}{2}\|\mathbf{p}-\overline{\mathbf{p}}\|_2^2
  + \frac{\rho}{2}\|K_n\mathbf{p}-\mathbf{s}\|_2^2.
\end{align*}
A simplified ADMM algorithm based on the Block--Marschak formulation is detailed in Algorithm \ref{alg:admm_BM}. A more detailed implementation, including over-relaxation and constraint-specific penalty coefficients, is provided in online Appendix \ref{section:bm_with_relaxation}.

We note that Step 2 introduces a proximal term with weighting matrix $\Omega^{\textrm{prox}}
=\lambda_{\max}(\Omega)I-\Omega
\in \mathbb{R}^{|\mathcal{J}|\times |\mathcal{J}|}$ to accommodate a possibly nonidentity weighting matrix $\Omega$. This proximal term transforms the Step 2 subproblem into a standard evaluation of proximal operators associated with unit simplices. Such proximal regularization is standard in variants of ADMM and preserves convergence \citep{Ryu_Yin_2022}.

\begin{algorithm}[htbp]
\caption{BM-ADMM}
\label{alg:admm_BM}
\begin{algorithmic}[1]
\setlength{\abovedisplayskip}{4pt}
\setlength{\belowdisplayskip}{4pt}
\setlength{\abovedisplayshortskip}{3pt}
\setlength{\belowdisplayshortskip}{3pt}
\Require Initial values $(\mathbf{p}_0,\overline{\mathbf{p}}_0,\mathbf{s}_0,\mathbf{u}_0,\mathbf{v}_0)$, matrix $K_n$, observed probabilities $\widehat{p}$, penalty parameter $\rho$
\For{$t=0,1,...,$}      
    \State \textbf{Step 1:} Update $\mathbf{p}_{t+1}$ by solving
\begin{equation*}\label{eqn:update_p}
\min_{\mathbf{p}}
\Bigg\{
\langle \mathbf{u}_t,\mathbf{p}-\overline{\mathbf{p}}_{t} \rangle 
+ \langle \mathbf{v}_t, K_n\mathbf{p}-\mathbf{s}_t\rangle 
+ \frac{\rho}{2}\|\mathbf{p}-\overline{\mathbf{p}}_t\|_2^2 
+ \frac{\rho}{2}\|K_n\mathbf{p}-\mathbf{s}_t\|_2^2 
\Bigg\}.
\end{equation*}

    \State \textbf{Step 2:} Update $\mathbf{s}_{t+1}$ and $\overline{\mathbf{p}}_{t+1}$:
\begin{align*}
\mathbf{s}_{t+1}
=&
\arg\min_{\mathbf{s}}
\Big\{
\iota_{\mathbb{R}_+^{n2^{n-1}}}(\mathbf{s})
+ \langle \mathbf{v}_t, K_n\mathbf{p}_{t+1}-\mathbf{s}\rangle
+ \frac{\rho}{2}\|K_n\mathbf{p}_{t+1}-\mathbf{s}\|_2^2
\Big\}\\
=&
\Pi_{\mathbb{R}_+^{n2^{n-1}}}\!\left(
K_n\mathbf{p}_{t+1}+ \frac{1}{\rho}\mathbf{v}_t
\right).    
\end{align*}
  \begin{equation*}
    \overline{\mathbf{p}}_{t+1}=\arg\min_{\overline{\mathbf{p}}} \left\{
    \begin{split}
    &\|\pcalD-\widehat{p}\|^2_{\Omega}+\sum_{D\in 2^X}\iota_{\Delta^D}\left(\pD\right)\\
    &+ \frac{\rho}{2}\|\mathbf{p}_{t+1}-\overline{\mathbf{p}}\|^2_2+\|\pcalD-\overline{\mathbf{p}}_{\mathcal{D},t}\|^2_{\Omega^\textrm{prox}}+ \langle \mathbf{u}_t,\mathbf{p}_{t+1}-\overline{\mathbf{p}} \rangle      
    \end{split}
 \right\}.
  \end{equation*}
    \State \textbf{Step 3:} Update the dual variables:
\begin{equation*}
\mathbf{u}_{t+1}
=
\mathbf{u}_t+\rho\bigl(\mathbf{p}_{t+1}-\overline{\mathbf{p}}_{t+1}\bigr),
\qquad
\mathbf{v}_{t+1}
=
\mathbf{v}_t+\rho\bigl(K_n\mathbf{p}_{t+1}-\mathbf{s}_{t+1}\bigr).
\end{equation*}  
\EndFor
\end{algorithmic}
\end{algorithm}

The appeal of the ADMM framework, when coupled with the BM formulation, lies in the fact that the resulting subproblems can be solved efficiently with relatively low memory requirements. Both computational efficiency and memory usage are critical in our setting, as the underlying formulations are exponential in size. Each iteration of Algorithm~\ref{alg:admm_BM} performs two primal-update steps followed by a dual update. Step 2 is relatively straightforward: it involves enforcing the nonnegativity constraint and evaluating the proximal operators associated with the simplices $\Delta^D$. The evaluation of the simplex proximal operator is standard, and we provide the details in Section \ref{section:proximal_operator_simplex}. Importantly, these operations consist only of elementary arithmetic and are highly parallelizable, making them particularly well suited to GPU computation. The main computational challenge arises in Step 1, which requires solving a quadratic program with a large number of decision variables.

Our key observation is that Step 1 can be solved in almost-linear time—specifically, $O(d\log d)$, where $d$ is the dimension of the decision variables (for $n$ alternatives,  $d=n2^{n-1}$). This efficiency is achieved by exploiting the structure of the Block--Marschak coefficient matrix.

\subsection{Efficiently Solving Step 1}\label{section:bm_efficient}
In the remainder of this section, we describe the near-linear-time method for solving Step 1 in Algorithm \ref{alg:admm_BM}. We note that the first-order condition in Step  1 is 
\begin{align}\label{eqn:admm_bm_step1}
 \left(
\rho I
+ \rho K_n^\top K_n\right)\mathbf{p}
= b_t,
\end{align}
with   
$b_t=  - \mathbf{u}_t
- K_n^\top \mathbf{v}_t
+ \rho \overline{\mathbf{p}}_t
+ \rho K_n^\top \mathbf{s}_t$.

The linear system above can be solved efficiently by exploiting the structure of $K_n$. To illustrate this point, we first provide a Kronecker-product representation of the Block--Marschak coefficient matrix. Denote the M\"obius matrix as
  \begin{equation}\label{M2}
    M=\begin{bmatrix}
    1 & 0 \\
    -1 & 1 \\
\end{bmatrix}.
\end{equation}
The M\"obius matrix $M$ can be used to calculate the values of BM polynomials with two alternatives $\{1,2\}$.\footnote{For example, the values of the BM polynomials associated
with alternative $1$ are
\begin{equation}\label{eqn:bm_n2}
M
\begin{bmatrix}
    p(1,\{1,2\})\\
    p(1,\{1\})
\end{bmatrix}
=
\begin{bmatrix}
    p(1,\{1,2\})\\
    p(1,\{1\})-p(1,\{1,2\})
\end{bmatrix}.
\end{equation}} 
With $n$ alternatives, the values of the BM polynomials involving alternative $k$ can be shown to be 
\begin{equation}\label{eqn:bm_kronecker}
M^{\otimes (n-1)} p_{k}
= \left(\underbrace{M \otimes M \otimes \cdots \otimes M}_{n-1 \text{ times}}\right) p_{k},\footnote{We have included a proof in Section \ref{section:BM}.}
\end{equation}
where $p_{k}\in \mathbb{R}^{2^{n-1}}$ denotes the choice probability vector associated with the $k$th alternative, indexed according to the rule introduced in Section~\ref{section:index}. We note that, with $n$ alternatives, the Block--Marschak inequalities can be written as
\begin{equation*}
    \underbrace{\begin{bmatrix}
        M^{\otimes (n-1)} & 0 & 0 & \cdots & 0 \\
        0 & M^{\otimes (n-1)} & 0 & \cdots & 0 \\
        \vdots & \vdots & \vdots & \ddots & \vdots \\
        0 & 0 & 0 & \cdots & M^{\otimes (n-1)}
    \end{bmatrix}}_{K_n}
    \begin{bmatrix}
        p_1 \\ p_2 \\ \vdots \\ p_n
    \end{bmatrix}
    \geq 0.
\end{equation*}
In words, $K_n$ is a block-diagonal matrix with $n$ identical blocks 
$M^{\otimes(n-1)}$. Consequently, $K_n^\top K_n$ is also block diagonal, 
with $n$ identical blocks $\left(M^{\otimes(n-1)}\right)^\top M^{\otimes(n-1)}$.

For our purposes, we need to characterize the spectral properties of $K_n^\top K_n$. Standard algebraic properties of Kronecker products yield 
a convenient spectral decomposition of this matrix. We collect the 
relevant results in the following lemma.

\begin{lemma}\label{lemma:mobius_eigen}
Define $M$ as in (\ref{M2}). We have the identity
\begin{equation*}
    \left(M^{\otimes (n-1)}\right)^\top M^{\otimes (n-1)}= \left(\begin{bmatrix}
    2 & -1 \\
    -1 & 1 \\
\end{bmatrix}\right)^{\otimes (n-1)}
=  Q^{\otimes \left(n-1\right) }\Lambda ^{\otimes\left(n-1\right)}\left(Q^{\otimes\left(n-1\right)} \right)^\top,
\end{equation*}
where
\begin{equation*}
    Q=\begin{bmatrix}
        -\sqrt{\frac{3+\sqrt{5}}{5+\sqrt{5}}} & \sqrt{\frac{3-\sqrt{5}}{5-\sqrt{5}}}  \\
        \sqrt{\frac{2}{5+\sqrt{5}}} & \sqrt{\frac{2}{5-\sqrt{5}}}
    \end{bmatrix} \textrm{ and }     \Lambda =\begin{bmatrix}
    \frac{3+\sqrt{5}}{2} & 0 \\
    0 &  \frac{3-\sqrt{5}}{2} \\
    \end{bmatrix}.
\end{equation*}
Moreover, for any scalars $\alpha$ and $\beta$, the matrix $ \alpha I_{2^{n-1}}+\beta \left(M^{\otimes (n-1)}\right)^\top M^{\otimes (n-1)}$ has spectral decomposition
\begin{equation*}
Q^{\otimes \left(n-1\right)}\left( \alpha I_{2^{n-1}}+\beta  \Lambda ^{\otimes\left(n-1\right)}\right)
  \left(Q^{\otimes\left(n-1\right)} \right)^\top.
\end{equation*}
\end{lemma}
A proof is included in Section \ref{section:BM}. 

Lemma \ref{lemma:mobius_eigen} and the block structure of $K^\top_n K_n$ imply that
the linear system \eqref{eqn:admm_bm_step1} decomposes into $n$ independent
systems, one for each alternative. We partition $\mathbf{p}$ and $b_t$ into blocks $\mathbf{p}_k$ and
$b_{k,t}$, respectively, which correspond to the $k$th alternative. Using Lemma
\ref{lemma:mobius_eigen}, the linear system can be written as
\begin{equation*}
\left[
\begin{array}{cccc}
Q^{\otimes(n-1)} D \left(Q^{\otimes(n-1)}\right)^\top
&  \cdots & 0 \\
\vdots &  \ddots & \vdots \\
0 &  \cdots &
Q^{\otimes(n-1)} D \left(Q^{\otimes(n-1)}\right)^\top
\end{array}
\right]
\left[
\begin{array}{c}
\mathbf{p}_{1} \\
\vdots \\
\mathbf{p}_{n}
\end{array}
\right]
=
\left[
\begin{array}{c}
b_{1,t} \\
\vdots \\
b_{n,t}
\end{array}
\right],
\end{equation*}
where $D
=
\rho I_{2^{n-1}}
+\rho\Lambda^{\otimes(n-1)}$.
Thus, \eqref{eqn:admm_bm_step1} reduces to solving
$n$ linear systems, each of which has a solution of the form
\begin{equation}\label{eqn:qvq}
  \mathbf{p}_k
  =\rho^{-1}Q^{\otimes(n-1)}
  \left(I_{2^{n-1}}+\Lambda^{\otimes(n-1)}\right)^{-1}
  \left(Q^{\otimes(n-1)}\right)^\top b_{k,t},\quad\forall k\in [n].
\end{equation}
Because the middle matrix is diagonal and $\left(Q^{\otimes(n-1)}\right)^\top=\left(Q^\top\right)^{\otimes (n-1)}$, solving a block amounts to applying
$\left(Q^\top\right)^{\otimes (n-1)}$, scaling each coordinate by the
reciprocal of the corresponding eigenvalue, and then applying
$Q^{\otimes(n-1)}$.

A key computational property we exploit is that multiplication by
$Q^{\otimes(n-1)}$ or $\left(Q^\top\right)^{\otimes(n-1)}$ can be
carried out efficiently using the \textit{butterfly operation} (e.g. see Section 30.3 of \cite{cormen2022introduction}). The butterfly operation computes products of the form
$U^{\otimes N}v$, where $U$ is an arbitrary $2$-by-$2$ matrix, $N$ is a positive integer and $v\in\mathbb{R}^{2^N}$ is an arbitrary vector.\footnote{When $U=\left(1 , 1 ; 1 , -1\right)$, the butterfly operation is referred to as the fast Walsh--Hadamard transform \citep{o2014analysis}.} A template is shown in Algorithm \ref{alg:bo}. The butterfly operation offers two primary advantages. First, for any vector \(v\), the product \(U^{\otimes N} v\) can be computed in \(O(N2^N)\) time, whereas a naive multiplication with a dense \(2^N \times 2^N\) matrix requires \(O(4^N)\) operations.\footnote{To see this, let $T(N)$ denote the number of arithmetic
operations required to compute $U^{\otimes N}v$ using
the butterfly operation. The algorithm recursively calls two butterfly operations with $U^{\otimes (N-1)}$ and performs four scalar--vector multiplications and two vector additions, each involving vectors of length $2^{N-1}$. Hence, $T(N) \le 2T(N-1) + \left(4 \cdot 2^{N-1} + 2 \cdot 2^{N-1}\right)
    = 2T(N-1) + 3 \cdot 2^N.$ 
It can then be shown by induction that $T(N)\le C N2^N$ for a sufficiently large constant $C$.} Second, the method is memory efficient, since the matrix \(U^{\otimes N}\) never needs to be stored explicitly.
\begin{algorithm}[h]
\caption{$\texttt{Butterfly}(v,N)$: compute $z = U^{\otimes N} v$; $U=(a,b;c,d)$}
\label{alg:bo}
\begin{algorithmic}[1]
\Require $v\in \mathbb{R}^{2^N}$, $N\in \mathbb{N}$, $N\ge 1$
\If{$N=1$}
    \State \Return $\left[\;(a\times v[1]+b\times v[2])\;;\;(c\times v[1]+d\times v[2])\;\right]$
\Else
    \State $v_1 \gets \texttt{Butterfly}\!\left(v[1:2^{N-1}],\,N-1\right)$
    \State $v_2 \gets \texttt{Butterfly}\!\left(v[(2^{N-1}+1):2^N],\,N-1\right)$
    \State \Return $\left[\;(a\times v_1+b\times v_2)\;;\; (c\times v_1+d\times v_2)\;\right]$
\EndIf
\end{algorithmic}
\end{algorithm}

We now use the butterfly operation to characterize the complexity of solving Step 1 as in \eqref{eqn:admm_bm_step1}. There are two computational components: forming the right-hand side
$b_t$ in \eqref{eqn:admm_bm_step1} and evaluating the block solutions in
\eqref{eqn:qvq}. Forming $b_t$ is dominated by computing
$K_n^\top \mathbf{v}_t$ and $K_n^\top \mathbf{s}_t$. Because $K_n$ consists of $n$
blocks of $M^{\otimes(n-1)}$, each multiplication can be carried out using
the same butterfly operation in $O\bigl(n^2 2^{n-1}\bigr)$ time: each of the
$n$ blocks requires $O\bigl((n-1)2^{n-1}\bigr)$ operations. Evaluating
\eqref{eqn:qvq} has the same $O\bigl(n^2 2^{n-1}\bigr)$ complexity: for each of the $n$
blocks, it requires two butterfly applications, together with a diagonal
scaling, and hence costs $O\bigl((n-1)2^{n-1}\bigr)$ operations per block.
Therefore, \eqref{eqn:admm_bm_step1} can be solved in
$O\bigl(n^2 2^{n-1}\bigr)$ time. Moreover, none of the matrices involved need
to be formed explicitly, so the procedure requires only
$O\bigl(n2^{n-1}\bigr)$ memory.

For comparison, and to illustrate the computational advantage of the
butterfly operation, let $d=n2^{n-1}$ denote the dimension of the coefficient
matrix in \eqref{eqn:admm_bm_step1}. A generic dense direct solve, such as
Cholesky factorization followed by triangular solves, requires $O(d^3)$ time
and $O(d^2)$ memory. Each subsequent solve
for a new right-hand side requires $O(d^2)$ operations. An iterative method, such as conjugate gradient, requires
$O(d\log d)$ operations per iteration by using the fast matrix--vector products described above. By contrast, our approach computes the
exact solution in $O(d\log d)$ time, which is comparable
to the cost of a single conjugate-gradient iteration, and requires only $O(d)$ memory. We establish a convergence guarantee and provide an optimality certificate for BM-ADMM in Proposition \ref{prop:admm_certificate}.

\medskip

%% file: paper_main/A_Flow_Formulation.tex
In this section, we apply the flow representation introduced
in Section~\ref{section:flow_rep} to column generation and ADMM. The \textit{flow formulation} with $n$ alternatives is
\begin{equation}\label{eqn:flow_formulation}
\min_{\mathbf f,\mathbf p}\quad 
\|\mathbf p_{\mathcal{D}}-\widehat{p}\|_\Omega^2
\qquad
\text{s.t.}\quad
\mathbf p=Z_n\mathbf f,
\qquad
A\mathbf f=b,
\qquad
\mathbf f\ge 0.
\end{equation}
Here, $\mathbf p,\mathbf f\in\mathbb{R}^{n2^{n-1}}$ are decision variables, $\widehat{p}$ denotes the observed choice probabilities, and $\mathbf p_{\mathcal{D}}$ denotes the restriction of $\mathbf p$ to the collection of observed choice sets $\mathcal{D}$. The linear mapping $\mathbf p=Z_n\mathbf f$, defined in \eqref{eqn:flow_to_p_matrix_form}, maps the flow variables to the corresponding choice probabilities, while the constraints
$A\mathbf f=b$ and $\mathbf f\ge 0$ characterize the flow polytope defined in \eqref{eqn:flow_polytope}.

%% file: paper_main/A_Bellman_Ford.tex
We observe that the pricing problem arising in the column-generation procedure can be formulated as a linear optimization problem over the unit-flow polytope. Replacing the integer programming pricing subroutine with a flow-based pricing subroutine gives the Flow-CG algorithm. We note that the Flow-CG algorithm extends directly to arbitrary differentiable convex objective functions, including likelihood-based criteria.

To see this, let $\mathcal{R}$ denote the set of all $\emptyset$--$X$ paths in the power-set digraph $G$, and let $\mathcal{F}_X$ denote the unit $\emptyset$--$X$ flow polytope defined in \eqref{eqn:flow_polytope}. The pricing problem \eqref{eqn:pricing} can be reformulated as
\begin{align}
\max_{\pi\in \Pi} \;
\left\langle
\widehat p-p,\, \Omega \left(p^\pi_\D-p\right)
\right\rangle
&\;\Leftrightarrow\;
\max_{\widetilde{p}\in \mathcal{P}^{\mathcal{D}}_{\textrm{RUM}}} \;
\left\langle
\widehat p-p,\, \Omega \left(\widetilde{p}-p\right)
\right\rangle
\label{eqn:vtop1}\\
&\;\Leftrightarrow\;
\max_{\widetilde{p},f}\;
\left\langle
\widehat p-p,\, \Omega \left(\widetilde{p}_{\mathcal D}-p\right)
\right\rangle
\quad
\text{s.t.}\quad
\widetilde{p}=Z_n f,\quad f\in\mathcal{F}_X.
\label{cg_flow2}
\end{align}

Writing $\widetilde{p}_{\mathcal D}=Z_{\mathcal D}f$, where $Z_{\mathcal D}$ is the submatrix of $Z_n$ consisting of the rows that map $f$ to the choice probabilities associated with the observed choice-set collection $\mathcal D$, we obtain
\begin{align}
&\eqref{cg_flow2}
\;\Leftrightarrow\;
\max_{f\in\mathcal{F}_X}
\left\langle
\widehat p-p,\,
\Omega\left(Z_{\mathcal D}f-p\right)
\right\rangle
\Leftrightarrow\;
\max_{f\in\mathcal{F}_X}
\left\langle
\widehat p-p,\,
\Omega Z_{\mathcal D}f
\right\rangle
-
\left\langle
\widehat p-p,\,
\Omega p
\right\rangle
\nonumber\\
&\;\Leftrightarrow\;
\max_{f\in\mathcal{F}_X}
\left\langle
f,\,
Z_{\mathcal D}^{\top}\Omega^\top
\left(\widehat p-p\right)
\right\rangle
-
\left\langle
\widehat p-p,\,
\Omega p
\right\rangle
\Leftrightarrow\;
\max_{r\in\mathcal{R}}
\left\langle
r,\,
Z_{\mathcal D}^{\top}\Omega^\top
\left(\widehat p-p\right)
\right\rangle
-
\left\langle
\widehat p-p,\,
\Omega p
\right\rangle .
\label{eqn:vtop2}
\end{align}

The first equivalence in \eqref{eqn:vtop1} follows because
$\mathcal{P}^{\mathcal D}_{\textrm{RUM}}$ is the convex hull of the deterministic choice vectors $p^\pi_\D$, and a linear objective function attains its maximum over a polytope at a vertex. Similarly, the final equivalence in \eqref{eqn:vtop2} follows because the vertices of the unit-flow polytope $\mathcal{F}_X$ are precisely the incidence vectors of $\emptyset$--$X$ paths in the power-set digraph.\footnote{This follows from the total unimodularity of the node--arc incidence matrix $A$ together with the integrality of $\mathbf b$. See Section \ref{section:node_arc_incidence_matrix} for a reference.}

The problem $\max_{r\in\mathcal{R}}
\left\langle r,\,
Z_{\D}^{\top}\Omega^\top \left(\widehat{p}-p\right)\right\rangle$ is the longest path problem on a power-set digraph and it can be readily solved using the Bellman--Ford algorithm with complexity $O(n^22^{n-1})$ when the weights are given.\footnote{There are $n2^{n-1}$ arcs and the maximum length of a path is $n$ in a power-set digraph with $n$ alternatives. The Bellman--Ford algorithm has complexity $O(k|E|)$, where $k$ is the maximum number of arcs in a directed
path and $|E|$ is the number of arcs \citep{schrijver2003combinatorial}.} For completeness, we include a version of the Bellman--Ford algorithm (\texttt{Flow\_Pricing}) specialized for our power-set digraph below in Algorithm \ref{alg:pricing_rolling_path}. We refer to the column generation algorithm that uses the \texttt{Flow\_Pricing} algorithm in the pricing step as the \textbf{Flow-CG} algorithm. We note that the sum
$\sum_{\substack{D'\in\mathcal D, x\in D'\subseteq D}}s(x,D')$
is an explicit expression for the component of
$Z_{\mathcal D}^{\top}\Omega(\widehat p-p^*)$
corresponding to the arc $(D\setminus\{x\},D)$.

\begin{algorithm}[h]
\caption{Bellman--Ford on a power-set digraph (\texttt{Flow\_Pricing})}
\label{alg:pricing_rolling_path}
\begin{algorithmic}[1]
\State \textbf{Input:} choice probabilities $\widehat p$, current solution $p^*$
\State \textbf{Output:} maximum path weight
$c_{\mathrm{opt}}$ and an optimal ranking $\pi$
\State Calculate price vector $s$, where $s(x,D)=\widehat{p}(x,D)-p^*(x,D), \forall x\in D, D\in \mathcal{D}$.
\State Update $s\gets\Omega s$.
\State $\texttt{d}(\emptyset)\gets 0$
\State Initialize $\texttt{pred}(D)$ for all $D\subseteq X$
\For{$k=1,\dots,|X|$} \Comment{$k$ denotes the cardinality layer}
    \State Initialize $\texttt{d}(D)$ for all $D\subseteq X$ with $|D|=k$
    \begin{align}
    \texttt{d}(D)&=
    \max_{x\in D}
    \left\{
    \texttt{d}(D\setminus\{x\})
    +
    \sum_{x\in D',\, D'\subseteq D,\, D'\in\mathcal D}
    s(x,D')
    \right\}\label{eqn:relaxation1}\\
    \texttt{pred}(D)&=
    \arg\max_{x\in D}
    \left\{
    \texttt{d}(D\setminus\{x\})
    +
    \sum_{x\in D',\, D'\subseteq D,\, D'\in\mathcal D}
    s(x,D')
    \right\}\label{eqn:relaxation2}
    \end{align}
\EndFor
\State $c_{\textrm{opt}}\gets \texttt{d}(X)$; $D\gets X$
\For{$k=n,n-1,\ldots,1$}\Comment{Return the optimal ranking}
    \State $x\gets\texttt{pred}(D)$; $\pi(x)\gets k$; $D\gets D\setminus\{x\}$
\EndFor
\State \Return $c_{\mathrm{opt}}$ and $\pi$
\end{algorithmic}
\end{algorithm}

\begin{algorithm}[h]
\caption{Flow-CG for the \DRUM Projection Problem \eqref{eqn:DC-RUM_Proj}}
\label{alg:Flow_CG}
\begin{algorithmic}[1]
\State \textbf{Input:} choice probabilities $\widehat{p}$ and an initial nonempty set of rankings $\Pi^r \subseteq \Pi$
\While{termination criterion is not met}
    \State Solve the restricted master problem $\left(v_{\mathrm{opt}},p\right)
        \gets\texttt{DRUM\_RM}\left(\widehat{p},\hspace{2pt}\Pi^r\right)$.
    \State Solve the pricing problem $\left(c_{\mathrm{opt}},\pi\right)\gets\texttt{Flow\_Pricing}\left(\widehat{p},\hspace{2pt} p\right)$.
    \State Update $\Pi^r \gets \Pi^r \cup \{\pi\}$
\EndWhile
\State \textbf{Return:} $v_{\mathrm{opt}}$ and $p$
\end{algorithmic}
\end{algorithm}

A key advantage of the \texttt{Flow\_Pricing} routine is that its computational structure is particularly amenable to GPU implementation. At layer $k$, the algorithm solves $\binom{n}{k}$
subproblems of the form \eqref{eqn:relaxation1}.
For example, when $n=30$ and $k=15$, there are
$\binom{30}{15}=155{,}117{,}520$ such subproblems. The number of subproblems is large. However, these subproblems can be solved independently using only simple arithmetic operations, primarily addition and maximization. This structure is well suited to GPUs, which can execute a large number of lightweight and homogeneous operations in parallel. Moreover, arc weights can be computed on the fly, avoiding the need to store all $n2^{n-1}$ arc weights. Figure~\ref{fig:flow_pricing} illustrates the algorithm on the power-set digraph for three alternatives. The algorithm proceeds layer by layer: once the values in one layer have been computed, the values at all nodes in the next layer can be computed independently and in parallel.

By contrast, the pricing problem in IP-CG is typically solved using branch-and-cut, in which the subproblems involve solving linear programs and other heterogeneous operations. Such computations are substantially less amenable to large-scale GPU parallelization. In practical terms, for $n=35$, flow-based pricing takes about 7 seconds per run, compared with approximately 2,067 seconds for the IP method—a nearly 300-fold speedup.
 
We establish a convergence guarantee and provide an optimality certificate for Flow-CG in Proposition \ref{prop:convergence_of_FW}.

\input{paper_main/A_Flow_Pricing_Tikz}

The flow representation also supports Flow-ADMM (Supplemental Material, Section~\ref{section:ADMM}).
Flow-CG applies whenever the RUM admits a network representation. We derive one next for the consumption-budget setting of \citet{kitamura2018nonparametric}.

%% file: paper_main/A_Flow_Pricing_Tikz.tex

\begin{figure}[htbp]
\centering

\captionsetup[subfigure]{
    font=small,
    skip=4pt,
    justification=centering
}

\tikzset{
    pricinggraph/.style={
        x=0.78cm,
        y=0.78cm,
        every node/.style={font=\scriptsize}
    },
    pricingnode/.style={
        circle,
        draw,
        inner sep=0.6pt,
        minimum size=5mm
    },
    pricingpending/.style={
        draw=black!55,
        fill=white
    },
    pricingdone/.style={
        draw=black!55,
        fill=black!16
    },
    pricingactive/.style={
        draw=blue!65!black,
        fill=blue!18,
        line width=0.6pt
    },
    pricingbase/.style={
        draw=gray!45,
        line width=0.35pt,
        -
    },
    pricingused/.style={
        draw=blue!65!black,
        line width=0.7pt,
        -
    }
}

\newcommand{\flowpricingpanel}[2]{%
    \begin{subfigure}[t]{0.31\linewidth}
        \centering
        \begin{adjustbox}{max width=\linewidth}
            \begin{tikzpicture}[pricinggraph]

                \tikzset{
                    layerone/.style={pricingpending},
                    layertwo/.style={pricingpending},
                    layerthree/.style={pricingpending}
                }

                \ifnum#1=1\relax
                    \tikzset{
                        layerone/.style={pricingactive}
                    }
                \fi
                \ifnum#1=2\relax
                    \tikzset{
                        layerone/.style={pricingdone},
                        layertwo/.style={pricingactive}
                    }
                \fi
                \ifnum#1=3\relax
                    \tikzset{
                        layerone/.style={pricingdone},
                        layertwo/.style={pricingdone},
                        layerthree/.style={pricingactive}
                    }
                \fi

                \node[pricingnode,pricingdone]
                    (e) at (0,0) {$\emptyset$};

                \node[pricingnode,layerone]
                    (1) at (1.5, 1.35) {$1$};
                \node[pricingnode,layerone]
                    (2) at (1.5, 0) {$2$};
                \node[pricingnode,layerone]
                    (3) at (1.5,-1.35) {$3$};

                \node[pricingnode,layertwo]
                    (12) at (3, 1.35) {$12$};
                \node[pricingnode,layertwo]
                    (13) at (3, 0) {$13$};
                \node[pricingnode,layertwo]
                    (23) at (3,-1.35) {$23$};

                \node[pricingnode,layerthree,font=\tiny]
                    (123) at (4.5,0) {$123$};

                \foreach \u/\v in {
                    e/1,e/2,e/3,
                    1/12,1/13,2/12,2/23,3/13,3/23,
                    12/123,13/123,23/123%
                }{
                    \draw[pricingbase] (\u)--(\v);
                }

                \ifnum#1=1\relax
                    \foreach \v in {1,2,3}{
                        \draw[pricingused] (e)--(\v);
                    }
                \fi
                \ifnum#1=2\relax
                    \foreach \u/\v in {
                        1/12,1/13,2/12,2/23,3/13,3/23%
                    }{
                        \draw[pricingused] (\u)--(\v);
                    }
                \fi
                \ifnum#1=3\relax
                    \foreach \u in {12,13,23}{
                        \draw[pricingused] (\u)--(123);
                    }
                \fi

            \end{tikzpicture}
        \end{adjustbox}
        \caption{#2}
    \end{subfigure}%
}

\flowpricingpanel{1}{Step 1: singletons}%
\hfill
\flowpricingpanel{2}{Step 2: pairs}%
\hfill
\flowpricingpanel{3}{Step 3: full set}

\par\smallskip

{\footnotesize
\tikz[baseline=-.55ex]
    \node[pricingnode,pricingdone,minimum size=3.5mm]{};
\ Completed
\qquad
\tikz[baseline=-.55ex]
    \node[pricingnode,pricingactive,minimum size=3.5mm]{};
\ Processing in parallel
\qquad
\tikz[baseline=-.55ex]
    \node[pricingnode,pricingpending,minimum size=3.5mm]{};
\ Pending
}

\caption{%
Parallel layer updates in \texttt{Flow\_Pricing} for $n=3$.
The value at $\emptyset$ is initialized before Step~1.
Blue arcs mark predecessor contributions. All arcs run left to right.
The label $12$ denotes $\{1,2\}$.
}
\label{fig:flow_pricing}
\end{figure}

%% file: paper_main/A_BudgetSet.tex
\subsection{Budgets and Patches in KS}\label{section:KS}
The material in this section is adapted from Section~3 of \citet{kitamura2018nonparametric} (hereafter KS).
The primitive data in KS consist of choice frequencies observed on $J$ linear budgets, each of the form $\B_j = \{y \in \mathbb{R}^K_{+}: q_j^\top y = 1\}$,
where $K$ denotes the number of commodities and $q_j$ is the normalized price vector.\footnote{The original price vector is divided by the corresponding expenditure level, thereby normalizing expenditure to one.} It is assumed that the price vectors $\{q_j\}_{j=1}^J$ are strictly positive and distinct so that they induce different budget sets. 

KS partition the union of the budget sets into \textit{patches}. Each budget is divided into patches according to whether its points lie on, strictly above, or strictly below each of the other budget planes.

\begin{definition}[(Definition~3.2 in KS)]\label{def:KS_3.2} Let $\mathcal{X}=\{x_1,\ldots,x_I\}$ be the coarsest partition of $\bigcup_{j=1}^J \B_j$ such that, for every $i\in\{1,\ldots,I\}$ and $j\in\{1,\ldots,J\}$, the patch $x_i$ lies entirely on, strictly above, or strictly below the budget plane $\B_j$. (Here, $I$ denotes the number of patches in the coarsest
partition, that is, the smallest number of patches
satisfying this condition.) Equivalently, two points $y_1,y_2\in\bigcup_{j=1}^J \B_j$ belong to the same patch if and only if $\operatorname{sg}(q_j^\top y_1-1)=\operatorname{sg}(q_j^\top y_2-1),
    \forall j\in [J]$, 
where $\operatorname{sg}$ denotes the sign function.
\end{definition}
Each budget $\B_j$ is a disjoint union of $I_j$ patches, which we enumerate as $x_{1|j},\ldots,x_{I_j|j}$. The data relevant for rationalizability in the KS sense are the probabilities of selecting each patch on each budget.

\begin{definition}[(Definition~3.4 in KS)]\label{def:cp_system}
The vector representation of a choice probability system is $p :=
    \left(
        p_{1|1},\ldots,p_{I_1|1},
        \ldots,
        p_{1|J},\ldots,p_{I_J|J}
    \right)^\top$,
where $p_{i|j}$ is the probability that the bundle chosen from budget $\B_j$ belongs to patch $x_{i|j}$. 
\end{definition}

\subsubsection{Rational Types}
The KS approach tests whether the observed vector of choice probabilities $p$ can be expressed as a convex combination of the deterministic choice vectors induced by \textit{rational types}. To define a rational type, we first define a \textit{choice type}. A choice type selects one patch from each budget. We represent choice type $t$ by the binary vector
\begin{equation}\label{def:type}
    a_t = \left(
        d_{1|1},\ldots,d_{I_1|1},\,
        d_{1|2},\ldots,d_{I_2|2},\,
        \ldots,\,
        d_{1|J},\ldots,d_{I_J|J}
    \right)^\top
    \in \{0,1\}^{\sum_{j=1}^J I_j},
\end{equation}
where $d_{i|j}=1$ if type $t$ selects patch $x_{i|j}$ from budget $\B_j$ and $d_{i|j}=0$ otherwise.

Rationality imposes additional restrictions on a choice type, which can be formulated using either SARP or GARP. Following KS, we impose the continuous demand assumption and focus on nonintersection patches. On such patches, SARP and GARP impose the same restrictions on choice types.

From now on, all definitions involving patches exclude intersection patches, and we use SARP to define rationality. We retain the notation $x_{1|j},\ldots,x_{I_j|j}$ for the remaining patches on budget $\B_j$. Following Section~3.4.1 of KS, a rational type is defined with respect to a directed graph on budgets.

\begin{definition}[Induced Revealed Preference Relation on Budgets]\label{def:rp}
Fix a choice type $a_t$ and the collection of budgets $\B=\{\B_j\}_{j=1}^J$. For any pair $j,k\in [J]$, we say that $\B_k$ is directly revealed preferred to $\B_j$ if 
there is an index $i\in [I_j]$ such that
\begin{equation*}
    d_{i|j}=1
    \quad\text{and}\quad
    q_k^\top y<1
    \quad\text{for all } y\in x_{i|j}.
\end{equation*}
We define a directed revealed-preference graph with node set $\B=\{\B_j\}_{j=1}^J$. For any distinct budgets $\B_j$ and $\B_k$, include an arc $\B_k\to\B_j$ if $\B_k$ is revealed preferred to $\B_j$.
\end{definition}

A choice type satisfies SARP if and only if its induced graph is acyclic, yielding the following definition.

\begin{definition}[Rational Type]\label{def:rationaltype}
A choice type $a_t$ is a \textit{rational type} if its induced revealed preference graph on budgets is acyclic.
\end{definition}

Given a collection of budgets, we can construct the collection of all distinct rational types $\{a_t\}_{t=1}^T$, where $T$ is the number of such types. 
Recall that $I$ is the number of patches in the coarsest partition in Definition \ref{def:KS_3.2}. Thus $I$ is the dimension of the choice probability vector. We stack the rational types columnwise to form a matrix $A\in \mathbb{R}^{I\times T}$. By Theorem~3.1 of KS, a choice probability vector $p$ is rationalizable if and only if
$\exists v\in \Delta^T$ such that $p = Av$,
where $\Delta^T$ is the unit simplex in $\mathbb{R}^T$.

We denote the polytope generated by the KS rational types by
$\mathcal{P}^{\mathrm{KS}}$:
\begin{equation}\label{def:KS}
    \mathcal{P}^{\mathrm{KS}}
    :=
    \{p\in\mathbb{R}^I:
      p=Av \text{ for some } v\in\Delta^T\}.
\end{equation}

\subsection{Multi-arc Network Flow Characterization}\label{section:f}

In the D-RUM setting, \cite{fiorini2004short}  establishes a correspondence between strict preference rankings over $n$ alternatives and directed paths from the empty set to the full set of alternatives in the corresponding power-set digraph.

This section shows that an analogous characterization holds in the consumption-budget setting, with some modifications. In particular, we construct a power-set digraph whose nodes correspond to subsets of budgets. Since the individual choice alternatives are patches, a given ordering of the budgets may be consistent with multiple patch choices. To represent these possibilities, we use a \textbf{multi-arc power-set digraph}, in which multiple arcs between the same pair of nodes represent distinct admissible patch choices.

\subsubsection{Example}

We illustrate the multi-arc power-set digraph with an example involving two commodities and three budgets.

\begin{figure}[htbp]
    \centering
    \tikzset{
        sequence/.style={
            x=0.36cm,y=0.32cm,
            >=stealth,
            line cap=round,
            line join=round,
            font=\scriptsize
        },
        eligible/.style={
            line width=1.5pt
        },
        remaining/.style={
            line width=0.65pt,
            opacity=0.5
        },
        ineligible/.style={
            draw=gray!65,
            densely dashed,
            line width=0.7pt
        },
        patchlabel/.style={
            fill=white,
            inner sep=0.8pt
        },
        intersection/.style={
            fill=white,
            draw=black!60,
            line width=0.45pt
        }
    }

    \newcommand{\budgetsequenceaxes}{%
        \path[use as bounding box]
            (-0.7,-1.0) rectangle (13.6,10.5);
        \draw[->,black!65,line width=0.4pt]
            (0,0) -- (12.7,0) node[right] {$y_1$};
        \draw[->,black!65,line width=0.4pt]
            (0,0) -- (0,9.6) node[above] {$y_2$};
        \node[below left,inner sep=1.5pt] at (0,0) {$0$};
        \coordinate (P12) at ({12/5},{21/5});
        \coordinate (P13) at ({24/7},{15/7});
        \coordinate (P23) at (6,{3/2});
    }

    \begin{subfigure}[t]{0.47\linewidth}
        \centering
        \begin{tikzpicture}[
            x=0.44cm,y=0.36cm,
            >=stealth,
            line cap=round,
            line join=round,
            font=\scriptsize,
            budget/.style={line width=1.3pt},
            patchlabel/.style={
                fill=white,
                inner sep=0.7pt,
                font=\scriptsize
            }
        ]
            \draw[->,black!75] (0,0) -- (12.85,0)
                node[right] {$y_1$};
            \draw[->,black!75] (0,0) -- (0,9.70)
                node[above] {$y_2$};
            \node[below left,inner sep=3pt] at (0,0) {$0$};

            \draw[budget,blue!75!black] (0,9) -- (4.5,0);
            \draw[budget,purple!75!black] (0,6) -- (8,0);
            \draw[budget,teal!80!black] (0,3) -- (12,0);
    
            \coordinate (P12) at ({12/5},{21/5});
            \coordinate (P13) at ({24/7},{15/7});
            \coordinate (P23) at (6,{3/2});
            \foreach \P in {P12,P13,P23} {
                \filldraw[
                    fill=white,
                    draw=black!65,
                    line width=0.65pt
                ] (\P) circle[radius=2.6pt];
            }
    
            \node[patchlabel,text=blue!75!black]
                at (1.13,7.28) {$x_{1\mid 1}$};
            \node[patchlabel,text=blue!75!black]
                at (2.58,3.12) {$x_{2\mid 1}$};
            \node[patchlabel,text=blue!75!black]
                at (3.54,1.02) {$x_{3\mid 1}$};
    
            \node[patchlabel,text=purple!75!black]
                at (0.95,4.73) {$x_{1\mid 2}$};
            \node[patchlabel,text=purple!75!black]
                at (4.25,3.22) {$x_{2\mid 2}$};
            \node[patchlabel,text=purple!75!black]
                at (6.98,0.36) {$x_{3\mid 2}$};
    
            \node[patchlabel,text=teal!80!black]
                at (1.38,2.28) {$x_{1\mid 3}$};
            \node[patchlabel,text=teal!80!black]
                at (4.90,1.44) {$x_{2\mid 3}$};
            \node[patchlabel,text=teal!80!black]
                at (9.08,1.11) {$x_{3\mid 3}$};
    
            \node[text=blue!75!black,anchor=west]
                at (4.62,0.27) {$\B_1$};
            \node[text=purple!75!black,anchor=west]
                at (8.10,0.27) {$\B_2$};
            \node[text=teal!80!black,anchor=west]
                at (12.10,0.27) {$\B_3$};
        \end{tikzpicture}
        \caption{Nine patches}\label{fig:patches}
    \end{subfigure}\hfill
    \begin{subfigure}[t]{0.47\linewidth}
        \centering
        \begin{tikzpicture}[sequence]
            \budgetsequenceaxes

            \draw[remaining,purple!75!black]
                (0,6) -- (8,0);
            \draw[remaining,teal!80!black]
                (0,3) -- (12,0);

            \draw[eligible,blue!75!black]
                (0,9)
                -- node[midway,right,patchlabel,text=blue!75!black]
                {$x_{1|1}$} (P12);

            \draw[ineligible]
                (P12)
                -- node[midway,left,patchlabel,text=gray!75]
                {$x_{2|1}$} (P13);
            \draw[ineligible]
                (P13)
                -- node[midway,left,patchlabel,text=gray!75]
                {$x_{3|1}$} (4.5,0);

            \foreach \P in {P12,P13,P23} {
                \filldraw[intersection]
                    (\P) circle[radius=1.6pt];
            }

            \node[below,text=blue!75!black]
                at (4.5,0) {$\B_1$};
            \node[below,text=purple!75!black]
                at (8,0) {$\B_2$};
            \node[below,text=teal!80!black]
                at (12,0) {$\B_3$};
        \end{tikzpicture}
        \caption{Choose from $\B_1$}\label{fig:budget_step_1}
    \end{subfigure}

    \medskip
    \begin{subfigure}[t]{0.47\linewidth}
        \centering
        \begin{tikzpicture}[sequence]
            \budgetsequenceaxes

            \draw[remaining,teal!80!black]
                (0,3) -- (12,0);

            \draw[eligible,purple!75!black]
                (0,6)
                -- node[
                    midway,above,xshift=2pt,
                    patchlabel,text=purple!75!black
                ] {$x_{1|2}$} (P12);
            \draw[eligible,purple!75!black]
                (P12)
                -- node[midway,above,patchlabel,text=purple!75!black]
                {$x_{2|2}$} (P23);

            \draw[ineligible] (P23) -- (8,0);
            \draw[gray!65,line width=0.3pt]
                (7,0.75) -- (7.6,1.8);
            \node[above,patchlabel,text=gray!75]
                at (7.6,1.8) {$x_{3|2}$};

            \foreach \P in {P12,P13,P23} {
                \filldraw[intersection]
                    (\P) circle[radius=1.6pt];
            }

            \node[below,text=purple!75!black]
                at (8,0) {$\B_2$};
            \node[below,text=teal!80!black]
                at (12,0) {$\B_3$};
        \end{tikzpicture}
        \caption{Choose from $\B_2$}\label{fig:budget_step_2}
    \end{subfigure}\hfill
    \begin{subfigure}[t]{0.47\linewidth}
        \centering
        \begin{tikzpicture}[sequence]
            \budgetsequenceaxes

            \draw[eligible,teal!80!black]
                (0,3)
                -- node[midway,above,patchlabel,text=teal!80!black]
                {$x_{1|3}$} (P13);
            \draw[eligible,teal!80!black]
                (P13)
                -- node[midway,above,patchlabel,text=teal!80!black]
                {$x_{2|3}$} (P23);
            \draw[eligible,teal!80!black]
                (P23)
                -- node[midway,above,patchlabel,text=teal!80!black]
                {$x_{3|3}$} (12,0);

            \foreach \P in {P13,P23} {
                \filldraw[intersection]
                    (\P) circle[radius=1.6pt];
            }

            \node[below,text=teal!80!black]
                at (12,0) {$\B_3$};
        \end{tikzpicture}
        \caption{Choose from $\B_3$}\label{fig:budget_step_3}
    \end{subfigure}
    \caption{Patches and sequential choices for $\B_1\succ\B_2\succ\B_3$.
    Hollow circles mark excluded intersections. In the three choice panels,
    thick colored segments indicate eligible patches, dashed gray segments
    indicate ineligible patches, and thin lines show the other remaining budgets.}
    \label{fig:budget_sequence}
\end{figure}
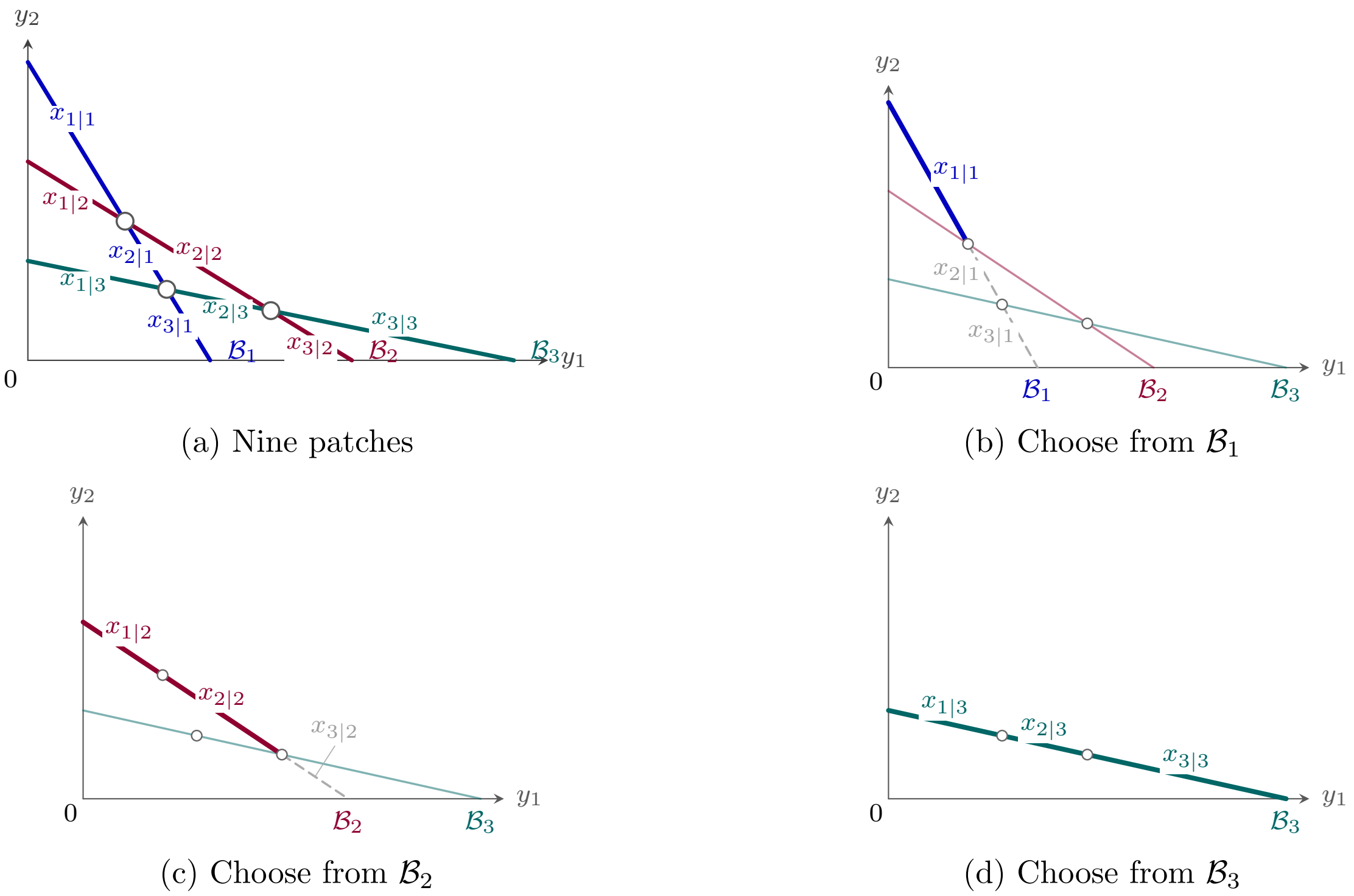

\paragraph{Two Commodities and Three Budgets}
Consider the following pairwise intersecting budgets in $\mathbb{R}_+^2$:
\begin{equation}\label{eq:budgets}
\begin{aligned}
    \B_1&=\left\{y\in\mathbb{R}_+^2:
        \frac{2}{9}y_1+\frac{1}{9}y_2=1\right\}, \B_2=\left\{y\in\mathbb{R}_+^2:
        \frac{1}{8}y_1+\frac{1}{6}y_2=1\right\},\\
    \B_3&=\left\{y\in\mathbb{R}_+^2:
        \frac{1}{12}y_1+\frac{1}{3}y_2=1\right\}.
\end{aligned}
\end{equation}
After excluding intersection patches, each budget $\B_j$ has three
patches, which we enumerate as $x_{1|j},x_{2|j},x_{3|j}$ in increasing
order of $y_1$. Figure~\ref{fig:patches} illustrates the resulting nine nonintersection patches.

Now consider an agent whose maximized utility is highest on $\B_1$,
followed by $\B_2$ and then $\B_3$. Thus, the agent must choose $x_{1|1}$ from $\B_1$,
since $x_{2|1}$ and $x_{3|1}$ lie strictly below $\B_2$. We then remove
$\B_1$ from the construction. On $\B_2$, both $x_{1|2}$ and $x_{2|2}$
lie above the remaining budget $\B_3$, whereas $x_{3|2}$ lies strictly
below it. Either $x_{1|2}$ or $x_{2|2}$ may therefore be selected.
After removing $\B_2$, any of the three original patches on $\B_3$
may be selected. Intuitively, the patch selected from the
highest-ranked remaining budget must lie strictly above every other
remaining budget. Figures~\ref{fig:budget_step_1}--\ref{fig:budget_step_3}
illustrate the three steps.

\begin{figure}[htbp]
    \centering
    \begin{tikzpicture}[
        >=stealth,
        state/.style={
            draw=black!60,fill=white,rounded corners=3pt,
            minimum width=10mm,minimum height=7mm,
            inner sep=4pt,font=\small
        },
        flow/.style={
            ->,line width=0.85pt,
            shorten >=1.4pt,shorten <=1.4pt
        },
        patchlabel/.style={
            font=\small,fill=white,inner sep=1.4pt
        },
        layerlabel/.style={
            font=\footnotesize,text=black!55
        }
    ]
        \node[state] (empty) at (0,0) {$\varnothing$};
        \node[state] (one) at (3.2,0) {$\{\B_1\}$};
        \node[state] (onetwo) at (6.4,0) {$\{\B_1,\B_2\}$};
        \node[state,minimum width=16mm] (all) at (9.6,0)
            {$\{\B_1,\B_2,\B_3\}$};

        \draw[flow,blue!75!black] (empty) --
            node[patchlabel,midway,above=1pt]
            {$x_{1|1}$} (one);

        \draw[flow,purple!75!black] (one) to[bend left=18]
            node[patchlabel,midway,above=1pt]
            {$x_{1|2}$} (onetwo);
        \draw[flow,purple!75!black] (one) to[bend right=18]
            node[patchlabel,midway,below=1pt]
            {$x_{2|2}$} (onetwo);

        \draw[flow,teal!80!black] (onetwo) to[bend left=28]
            node[patchlabel,midway,above=1pt]
            {$x_{1|3}$} (all);
        \draw[flow,teal!80!black] (onetwo) --
            node[patchlabel,midway]
            {$x_{2|3}$} (all);
        \draw[flow,teal!80!black] (onetwo) to[bend right=28]
            node[patchlabel,midway,below=1pt]
            {$x_{3|3}$} (all);

        \node[layerlabel] at (0,-0.78) {$|S|=0$};
        \node[layerlabel] at (3.2,-0.78) {$|S|=1$};
        \node[layerlabel] at (6.4,-0.78) {$|S|=2$};
        \node[layerlabel] at (9.6,-0.78) {$|S|=3$};
    \end{tikzpicture}
    \caption{The subnetwork corresponding to
    $\B_1\succ\B_2\succ\B_3$. Nodes record the budgets already removed.
    Arc colors identify the budget being removed, and arc labels
    identify the chosen patch.}
    \label{fig:ordered-flow}
\end{figure}
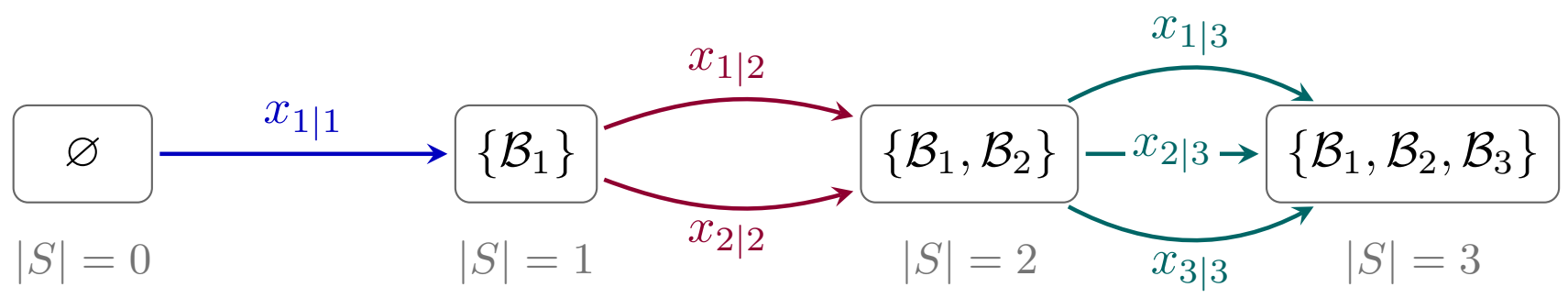

Figure~\ref{fig:ordered-flow} illustrates the subnetwork corresponding to
the budget ordering $\B_1\succ\B_2\succ\B_3$. Each node $S$ records the
indices of the budgets already removed. An arc $a=(S,S\cup\{\B_j\},i)$
represents choosing patch $x_{i\mid j}$ from budget $\B_j$ and then
removing $\B_j$.

The first transition has a single arc, labeled $x_{1|1}$. The second
transition has two parallel arcs, labeled $x_{1|2}$ and $x_{2|2}$,
and the final transition has three parallel arcs, one for each patch
on $\B_3$. Each path from $\varnothing$ to $\{\B_1,\B_2,\B_3\}$ selects one
patch from each budget and induces a rational type consistent with
the given ordering. The parallel arcs
capture the fact that a budget ordering can be consistent with
multiple patch choices. A deterministic rational type can thus be represented by sending one
unit of flow along a single path.

\begin{figure}[htbp]
    \centering
    \begin{tikzpicture}[
        x=0.72cm,y=0.72cm,
        >=stealth,
        state/.style={
            draw=black!60,fill=white,rounded corners=3pt,
            minimum width=10mm,minimum height=7mm,
            inner sep=4pt,font=\small
        },
        flow/.style={
            ->,line width=0.85pt,
            shorten >=1.4pt,shorten <=1.4pt
        },
        bridge/.style={
            preaction={draw=white,line width=3.5pt}
        },
        patchlabel/.style={
            font=\small,fill=white,inner sep=1.4pt,
            text opacity=1,fill opacity=0.96
        },
        layerlabel/.style={
            font=\footnotesize,text=black!55
        }
    ]
        \node[state] (empty) at (0,0) {$\varnothing$};
        \node[state] (one) at (4.6,3.6) {$\{\B_1\}$};
        \node[state] (two) at (4.6,0) {$\{\B_2\}$};
        \node[state] (three) at (4.6,-3.6) {$\{\B_3\}$};

        \node[state] (onetwo) at (9.2,3.6) {$\{\B_1,\B_2\}$};
        \node[state] (onethree) at (9.2,0) {$\{\B_1,\B_3\}$};
        \node[state] (twothree) at (9.2,-3.6) {$\{\B_2,\B_3\}$};

        \node[state,minimum width=16mm] (all) at (13.8,0)
            {$\{\B_1,\B_2,\B_3\}$};

        \draw[flow,blue!75!black] (empty) --
            node[patchlabel,pos=0.52,sloped,above=1pt]
            {$x_{1|1}$} (one);
        \draw[flow,purple!75!black] (empty) --
            node[patchlabel,pos=0.52,above=1pt]
            {$x_{2|2}$} (two);
        \draw[flow,teal!80!black] (empty) --
            node[patchlabel,pos=0.52,sloped,below=1pt]
            {$x_{3|3}$} (three);

        \draw[flow,teal!80!black] (one) --
            node[patchlabel,pos=0.25,sloped,above=1pt]
            {$x_{3|3}$} (onethree);
        \draw[flow,blue!75!black] (three) --
            node[patchlabel,pos=0.25,sloped,above=1pt]
            {$x_{1|1}$} (onethree);

        \draw[flow,purple!75!black] (one) to[bend left=12]
            node[patchlabel,midway,above=1pt]
            {$x_{1|2}$} (onetwo);
        \draw[flow,purple!75!black] (one) to[bend right=12]
            node[patchlabel,midway]
            {$x_{2|2}$} (onetwo);

        \draw[flow,bridge,blue!75!black] (two) to[bend left=9]
            node[patchlabel,pos=0.73,sloped,above=1pt]
            {$x_{1|1}$} (onetwo);
        \draw[flow,bridge,blue!75!black] (two) to[bend right=9]
            node[patchlabel,pos=0.73,sloped,below=1pt]
            {$x_{2|1}$} (onetwo);

        \draw[flow,bridge,teal!80!black] (two) to[bend left=9]
            node[patchlabel,pos=0.73,sloped,above=1pt]
            {$x_{2|3}$} (twothree);
        \draw[flow,bridge,teal!80!black] (two) to[bend right=9]
            node[patchlabel,pos=0.73,sloped,below=1pt]
            {$x_{3|3}$} (twothree);

        \draw[flow,purple!75!black] (three) to[bend left=12]
            node[patchlabel,midway]
            {$x_{2|2}$} (twothree);
        \draw[flow,purple!75!black] (three) to[bend right=12]
            node[patchlabel,midway,below=1pt]
            {$x_{3|2}$} (twothree);


        \draw[flow,teal!80!black] (onetwo) to[bend left=17]
            node[patchlabel,pos=0.49,sloped,above=1pt]
            {$x_{1|3}$} (all.north west);
        \draw[flow,teal!80!black] (onetwo) --
            node[patchlabel,pos=0.49,sloped]
            {$x_{2|3}$} (all.north west);
        \draw[flow,teal!80!black] (onetwo) to[bend right=17]
            node[patchlabel,pos=0.49,sloped,below=1pt]
            {$x_{3|3}$} (all.north west);

        \draw[flow,purple!75!black] (onethree) to[bend left=17]
            node[patchlabel,pos=0.49,above=1pt]
            {$x_{1|2}$} (all);
        \draw[flow,purple!75!black] (onethree) --
            node[patchlabel,pos=0.49]
            {$x_{2|2}$} (all);
        \draw[flow,purple!75!black] (onethree) to[bend right=17]
            node[patchlabel,pos=0.49,below=1pt]
            {$x_{3|2}$} (all);

        \draw[flow,blue!75!black] (twothree) to[bend left=17]
            node[patchlabel,pos=0.49,sloped,above=1pt]
            {$x_{1|1}$} (all.south west);
        \draw[flow,blue!75!black] (twothree) --
            node[patchlabel,pos=0.49,sloped]
            {$x_{2|1}$} (all.south west);
        \draw[flow,blue!75!black] (twothree) to[bend right=17]
            node[patchlabel,pos=0.49,sloped,below=1pt]
            {$x_{3|1}$} (all.south west);

        \node[layerlabel] at (0,-4.6) {$|S|=0$};
        \node[layerlabel] at (4.6,-4.6) {$|S|=1$};
        \node[layerlabel] at (9.2,-4.6) {$|S|=2$};
        \node[layerlabel] at (13.8,-4.6) {$|S|=3$};
    \end{tikzpicture}
    \caption{The complete multi-arc power-set digraph for the
    three-budget example. Nodes record the budgets already removed.
    An arc's color identifies the budget being removed, and its
    label identifies the chosen patch. Crossings of arcs are not
    additional nodes.}
    \label{fig:full-flow}
\end{figure}
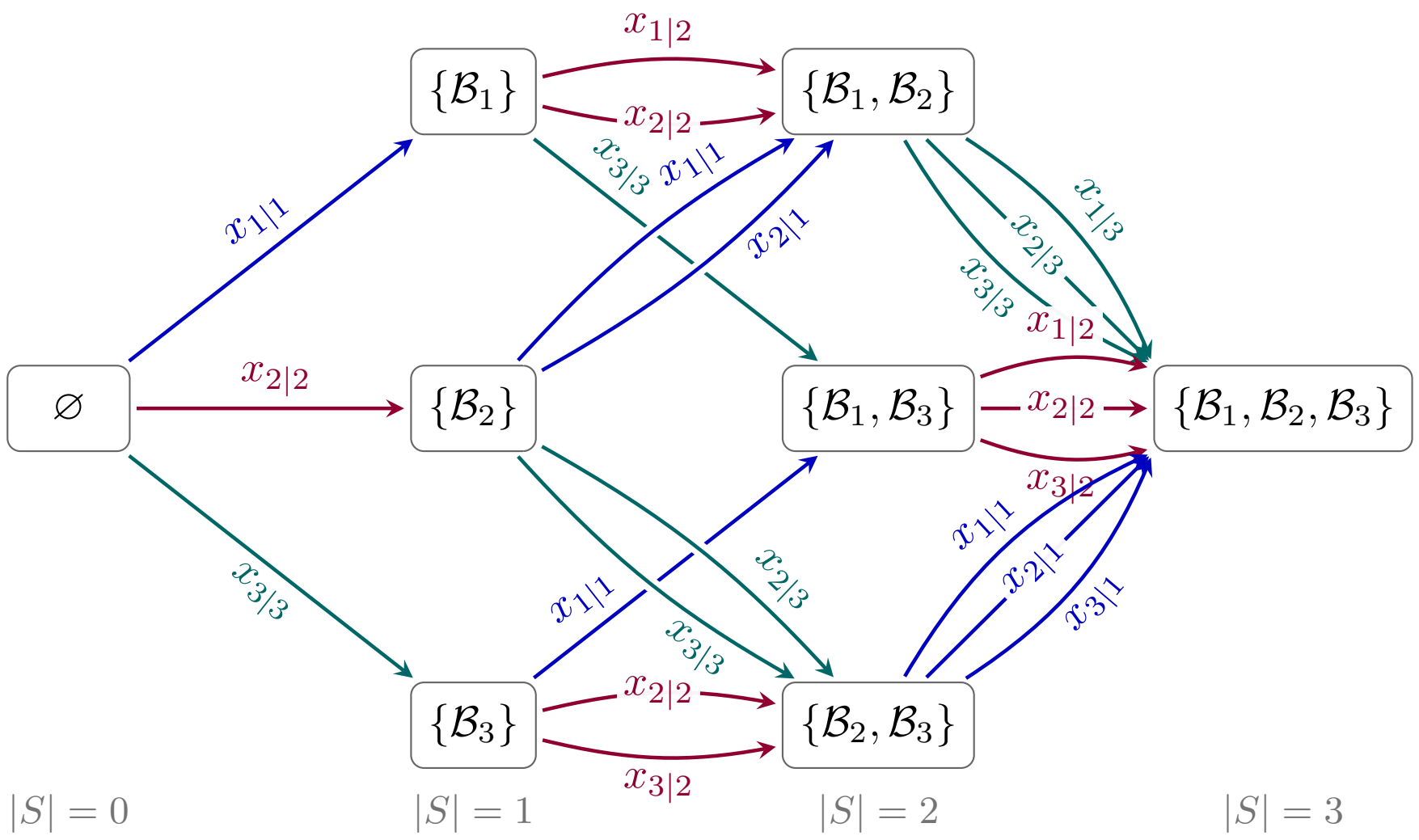

We complete the network by repeating this construction for the remaining
five strict rankings of the budgets. Figure~\ref{fig:full-flow} shows the complete multi-arc power-set
digraph, which has eight nodes and 22 arcs. Each path from $\varnothing$
to $\{\B_1,\B_2,\B_3\}$ specifies a budget ordering and one admissible patch choice
on each budget, thereby inducing a rational type. Different paths may
induce the same rational type.

Mapping arc flows to choice probabilities is particularly simple. For example, the probability of choosing patch $x_{1|1}$ from budget $\B_1$ equals the sum of the flows on all arcs labeled $x_{1|1}$ in Figure~\ref{fig:full-flow}.

\subsubsection{Formal Definitions}

Let $\mathcal{B}=\{\B_j\}_{j=1}^J$ denote the collection of budgets. We first introduce the notion of undominance, following \cite{shiraikoida}.

\begin{definition}[Undominated patch]\label{def:undominated}
Let $x_{i\mid j}$ be a patch on budget $\B_j$, and let
$\widetilde{\mathcal{B}}\subseteq\mathcal{B}\setminus\{\B_j\}$ be a
collection of budgets. We say that $x_{i\mid j}$ is
\emph{undominated} by $\widetilde{\mathcal{B}}$ if it lies strictly
above every budget hyperplane in $\widetilde{\mathcal{B}}$, that is,
\begin{equation}
    q_\ell^\top y>1
    \qquad
    \text{for all }y\in x_{i\mid j}
    \text{ and all }
    \B_\ell\in\widetilde{\mathcal{B}}.
    \label{eq:undominated}
\end{equation}
\end{definition}

\begin{definition}[Multi-arc power-set digraph]\label{def:ma_digraph}
Given a collection of $J$ budgets $\B=\{\B_j\}_{j=1}^J$, the initial node set is $2^{\mathcal{B}}$, with source node
$\varnothing$ and terminal node $\mathcal{B}$. For each
$k=0,\ldots,J$, define the layer $\mathcal{L}_k
    :=
    \{S\subseteq\mathcal{B}:|S|=k\}$.

The initial layer is $\mathcal{L}_0=\{\varnothing\}$. We construct
the graph iteratively, proceeding through the layers in increasing
order of cardinality.

For each $k=0,\ldots,J-1$, each retained node
$S\in\mathcal{L}_k$, each $\B_j\in\mathcal{B}\setminus S$,
and each patch $x_{i\mid j}$ that is undominated by the other remaining
budgets $\mathcal{B}\setminus(S\cup\{\B_j\})$, add an arc from $S$ to
$S\cup\{\B_j\}$ with label $i$. Formally, we write this arc as $a=(S,S\cup\{\B_j\},i)$.
After all candidate arcs from layer $\mathcal{L}_k$
to layer $\mathcal{L}_{k+1}$ have been considered, remove every
node in $\mathcal{L}_{k+1}$ with no incoming arcs.

Let $\mathcal{G}=(\mathcal{V},\mathcal{A})$ denote the resulting
directed multigraph.
\end{definition}
We note that the power-set digraph in \citet{fiorini2004short} includes every node, whereas some nodes may be absent in the consumption-budget setting. For example, if the feasible set of budget $\B_j$ lies entirely in the interior of another budget's feasible set, $\B_j$ cannot be ranked first, so the singleton node $\{\B_j\}$ is absent.

Now we define the unit-flow polytope.
\begin{definition}[Unit-flow polytope]\label{def:ufp}
Let $\mathcal{B}=\{\B_j\}_{j=1}^J$ denote the collection of
budgets, and let $\mathcal{G}=(\mathcal{V},\mathcal{A})$ be its
associated multi-arc power-set digraph. For an arc
$a=(S,S',i)\in\mathcal{A}$, write its flow as $f_a=f_{S,S',i}$.

The \emph{unit-flow polytope} $\mathcal{F}_{\mathcal{G}}$ is the
set of all nonnegative flow vectors $f\in\mathbb{R}_+^{\mathcal{A}}$
satisfying, for every node $S\in\mathcal{V}$,
\begin{equation}
    \sum_{(S',S,i)\in\mathcal{A}} f_{S',S,i}
    -
    \sum_{(S,S',i)\in\mathcal{A}} f_{S,S',i}
    =
    \begin{cases}
        -1, & S=\varnothing,\\
        1,  & S=\mathcal{B},\\
        0,  & \text{otherwise}.
    \end{cases}
    \label{eq:budget-unit-flow-polytope}
\end{equation}
The first sum is the total inflow into $S$. The second sum is
the total outflow from $S$.
A sum over an empty set of arcs is defined to be zero.

We write these constraints compactly as $\mathcal{F}_{\mathcal{G}}=\{f\in\mathbb{R}_+^{\mathcal{A}}:A_{\mathcal{G}}f=b_{\mathcal{G}}\}$, where $A_{\mathcal{G}}$ is the node--arc incidence matrix, and $b_{\mathcal{G}}$ has entry $-1$ at the source, $1$ at the terminal node, and $0$ elsewhere.
\end{definition}

The node--arc incidence matrix $A_{\mathcal{G}}$ is totally unimodular, and $b_{\mathcal{G}}$ is integral. Hence, the unit-flow polytope $\mathcal{F}_{\mathcal{G}}$ is integral. Since $\mathcal{G}$ is acyclic, the vertices of $\mathcal{F}_{\mathcal{G}}$ are precisely the incidence vectors of directed paths from $\varnothing$ to $\mathcal{B}$.

We define the following linear mapping from the flow space to the choice probability space.

\begin{definition}[Mapping flows to choice probabilities]\label{def:map}
Let $\mathcal{B}=\{\B_j\}_{j=1}^J$ denote the collection of
budgets, and let $\mathcal{G}=(\mathcal{V},\mathcal{A})$ be its
associated multi-arc power-set digraph, with unit-flow polytope
$\mathcal{F}_{\mathcal{G}}$. For a flow
$f\in\mathcal{F}_{\mathcal{G}}$, define the induced choice-probability
vector $p_f$ by
\begin{equation}
    (p_f)_{i\mid j}
    :=
    \sum_{\substack{S\subseteq\mathcal{B}\\
        (S,S\cup\{\B_j\},i)\in\mathcal{A}}}
        f_{S,S\cup\{\B_j\},i},
    \label{eq:budget-flow-to-choice}
\end{equation}
for $j\in [J]$ and $i\in [I_j]$. We denote
the corresponding linear map by $L_{\mathcal{G}}$, so that
$p_f=L_{\mathcal{G}}f$.
\end{definition}
In words, $(p_f)_{i\mid j}$ is the probability of choosing $x_{i\mid j}$ from budget $\B_j$. It equals the sum of the flows over all arcs associated with $x_{i\mid j}$.

Now we are ready to state the network characterization theorem. Recall the definition of $\mathcal{P}^{\textrm{KS}}$ from \eqref{def:KS}.

\begin{theorem}\label{thm:network}
Given a collection of budgets $\mathcal{B}=\{\B_j\}_{j=1}^J$,
let $\mathcal{G}=(\mathcal{V},\mathcal{A})$ be its associated
multi-arc power-set digraph from Definition~\ref{def:ma_digraph},
$\mathcal{F}_{\mathcal{G}}$ its associated unit-flow polytope from
Definition~\ref{def:ufp}, and $L_{\mathcal{G}}$ the linear map from
Definition~\ref{def:map}. Then
\begin{equation*}
    \mathcal{P}^{\mathrm{KS}}
    =
    \left\{
        L_{\mathcal{G}}f:
        f\in\mathcal{F}_{\mathcal{G}}
    \right\}.
\end{equation*}
\end{theorem}
Beyond its theoretical interest, this theorem allows us to adapt the Flow-CG algorithm developed for the \DRUM setting to the consumption-budget setting, since the Bellman--Ford algorithm also applies to multi-arc digraphs.

%% file: paper_main/A_Numerical_Final.tex
Section~\ref{section:compare_with_in} compares the three methods across simulation designs with varying \textit{density}, defined as the ratio of all observed choice probabilities over all possible choice probabilities.\footnote{With $n$ alternatives, there are $n2^{n-1}$ possible choice probabilities.} Flow-CG generally performs best at low densities and BM-ADMM at high densities. Flow-ADMM is generally less competitive, although it occasionally outperforms Flow-CG at low densities.

To assess the gains from our algorithms beyond the underlying formulations, we also implement the flow and BM formulations directly in Gurobi. For the instances considered, the BM formulation exceeds
available memory, while the direct flow implementation
already takes hours for binary-menu instances, compared
with seconds for our methods. These limitations are to be
expected given the exponential number of variables in both
formulations.

Guided by these results, Sections~\ref{section:compare_binary} and~\ref{section:compare_dense} compare the best-performing proposed methods with existing methods in two more demanding settings. Section~\ref{section:compare_binary} considers binary menus with an increasing number of alternatives, while Section~\ref{section:compare_dense} fixes $n=20$ and increases menu density.\footnote{All GPU experiments are run on a workstation equipped with an AMD Ryzen~9~9950X CPU (16 physical cores), 64 GB of DDR5 memory, and one NVIDIA GeForce RTX~5090 GPU (32 GB of memory and 21{,}760 CUDA cores). Each GPU solver uses the full GPU and at most one CPU thread for coordination. Given their greater computational demands, IP-CG and IP-Heuristic-CG are run on an HPC cluster, with one problem instance per job. Each job has exclusive access to a dual-socket Intel Xeon Gold node (Cascade Lake 6240 or Ice Lake 6342), with resources limited to 24 CPU cores and 96 GB of RAM. The workstation runs Windows~11 with a WDDM driver, CUDA~13.3, Python~3.12, and CuPy~14.}

Throughout this section, let $v$ denote the current feasible objective value and $\texttt{opt}$ a precomputed reference value for the optimal objective value. Flow-CG, Flow-ADMM, BM-ADMM, IP-CG, and IP-Heuristic-CG terminate when
\begin{equation}\label{eqn:termination_criterion}
v-\texttt{opt}<10^{-3}
\quad\text{or}\quad
\frac{v-\texttt{opt}}{\texttt{opt}}<10^{-2}.
\end{equation}
For choice-probability vectors inside the polytope, we set $\texttt{opt}=0$. For vectors outside the polytope, we compute \texttt{opt} to within $0.1\%$ of the true optimum. The absolute-gap criterion accommodates cases in which \texttt{opt} is close to zero and is the sole criterion used when $\texttt{opt}=0$. The weighting matrix is set to the identity, $\Omega=I$.

\subsection{Comparing Flow-CG, Flow-ADMM, and BM-ADMM}\label{section:compare_with_in}
We compare the performance of our methods in settings with $n\in\{18,19,20\}$ and varying menu densities. We first describe the simulation design. We generate menus sequentially, drawing each menu uniformly at random from those not yet selected until the desired menu density is reached. The only exception is the pairwise setting, in which we include all binary menus. Given the sampled menus, we draw an independent uniform random variable for each alternative in each menu. Then, we construct choice probabilities within each menu by normalizing these draws to sum to one. Because the draws are menu-specific, the resulting choice probabilities are unlikely to be \DRUM-rationalizable, as is indeed the case in our simulations.

For comparison, we also report results for IP-CG in all cases. In addition, we include results for the commercial solver Gurobi using the flow formulation in the pairwise cases.\footnote{We do not consider the BM formulation because its constraint matrix is too large to handle even in sparse format. For $n=18$, the BM coefficient matrix requires approximately 28 GB of memory, compared with 0.3 GB for the flow formulation. We do not consider denser instances with the flow formulation because Gurobi is already extremely slow in the pairwise cases relative to the other methods.} For Gurobi, we supply only the constraint matrices and objective coefficients and solve the flow formulation \eqref{eqn:flow_formulation} in sparse format. We consider the available Gurobi optimization methods, including simplex-based and barrier methods, and report the fastest runtime.

Table \ref{tab:density18} reports the results for $n=18$, with menu densities ranging from the pairwise case to 10\%. Results for $n=19$ and $n=20$, reported in Section \ref{sec:supp_simulations}, are qualitatively similar. Four conclusions emerge from Table \ref{tab:density18}, Table \ref{tab:density19} and Table \ref{tab:density20}. First, when menu density is low, Flow-CG is competitive, with Flow-ADMM occasionally outperforming Flow-CG.\footnote{We do not include the cache-construction time required by Flow-ADMM, as our focus is on its convergence behavior. When the algorithm is applied repeatedly, this one-time setup cost is amortized across runs, although it can be substantial for a single run.} Second, as menu density increases, BM-ADMM becomes competitive and typically performs best once density exceeds approximately $0.5\%$. Third, our new methods deliver substantial speed improvements relative to IP-CG. Fourth, direct ``plug-in'' implementations using commercial solvers are slower by large margins than both our new methods and IP-CG.

\input{paper_main/A_table_n18_numerical1}

\subsection{Comparing Flow-CG, IP-CG, and IP-Heuristic-CG on binary menus}\label{section:compare_binary}
In this section, we focus on a class of harder problems involving binary menus only but an increasingly large number of alternatives. Among our proposed methods, Flow-CG emerges as the most competitive in this setting. We therefore compare Flow-CG with IP-CG and IP-Heuristic-CG.

For each $n\in\{25,27,30,32,35\}$, we first generate a choice-probability vector $p^n_{\textrm{out}}$ outside the \DRUM polytope using the same procedure as in Section \ref{section:compare_with_in}. We then construct a choice-probability vector $p^n_{\textrm{in}}$ inside the polytope by assigning equal probability (i.e., $0.5$) to each alternative within every menu. Finally, we define $p^n_{\textrm{half}}$ as an equal-weight convex combination of $p^n_{\textrm{in}}$ and $p^n_{\textrm{out}}$. We expect $p^n_{\textrm{out}}$ and $p^n_{\textrm{half}}$ to be harder than $p^n_{\textrm{in}}$, because $p^n_{\textrm{in}}$ should admit many representations as mixtures of distinct  vertices. In total, we consider 15 instances, with three instances for each value of $n$. 

Table \ref{tab:binary} reports the results for Flow-CG, IP-CG, and IP-Heuristic-CG (IP-H-CG). When the choice probabilities lie outside the polytope,
as in our cases with $p^n_{\textrm{out}}$ and
$p^n_{\textrm{half}}$, Flow-CG is often hundreds of times
faster than the best existing method. When the choice probabilities lie inside the polytope, existing methods can be faster than Flow-CG. In absolute terms, however, all three methods solve these inside-polytope instances quickly, so these differences are of limited practical significance.

\input{paper_main/A_table_binary}

\subsection{Comparing BM-ADMM, IP-CG, and IP-Heuristic-CG on dense menus}\label{section:compare_dense}

In this section, we provide numerical evidence on the performance of BM-ADMM, IP-CG, and IP-Heuristic-CG on a class of harder problems with a fixed number of alternatives and increasingly dense menu collections. We focus on BM-ADMM, which performs best at higher menu
densities in the preceding comparison. We fix $n=20$ and increase the menu density from $0.1\%$ to $5\%$. We follow the same numerical design as in Section \ref{section:compare_binary}: we generate an outside choice-probability vector $p^n_{\textrm{out}}$, an inside choice-probability vector $p^n_{\textrm{in}}$, and their equal-weight convex combination $p^n_{\textrm{half}}$. The menu collections and the outside choice-probability vectors are generated using the same procedures described above. Table \ref{tab:dense} reports the results. In this setting, BM-ADMM performs substantially better than the existing methods. It typically solves the problems in less than 15 seconds, whereas the best existing methods often fail to converge within 24 hours.

\input{paper_main/A_table_dense}

%% file: paper_main/A_table_n18_numerical1.tex
\begin{table}[t]
\caption{Runtime by Menu Density for $n=18$}
\label{tab:density18}
\centering
\setlength{\tabcolsep}{4pt}
\begin{adjustbox}{max width=\linewidth}
\begin{tabular}{@{}l*{10}{r}@{}}
\hline
& \multicolumn{10}{c}{Density} \\
\cline{2-11}
Method
& Pairwise 
& 0.025\% 
& 0.05\% 
& 0.1\% 
& 0.2\% 
& 0.3\% 
& 0.5\% 
& 0.7\% 
& 1\% 
& 10\% \\
\hline
BM-ADMM 
& 2.97 
& 5.72 
& 4.54 
& 0.98 
& 0.62 
& 0.63 
& \textbf{\underline{0.43}} 
& \textbf{\underline{0.34}} 
& \textbf{\underline{0.40}} 
& \textbf{\underline{0.14}} \\

Flow-CG 
& \textbf{\underline{0.04}} 
& 0.133 
& \textbf{\underline{0.12}} 
& 0.21 
& \textbf{\underline{0.30}} 
& \textbf{\underline{0.41}} 
& 0.65 
& 0.89 
& 1.23 
& 12.6 \\

Flow-ADMM 
& 0.36 
& \textbf{\underline{0.129}} 
& 0.15 
& \textbf{\underline{0.20}} 
& 0.33 
& 0.56 
& 1.50 
& 4.61 
& 8.26 
& -- \\

IP-CG 
& 4.38 
& {\footnotesize $\geq 24$ h}
& {\footnotesize $\geq 24$ h}
& {\footnotesize $\geq 24$ h}
& {\footnotesize $\geq 24$ h}
& {\footnotesize $\geq 24$ h}
& {\footnotesize $\geq 24$ h}
& {\footnotesize $\geq 24$ h}
& {\footnotesize $\geq 24$ h}
& {\footnotesize $\geq 24$ h} \\

Gurobi 
& 4.5 h 
& -- 
& -- 
& -- 
& -- 
& -- 
& -- 
& -- 
& -- 
& -- \\
\hline
\end{tabular}
\end{adjustbox}
\par
\ifdefined\legend
  \legend{Runtime in seconds to satisfy the termination criterion in \eqref{eqn:termination_criterion}. The fastest method for each instance is shown in bold and underlined. ``--'' denotes an unreported result, and ``$\geq 24$ h'' indicates that the method did not converge within 24 hours. The Gurobi result for the pairwise case uses the barrier method. The direct Gurobi implementation uses Gurobi's own
termination criterion; all other methods use the criterion
in \eqref{eqn:termination_criterion}.}
\else
  \parbox{\linewidth}{\footnotesize\textit{Note:} Runtime in seconds to satisfy the termination criterion in \eqref{eqn:termination_criterion}. The fastest method for each instance is shown in bold and underlined. ``--'' denotes an unreported result, and ``$\geq 24$ h'' indicates that the method did not converge within 24 hours. The Gurobi result for the pairwise case uses the barrier method. The direct Gurobi implementation uses Gurobi's own
termination criterion; all other methods use the criterion
in \eqref{eqn:termination_criterion}.}
\fi
\end{table}

%% file: paper_main/A_table_binary.tex
\begin{table}[t]
\caption{Runtime for Binary Menus}
\label{tab:binary}
\centering
    \begin{adjustbox}{max width=\linewidth}
    \begin{tabular}{@{}l*{9}{r}@{}}
    \hline 
     &\multicolumn{3}{c}{$p^n_{\textrm{in}}$} & \multicolumn{3}{c}{$p^n_{\textrm{half}}$} & \multicolumn{3}{c}{$p^n_{\textrm{out}}$}\\
    \cline{2-4}\cline{5-7}\cline{8-10}
     $n$ & {\footnotesize Flow-CG} & {\footnotesize IP-CG} & {\footnotesize IP-H-CG}  & {\footnotesize Flow-CG} & {\footnotesize IP-CG} & {\footnotesize IP-H-CG} & {\footnotesize Flow-CG} & {\footnotesize IP-CG} & {\footnotesize IP-H-CG}\\ 
     \hline
    25 & \textbf{0.01} & 0.12 & 0.35 & \textbf{0.79} & 440 & 43 & \textbf{0.29} & 199 & 124 \\
    27 & \textbf{0.03} & 0.13 & 0.43 & \textbf{4.63} & 1,817 & 726 & \textbf{1.17} & 1,116 & 854 \\
    30 & 0.30 & \textbf{0.17} & 0.43 & \textbf{51} & 10,057 & 6,310 & \textbf{14.6} & 3,480 & 5,861 \\
    32 & 1.33 & \textbf{0.20} & 0.46 & \textbf{409} & 76,199 & 58,307 & \textbf{67.0} & 15,891 & 11,206 \\
35 & 13.6 &\textbf{0.28} & 0.45& \textbf{7,305} &  {\footnotesize $\geq 7$ d}&{\footnotesize $\geq 7$ d} & \textbf{721} & 214,936 & 196,874\\

    \hline
    \end{tabular}
    \end{adjustbox}
\par
\ifdefined\legend
  \legend{Time to satisfy the termination criterion \eqref{eqn:termination_criterion}, in seconds. For each instance, the fastest method is shown in bold. ``$\geq 7$ d'' indicates that the algorithm did not converge within 7 days. IP-H-CG is IP-Heuristic-CG with 10 random-insertion restarts per pricing call.}
\else
  \parbox{\linewidth}{\footnotesize\textit{Note:} Time to satisfy the termination criterion \eqref{eqn:termination_criterion}, in seconds. For each instance, the fastest method is shown in bold. ``$\geq 7$ d'' indicates that the algorithm did not converge within 7 days. IP-H-CG is IP-Heuristic-CG with 10 random-insertion restarts per pricing call.}
\fi
\end{table}

%% file: paper_main/A_table_dense.tex
\begin{table}[t]
\caption{Runtime for Dense Menus}
\label{tab:dense}
\centering
    \setlength{\tabcolsep}{4pt}
    \begin{adjustbox}{max width=\linewidth}
    \begin{tabular}{@{}l*{9}{r}@{}}
    \hline 
     & \multicolumn{3}{c}{$p^n_{\textrm{in}}$} 
     & \multicolumn{3}{c}{$p^n_{\textrm{half}}$} 
     & \multicolumn{3}{c}{$p^n_{\textrm{out}}$}\\
    \cline{2-4}\cline{5-7}\cline{8-10}
     density 
     & {\footnotesize BM-ADMM} 
     & {\footnotesize IP-CG} 
     & {\footnotesize IP-H-CG}
     & {\footnotesize BM-ADMM} 
     & {\footnotesize IP-CG} 
     & {\footnotesize IP-H-CG}
     & {\footnotesize BM-ADMM} 
     & {\footnotesize IP-CG} 
     & {\footnotesize IP-H-CG}\\ 
     \hline
    0.1\% 
    & \textbf{2.83} 
    & {\footnotesize $\geq 24$ h} 
    & {\footnotesize $\geq 24$ h}
    & \textbf{14.7} 
    & {\footnotesize $\geq 24$ h} 
    & {\footnotesize $\geq 24$ h}
    & \textbf{13.8} 
    & {\footnotesize $\geq 24$ h} 
    & {\footnotesize $\geq 24$ h} \\

    0.5\% 
    & \textbf{3.82} 
    & {\footnotesize $\geq 24$ h} 
    & {\footnotesize $\geq 24$ h}
    & \textbf{4.32} 
    & {\footnotesize $\geq 24$ h}
    & {\footnotesize $\geq 24$ h} 
    & \textbf{2.69} 
    & {\footnotesize $\geq 24$ h} 
    & {\footnotesize $\geq 24$ h} \\

    1\% 
    & \textbf{3.24} 
    & {\footnotesize $\geq 24$ h} 
    & {\footnotesize $\geq 24$ h} 
    & \textbf{3.14} 
    & {\footnotesize $\geq 24$ h} 
    & {\footnotesize $\geq 24$ h}
    & \textbf{1.87} 
    & {\footnotesize $\geq 24$ h}
    & 860\\

    5\% 
    & \textbf{3.27} 
    & {\footnotesize OOM}  
    & {\footnotesize $\geq 24$ h}
    & \textbf{2.02} 
    & {\footnotesize OOM} 
    & 12,121
    & \textbf{0.82} 
    & {\footnotesize OOM}
    & 3,373\\
    \hline
    \end{tabular}
    \end{adjustbox}
\par
\ifdefined\legend
  \legend{Time to satisfy the termination criterion \eqref{eqn:termination_criterion}, in seconds. For each instance, the fastest method is shown in bold. ``$\geq 24$ h'' indicates that the algorithm did not converge within 24 hours. OOM stands for out-of-memory. IP-H-CG is IP-Heuristic-CG with 10 random-insertion restarts per pricing call.}
\else
  \parbox{\linewidth}{\footnotesize\textit{Note:} Time to satisfy the termination criterion \eqref{eqn:termination_criterion}, in seconds. For each instance, the fastest method is shown in bold. ``$\geq 24$ h'' indicates that the algorithm did not converge within 24 hours. OOM stands for out-of-memory. IP-H-CG is IP-Heuristic-CG with 10 random-insertion restarts per pricing call.}
\fi
\end{table}

%% file: paper_main/A_Application.tex
We apply our algorithm to a pilot dataset from \cite{dean2022better} with 25 alternatives. Section~\ref{sec:additional_applications} presents further applications, including tests of stochastic rationality using AI-generated choices and consumption budget data.

\citet{dean2022better} propose a test for choice overload based on whether the vector of default-choice probabilities across menus lies in the RUM polytope. They interpret rejection of RUM, together with nonrejection of a model that allows individuals to switch to the default as menus expand, as evidence of choice overload. The pilot dataset contains choices from menus constructed from 24 nondefault alternatives and one default alternative.\footnote{The analysis in their paper uses 12 nondefault alternatives and one default alternative.} It includes all 24 binary menus containing the default, all 276 ternary menus containing the default, and the full menu, for a total of 301 menus. In their empirical analysis, the authors implement the bootstrap procedure using 10,000 simulated draws.

We consider two specifications: a default-only specification that uses only the choice probability of the default alternative in each menu, and a full-choice specification that uses the entire vector of choice probabilities over all available alternatives. Table~\ref{tab:dsg} reports the $p$-values and projection distances for both specifications using all 301 menus,
together with the computation times for Flow-CG, IP-CG, and IP-Heuristic-CG (IP-H-CG).

\begin{table}[t]
\caption{Choice Overload Test Results}
\label{tab:dsg}
\centering
\begin{tabular}{@{}lrrrrr@{}}
\hline
Specification & $p$-value & Distance & Flow-CG & IP-CG & IP-H-CG \\
\hline
Default Only     & 0.108 & 1.489 &   735 &  84,747 &  28,449 \\
All Alternatives & 0.139 & 2.299 & 1,503 & 348,381 & 204,779 \\
\hline
\end{tabular}
\par
\providecommand{\legend}[1]{\parbox{\linewidth}{\footnotesize\textit{Note:} #1}}
\legend{Computation times are in seconds and include calculating the distance and conducting statistical inference. We use 10,000 bootstrap draws. IP-H-CG denotes IP-Heuristic-CG with 10 random-insertion restarts per pricing call. We use an absolute optimality-gap tolerance of
$5\times10^{-4}$ and allow early stopping for bootstrap
projection problems once the relevant comparison is
certified; see Section~\ref{sec:statistical}.}
\end{table}

In the full-choice specification, Flow-CG is 136 times faster than the best existing method, compared with 39 times in the default-only specification. These speedups are smaller than those reported in Section \ref{section:numerical_evidence} because many of the bootstrap projection problems do not need to be solved to high accuracy, reducing the computational advantage of Flow-CG over existing methods.

As an exercise, we investigate whether including more alternatives can yield stronger evidence against the RUM restrictions in this dataset. To this end, we compute the $p$-value for the full dataset of 25 alternatives using all observed binary and ternary menus. We then compare this $p$-value with those obtained from 1,000 randomly sampled subsets, each consisting of 12 nondefault alternatives plus the
default option and their associated binary and ternary menus. To ensure comparability, we exclude the menu containing all 25 alternatives, as this menu cannot appear in any of the subsamples.

\begin{figure}[htbp]
\centering
\begin{subfigure}{.5\textwidth}
  \centering
  \includegraphics[width=.8\linewidth]{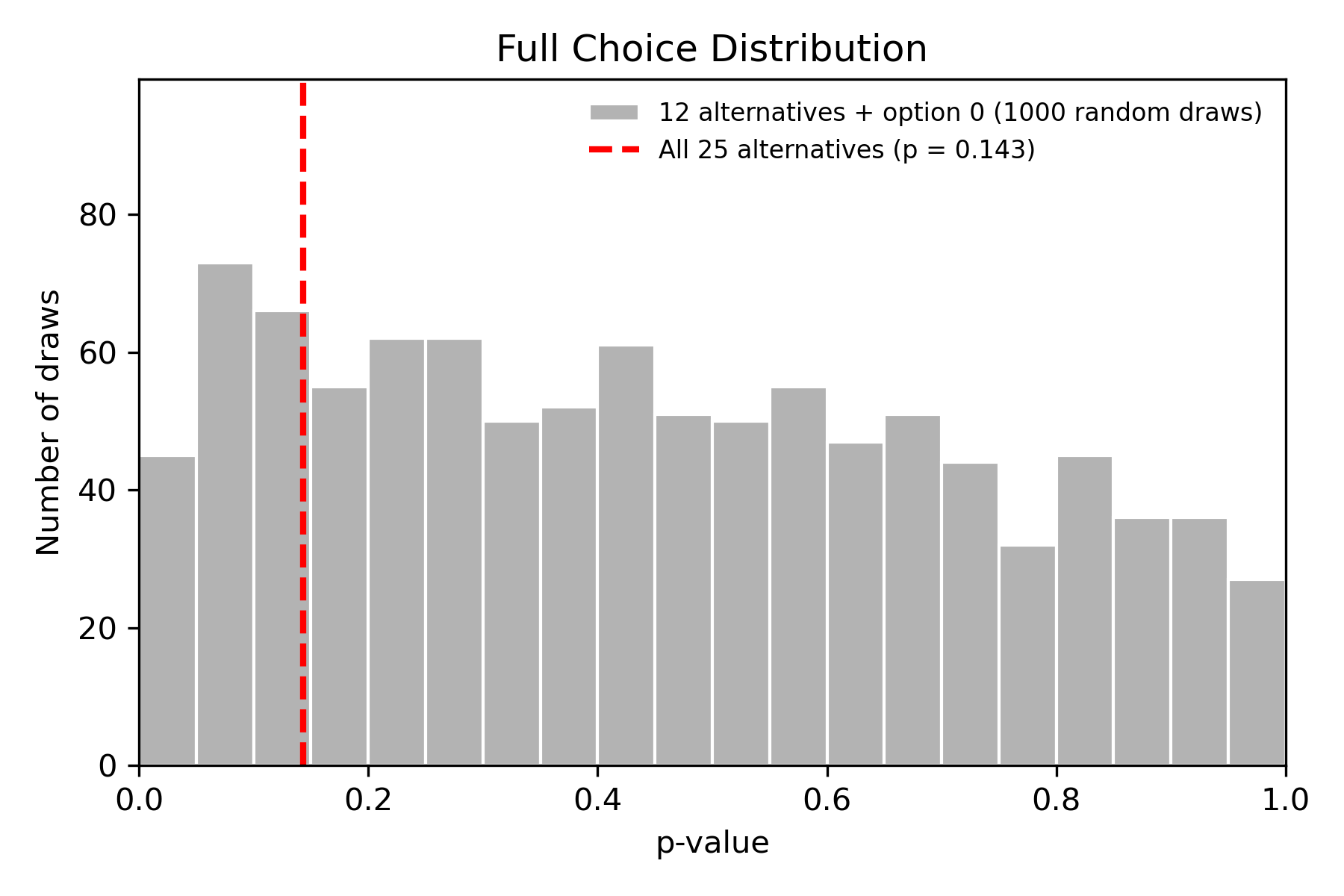}
  \caption{Full-choice data}
  \label{fig:pval_full_choice}
\end{subfigure}%
\begin{subfigure}{.5\textwidth}
  \centering
  \includegraphics[width=.8\linewidth]{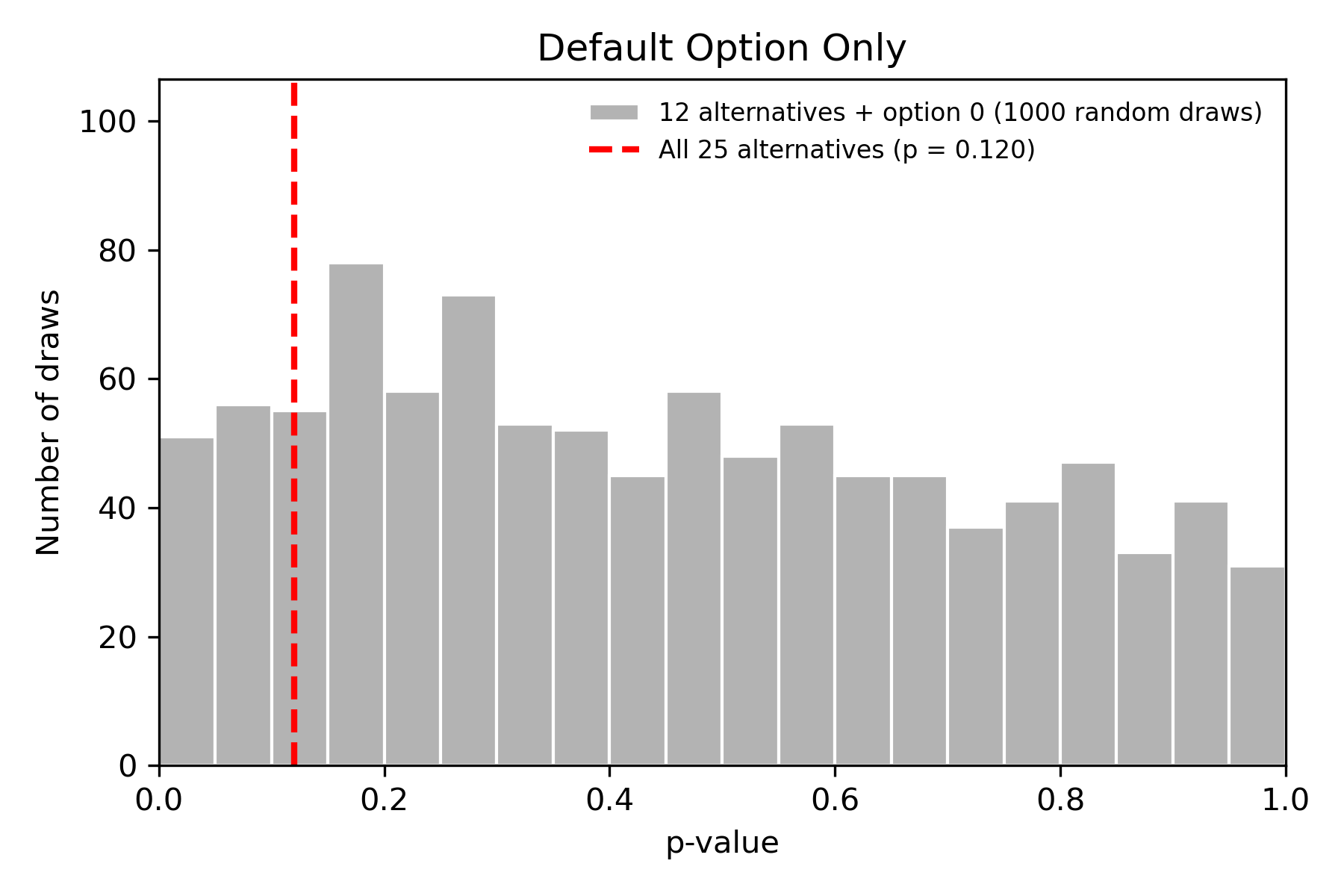}
  \caption{Default-only data}
  \label{fig:pval_default_only}
\end{subfigure}
\caption{RUM test $p$-values across 1,000 random subsamples.
Each subsample includes the binary and ternary menus for 12
nondefault alternatives and the default option.}
\label{fig:pval}
\end{figure}

Figure \ref{fig:pval} plots the distribution of the subsample $p$-values. Using the full choice distribution, the full-sample $p$-value is 0.143, while the first, second, and third quartiles of the subsample $p$-values are 0.212, 0.423, and 0.666, respectively. Using only the choice probabilities of the default option, the full-sample $p$-value is 0.120, and the corresponding quartiles are 0.210, 0.431, and 0.676. Thus, in both cases the full-sample $p$-value lies below the first quartile of the subsample distribution. This suggests that including more alternatives can strengthen the evidence against the RUM restrictions in this application.

%% file: paper_appendix/Z_NP_Hardness.tex
To describe the computational hardness of the \textbf{\DRUM Projection} problem, 
we introduce the following \textbf{\DRUM Strong Membership} problem \citep{grotschel2012geometric}.  Recall that $\mathbb{R}^{|\mathcal{J}|}$ is the ambient space of $\mathcal{P}^\D$.
\begin{definition}[\DRUM Strong Membership Problem]
Given a collection of menus $\D$ and a vector 
$\widehat{p}\in \mathbb{Q}^{\left|\mathcal{J}\right|}$, decide whether 
$\widehat{p}\in \mathcal{P}^{\D}_{\mathrm{RUM}}$.\footnote{The term \textit{strong} follows the terminology of \cite{grotschel2012geometric} and contrasts with the \textit{weak membership problem}, where membership is decided up to a tolerance $\varepsilon$. The term \textit{strong} carries no additional implications in this paper.}
\end{definition}
Clearly, if one can solve the \DRUM Projection problem, then one can also solve the \DRUM Strong Membership problem: given $\widehat{p}\in \mathbb{Q}^{|\mathcal{J}|}$, we can quickly decide whether $\widehat{p}$ belongs to $\mathcal{P}^{\mathcal{D}}$; if $\widehat{p}\in \mathcal{P}^{\mathcal{D}}$, we can apply the algorithm for the \DRUM Projection problem with $\Omega$ equal to the identity matrix to determine whether $\widehat{p}\in \mathcal{P}^{\mathcal{D}}_{\mathrm{RUM}}$ by checking whether the optimal objective value is zero.

We have the following result describing the hardness of the 
\DRUM Projection problem and the \DRUM Strong Membership problem.  \citet{smeulders2021nonparametric} showed a related NP-hardness result for SARP-based rational choice with consumption budgets, not for the \DRUM Strong Membership problem.

\begin{proposition}\label{prop:NPhard}
The \DRUM Strong Membership problem is NP-hard. Consequently, the 
\DRUM Projection problem is NP-hard.\footnote{We define NP-hardness as in Chapter 5 of \cite{garey2002computers} or Section 1.2 of \cite{grotschel2012geometric}. }    
\end{proposition}

The computational hardness associated with the \DRUM polytope has been noted previously \citep{smeulders2018testing,ok2023measuring}, although, to our knowledge, no detailed proof has been provided. We provide a complete proof in Section~\ref{section:NP_proof}. The argument builds on the linear ordering problem \citep{charon2010updated,marti2011linear}: when the observed menus consist of all binary choice sets, the \DRUM polytope can be identified with the linear ordering polytope, over which linear optimization is NP-hard. The main technical step is to verify the geometric and encoding conditions required to invoke the equivalence between linear optimization and strong membership \citep{grotschel2012geometric}. 

%% file: paper_main/A_ADMM_algorithm_flow_mingxi_08232026.tex
In this section, we briefly discuss the ADMM algorithm based on the flow formulation. The linear systems arising in Flow-ADMM can be solved using
techniques similar to those used for BM-ADMM. We therefore
present only the algorithm in the main text and defer the
computational details to Section~\ref{section:computation_efficiency}.
The key difference is that Flow-ADMM exploits the spectral
properties of the graph Laplacian associated with the
power-set digraph, rather than those of the BM coefficient matrix used in the BM-ADMM algorithm.

Under the flow formulation, the \DRUM projection problem
\eqref{eqn:proj3} can be written as
\begin{equation}\label{eqn:flow_formulation2}
\min_{\mathbf f}\quad 
\|Z_{\mathcal D}\mathbf f-\widehat{p}\|_\Omega^2
\qquad
\text{s.t.}\quad
A\mathbf f=b,\;\mathbf f\ge 0,
\end{equation}
where $Z_{\mathcal D}\mathbf f$ gives the implied choice
probabilities over the observed menu collection $\mathcal D$. We introduce an auxiliary variable \(\mathbf s\) and rewrite the problem as
\begin{equation}\label{eqn:problem_split}
\begin{aligned}
\min_{\mathbf f\in\mathbb{R}^{d},\mathbf s \in\mathbb{R}^{d}}\quad &
\|Z_{\mathcal D}\mathbf f-\widehat p\|_\Omega^2+\iota_{\mathbb R_+^d}(\mathbf s) \qquad
\text{s.t.}
\begin{bmatrix}
A\\ I
\end{bmatrix}\mathbf f
-
\begin{bmatrix}
0\\ I
\end{bmatrix}\mathbf s
=
\begin{bmatrix}
b\\ \mathbf 0
\end{bmatrix},
\end{aligned}
\end{equation}
where $d=n2^{n-1}$ is the dimension of $\mathbf{f}$ and $\mathbf{s}$ and \(\iota_{\mathbb R_+^d}(\cdot)\) denotes the indicator function of the nonnegative orthant. The associated augmented Lagrangian is
\begin{equation}\label{eq:admm_main}
\begin{aligned}
\mathcal{L}_\rho(\mathbf f,\mathbf s,\mathbf u,\mathbf v)
=&
\|Z_{\mathcal D}\mathbf f-\widehat p\|_\Omega^2
+\iota_{\mathbb R_+^d}(\mathbf s)
+\langle \mathbf v,A\mathbf f-b\rangle +\langle \mathbf u,\mathbf f-\mathbf s\rangle 
+\frac{\rho}{2}\|A\mathbf f-b\|_2^2
+\frac{\rho}{2}\|\mathbf f-\mathbf s\|_2^2,
\end{aligned}
\end{equation}
where $\rho>0$ is fixed. We introduce the dual variable $\mathbf{v}$ for the constraint $A\mathbf f-b=0$ and the dual variable $\mathbf{u}$ for the constraint $\mathbf f-\mathbf s=0$. A simplified template for our ADMM algorithm with the flow formulation is included in Algorithm \ref{alg:admm_simplify} and a more detailed implementation is included in Section \ref{section:flow_with_relaxation}.

\begin{algorithm}[htbp]
\caption{Flow-ADMM}
\label{alg:admm_simplify}
\begin{algorithmic}[1]
\Require Initial values $(\mathbf{f}_0,\mathbf{s}_0,\mathbf{u}_0,\mathbf{v}_0)$, operators $A,Z_{\mathcal D}$, vector $b$, observed probabilities $\widehat{p}$, penalty $\rho$
\For{$t=0,1,...,$}
    \State \textbf{Step 1:} Update $\mathbf{f}_{t+1}$ by solving
\begin{equation}\label{eqn:update_r}
\min_{\mathbf{f}}
\Bigg\{
\begin{aligned}
&\|Z_{\mathcal D}\mathbf{f}-\widehat p\|_\Omega^2
+\langle \mathbf{u}_t,\mathbf{f}-\mathbf{s}_t\rangle  +\langle \mathbf{v}_t,A\mathbf{f}-b\rangle
+\frac{\rho}{2}\|\mathbf{f}-\mathbf{s}_t\|_2^2
+\frac{\rho}{2}\|A\mathbf{f}-b\|_2^2
\end{aligned}
\Bigg\}    
\end{equation}
    \State \textbf{Step 2:} Update $\mathbf{s}_{t+1}$ by
    \begin{equation}\label{eqn:update_s}
        \begin{split}
    \mathbf{s}_{t+1}\gets&\arg\min_{\mathbf{s}}
    \Big\{
    \iota_{\mathbb{R}_+^d}(\mathbf{s})
    +\langle \mathbf{u}_t,\mathbf{f}_{t+1}-\mathbf{s}\rangle
    +\frac{\rho}{2}\|\mathbf{f}_{t+1}-\mathbf{s}\|_2^2
    \Big\}\\
    = & 
    \Pi_{\mathbb{R}_+^d}\!\left(
    \mathbf{f}_{t+1}+\frac{1}{\rho}\mathbf{u}_t
    \right)               
        \end{split}
    \end{equation}    
    \State \textbf{Step 3:} Update the dual variables
    \begin{equation}
       \mathbf{u}_{t+1}
    =
    \mathbf{u}_t+\rho\bigl(\mathbf{f}_{t+1}-\mathbf{s}_{t+1}\bigr),\quad          \mathbf{v}_{t+1}
    =
    \mathbf{v}_t+\rho\bigl(A\textbf{f}_{t+1}-b\bigr).
    \end{equation}  
\EndFor
\end{algorithmic}
\end{algorithm}

As in BM-ADMM, the main computational advantage of
Algorithm~\ref{alg:admm_simplify} comes from exploiting
the structure of the linear system arising from the
subproblem \eqref{eqn:update_r}. After preprocessing,
each solve requires $O(d\log d+|\mathcal J|^2)$ operations.
This cost is nearly linear in $d$ when the number of
observed choice probabilities is sufficiently small.
We provide the computational details in
Section~\ref{section:computation_efficiency} and establish
a convergence guarantee and an optimality certificate
for Flow-ADMM in Proposition~\ref{prop:admm_certificate}.

%% file: paper_main/A_FW_Guarantee.tex
This section establishes convergence guarantees and optimality certificates for the three algorithms (Flow-CG, Flow-ADMM, and BM-ADMM).
By an optimality certificate, we mean an interval that always contains the optimal value.
In all cases, we also show that the widths of these intervals converge to zero as the algorithms converge.
We start with the Flow-CG algorithm, for which the argument follows directly from standard Frank--Wolfe duality-gap arguments.
For the BM-ADMM and Flow-ADMM algorithms, we first introduce a pushback procedure that restores primal feasibility of the ADMM iterates and then establish the corresponding optimality certificates and convergence guarantees. The pushback procedure and its distance guarantee (Proposition \ref{prop:repair_flow})  appear novel.

\subsection{Flow-CG}
Proposition \ref{prop:convergence_of_FW} provides the convergence guarantee and optimality certificate for the Flow-CG algorithm. In particular, the lower bound $\|p_t-\widehat p\|_\Omega^2-2c(p_t)$ and the upper bound $\|p_t-\widehat p\|_\Omega^2$
are both computable from the quantities generated by the algorithm and therefore
provide an optimality certificate with a user-specified tolerance.
A proof is provided in
Section~\ref{proof:IP_certificate}.
\begin{proposition}\label{prop:convergence_of_FW}
Let $\{p_t\}_{t\in\mathbb{N}}$ be the sequence of iterates generated by the
Flow-CG algorithm in Algorithm~\ref{alg:Flow_CG}. Let $\mathcal{P}^*$ denote
the set of optimizers of the projection problem \eqref{eqn:DC-RUM_Proj}, and
let $v^*$ denote its optimal value. Then $\operatorname{dist}(p_t,\mathcal{P}^*)\to 0$.

Define $c:\mathcal{P}_{\textrm{RUM}}^\D\to\mathbb{R}$ by
\begin{equation}\label{eqn:IP_value}
c(p)
=
\max_{ \mathbf{\widetilde p}\in\mathcal{P}^\D_{\textrm{RUM}}}
(\widehat p-p)^\top\Omega(\mathbf{\widetilde p}-p).
\end{equation}
Then, for every $t$,
\begin{equation}\label{eqn:lb}
\|p_t-\widehat p\|_\Omega^2-2c(p_t)
\leq
v^*
\leq
\|p_t-\widehat p\|_\Omega^2.
\end{equation}
Moreover, $c(p_t)
=
c_{\mathrm{opt}}^t
-
(\widehat p-p_t)^\top\Omega p_t$,
where $c_{\mathrm{opt}}^t$ denotes the maximum path weight
returned by \texttt{Flow\_Pricing} (Algorithm \ref{alg:pricing_rolling_path}) at $p_t$. Finally,
\[
c(p_t)
\leq
6\sqrt{|\D|}\,
\lambda_{\max}(\Omega)\,
\operatorname{dist}(p_t,\mathcal{P}^*),
\]
and hence $\lim_{t\to\infty} c(p_t)=0$.
\end{proposition}
The bounds in \eqref{eqn:lb} hold for any feasible point
$p_t\in\mathcal{P}_{\textrm{RUM}}^\D$,
regardless of whether it is generated by the Flow-CG algorithm.

%% file: paper_main/A_ADMM_certificate_flow_08232026.tex
\subsection{BM-ADMM and Flow-ADMM}

A limitation of the general ADMM framework, however, is that it does not naturally provide certificates of optimality. Standard stopping criteria are typically based on quantities such as primal and dual residuals \citep{neal2011distributed}, which do not directly translate into guarantees on suboptimality. Interestingly, the flow representation allows us to construct explicit suboptimality certificates for the ADMM iterates in the form of lower and upper bounds on the true optimal value. Moreover, these bounds can be shown to converge to the optimal value as the number of ADMM iterations increases and are hence asymptotically nonslack. The remainder of this section develops bounds on the optimal value of problems \eqref{eqn:bm_split} and \eqref{eqn:problem_split} based on the iterates produced by the ADMM algorithm. The procedure can be applied to both BM-ADMM and Flow-ADMM. We use the iterates of Flow-ADMM for illustration.


Unlike the CG iterates, the iterates \((\mathbf{f}_t,\mathbf{s}_t)\) in the ADMM algorithm need not be feasible. In particular, \(\mathbf{f}_t\) may not lie in the unit flow polytope and the induced choice probability vector \(Z_{\mathcal{D}}\mathbf{f}_t\) may not correspond to a feasible (i.e., rationalizable) vector of observed choice probabilities. As a result, the pricing certificate in Proposition~\ref{prop:convergence_of_FW}, which is only applicable to feasible choice probabilities, cannot be directly used.

To address this issue, we propose a pushback procedure to restore the feasibility of $\mathbf f_t$.\footnote{Directly solving the $\ell_2$ projection problem
\begin{equation*}
\min_{\mathbf f}\|\mathbf{f}_t - \mathbf{f}\|_2^2 \quad \text{subject to } A\mathbf{f} = b, \mathbf{f} \geq 0,
\end{equation*}
is challenging in high dimensions (e.g., $\mathbf f \in \mathbb{R}^{n2^{n-1}}$ with $n$ alternatives). First-order methods based on CG/FW may require generating a large number of vertices before attaining the desired projection accuracy. Second-order methods, such as interior-point algorithms, are also difficult to apply as solving the associated KKT systems becomes computationally prohibitive. Meanwhile, first-order methods such as ADMM or PDHG do not guarantee feasibility of the iterates.
} The details of the pushback procedure are provided in Algorithm~\ref{alg:pushback_infeasible_flow}, and a graphical illustration for the three-alternative case is provided in Figure~\ref{fig:backward_adjustment_panel}. For Flow-ADMM, this procedure can be applied directly to the iterates. For BM-ADMM, one first maps the potentially non-rationalizable choice probabilities to flows using $K_n$, applies the pushback procedure, and then maps the resulting feasible flows back to the choice-probability space using $Z_n$.

We briefly describe the procedure. Given a possibly infeasible flow $\mathbf{f}$ on the power-set digraph $G=(N,E)$, the algorithm first enforces nonnegativity by taking its componentwise positive part, $\mathbf{f}^+=\max\{\mathbf{f},0\}$. It then proceeds backward through the graph, constructing a feasible flow layer by layer. Specifically, the flows from layer $k-1$ to layer $k$ are determined by the already-constructed feasible flows from layer $k$ to layer $k+1$.

When constructing feasible flows from layer $k-1$ to layer $k$, the algorithm repairs the flows based on the nodes in layer $k$. At the terminal node $X$, we set $s(X)=1$.
At each other nonempty node $D$ in layer $k$, the
algorithm computes the total outflow
\[
s(D)=\sum_{D'\in\delta^+(D)}
\mathbf{f}^{\textrm{feas}}(D,D').
\] 
Here, $\delta^+(D)$ denotes the set of child nodes of $D$ in layer $k+1$, and $\{\mathbf{f}^{\textrm{feas}}(D,D')\}_{D'\in\delta^+(D)}$ denotes the already repaired flows on the arcs from $D$ to these child nodes. 

The algorithm then allocates $s(D)$ among the incoming arcs to $D$. These arcs are of the form $(D',D)$ with $D'\in\delta^-(D)$, where $\delta^-(D)$ denotes the set of parent nodes of $D$. The allocation is proportional to the truncated input
flows $\mathbf f^+(D',D)$ when their sum is positive,
and uniform otherwise. The rule is
\begin{align*}
\mathbf{f}^{\textrm{feas}}\left(D',D\right)
&=
\begin{cases}
\dfrac{s\left(D\right)}{|\delta^-\left(D\right)|}, & m\left(D\right)=0,\\[1mm]
\dfrac{s\left(D\right)\,\mathbf{f}^+\left(D',D\right)}{m\left(D\right)}, & m\left(D\right)>0,
\end{cases}
\qquad 
m\left(D\right)
=\sum_{D'\in\delta^-(D)} \mathbf{f}^+\left(D',D\right),
\end{align*}
for all $D'\in \delta^-(D)$. When $m(D)>0$, each incoming arc $(D',D)$ receives
the fraction $\mathbf f^+(D',D)/m(D)$ of the total
flow $s(D)$. If $m(D)=0$, this flow is distributed
uniformly across the incoming arcs.

Surprisingly, this simple flow-repair heuristic admits a provable distance guarantee: the discrepancy between the pushed-back feasible flow and the original infeasible flow is bounded above by the infeasibility of the original flow. In the context of our ADMM algorithms, this guarantee implies that the corresponding pushed-back feasible flows also approach the optimal set as the iterates approach feasibility and optimality. 

\begin{algorithm}[h!]
\caption{Backward adjustment of an infeasible flow on the power-set digraph}
\label{alg:pushback_infeasible_flow}
\begin{algorithmic}[1]
\Require Arc flows $\{\mathbf{f}\left(D',D\right)\}_{(D',D)\in E}$ on the power-set digraph $G=\left(N,E\right)$ with $n$ alternatives.
\For{each arc $(D',D)\in E$}
\Comment{Initial nonnegativity step}
    \State $\mathbf{f}^+(D',D) \gets \max\{\mathbf{f}(D',D),0\}$
\EndFor
\For{$k=n,n-1,\dots,1$}
    \For{each choice set $D$ with cardinality $k$}
        \State Let $ \delta^-(D):=\{D'\subset D: |D'|=k-1\}$.
\Comment{predecessors of $D$}
        \If{$k=n$}
            \Comment{top layer: normalize total incoming flow to $1$}
             \State Define $s(D)=1$.
        \Else
            \Comment{lower layers: preserve total mass at $D$}
    \State Let $\delta^+(D):=\{D'\supset D: |D'|=k+1\}$.
\Comment{successors of $D$}
    \State Compute $s(D)\gets \sum_{D'\in \delta^+(D)}\mathbf{f}^\textrm{feas}\left(D,D'\right) $.
            \Comment{total flow out of $D$}
        \EndIf
        \State Compute $m(D)\gets \sum_{D'\in\delta^-(D)} \mathbf{f}^+\left(D',D\right)$.
        \If{$m(D)=0$}
            \State Set $\mathbf{f}^\textrm{feas}\left(D',D\right)\gets s(D)/|\delta^-(D)|$ for all $D'\in \delta^-(D)$.
        \Else
            \State Set $\mathbf{f}^\textrm{feas}\left(D',D\right)\gets s(D)\mathbf{f}^+\left(D',D\right)/m(D)$ for all $D'\in \delta^-(D)$.
        \EndIf
    \EndFor
\EndFor

\State \Return $\{\mathbf{f}^\textrm{feas}\left(D',D\right)\}_{(D',D)\in E}$
\end{algorithmic}
\end{algorithm}

\begin{figure}[htbp]
\centering

\captionsetup[subfigure]{
    font=small,
    skip=3pt,
    justification=centering
}

\tikzset{
    powersetgraph/.style={
        x=0.78cm,
        y=0.78cm,
        every node/.style={font=\tiny}
    },
    setnode/.style={
        circle,
        draw,
        inner sep=0.8pt,
        minimum size=3.8mm
    },
    active/.style={
        setnode,
        fill=blue!18
    },
    smallsetnode/.style={
        circle,
        draw,
        inner sep=0.2pt,
        minimum size=2.8mm,
        font=\tiny
    },
    smallactive/.style={
        smallsetnode,
        fill=blue!18
    },
    base/.style={gray!45,thin,-},
    useedge/.style={blue,thick,-},
    bad/.style={red,dashed,thick,-}
}

\newcommand{\powersetgraphbase}{%
    \node[setnode] (e) at (0,0) {$\emptyset$};

    \node[setnode] (1) at (1.5, 1.35) {$1$};
    \node[setnode] (2) at (1.5, 0)    {$2$};
    \node[setnode] (3) at (1.5,-1.35) {$3$};

    \node[setnode] (12) at (3, 1.35) {$12$};
    \node[setnode] (13) at (3, 0)    {$13$};
    \node[setnode] (23) at (3,-1.35) {$23$};

    \node[smallsetnode] (123) at (4.5,0) {$123$};

    \foreach \u/\v in {
        e/1,e/2,e/3,
        1/12,1/13,2/12,2/23,3/13,3/23,
        12/123,13/123,23/123%
    }{
        \draw[base] (\u)--(\v);
    }
}

\newcommand{\powersetpanel}[2]{%
    \begin{subfigure}[t]{0.31\linewidth}
        \centering
        \begin{adjustbox}{max width=\linewidth}
            \begin{tikzpicture}[powersetgraph]
                \powersetgraphbase
                #1
            \end{tikzpicture}
        \end{adjustbox}
        \caption{#2}
    \end{subfigure}%
}


\powersetpanel{
    \foreach \u/\v in {
        e/1,e/2,e/3,
        1/12,1/13,2/12,2/23,3/13,3/23,
        12/123,13/123,23/123%
    }{
        \draw[bad] (\u)--(\v);
    }
}{Truncate negatives}%
\hfill
\powersetpanel{
    \node[smallactive] at (123) {$123$};

    \draw[useedge] (12)--(123);
    \draw[useedge] (13)--(123);
    \draw[useedge] (23)--(123);
}{Normalize inflow at $123$}%
\hfill
\powersetpanel{
    \node[active] at (12) {$12$};

    \draw[useedge] (1)--(12);
    \draw[useedge] (2)--(12);
}{Update $12$}

\par\medskip


\powersetpanel{
    \node[active] at (13) {$13$};

    \draw[useedge] (1)--(13);
    \draw[useedge] (3)--(13);
}{Update $13$}%
\hfill
\powersetpanel{
    \node[active] at (23) {$23$};

    \draw[useedge] (2)--(23);
    \draw[useedge] (3)--(23);
}{Update $23$}%
\hfill
\powersetpanel{
    \node[active] at (1) {$1$};

    \draw[useedge] (e)--(1);
}{Update $1$}

\par\medskip


\powersetpanel{
    \node[active] at (2) {$2$};

    \draw[useedge] (e)--(2);
}{Update $2$}%
\hspace{0.035\linewidth}%
\powersetpanel{
    \node[active] at (3) {$3$};

    \draw[useedge] (e)--(3);
}{Update $3$}

\par\medskip


\begin{subfigure}[t]{\linewidth}
    \centering
    \begingroup
    \footnotesize
    \begin{adjustbox}{max width=\linewidth}
        $\displaystyle
        \begin{gathered}
            f_{D',D}=
            \begin{cases}
                \dfrac{s(D)}{|\delta^-(D)|},
                    & m(D)=0,\\[1mm]
                \dfrac{s(D)\,f^+_{D',D}}{m(D)},
                    & m(D)>0
            \end{cases}\\
                        m(D)=
            \sum_{D'\in\delta^-(D)} f^+_{D',D},
        \end{gathered}
        \qquad
        \begin{gathered}
            s(D)=
            \begin{cases}
                1,
                    & D=X,\\[1mm]
                \displaystyle
                \sum_{D'\in\delta^+(D)} f_{D,D'},
                    & D\ne X.
            \end{cases}
        \end{gathered}
        $
    \end{adjustbox}
    \endgroup
    \caption{Backward rescaling rule}
\end{subfigure}

\caption{Backward adjustment on the 3-element power-set digraph.
All arcs are directed from left to right; arrowheads are omitted.
After truncating negative arc flows, the incoming flow to the
terminal node is normalized to one. The algorithm then proceeds
backward layer by layer, rescaling the incoming arcs at each
nonempty node $D$ so that they sum to the prescribed flow $s(D)$.
In Panel~(i), $f_{D',D}$ denotes the repaired flow on arc $(D',D)$,
while $f^+_{D',D}=\max\{f_{D',D}^{\mathrm{in}},0\}$ denotes the
nonnegative part of the input flow $f_{D',D}^{\mathrm{in}}$.}
\label{fig:backward_adjustment_panel}
\end{figure}

\begin{proposition}\label{prop:repair_flow}
Let ${\mathbf{f}}=\{\mathbf{f}\left(D',D\right)\}_{\left(D',D\right)\in E}$ be a collection of possibly infeasible arc flows on the power-set digraph $G=(N,E)$ with $n$ alternatives. Let $\mathbf{f}^{\textrm{feas}}=\{\mathbf{f}^{\textrm{feas}}\left(D',D\right)\}_{\left(D',D\right)\in E}$ be the output of Algorithm~\ref{alg:pushback_infeasible_flow} with input $\mathbf{f}$. Let $\|\cdot\|_1$ denote the $\ell_1$ norm. Then
\begin{equation}
\|\mathbf{f}-\mathbf{f}^{\textrm{feas}}\|_1
\le
n\|A\mathbf{f}-b\|_1
+
2n\,\|\min\{\mathbf{f},0\}\|_1,
\end{equation}
where $\min\{\mathbf{f},0\}=\{\min\{\mathbf{f}\left(D',D\right),0\}\}_{(D',D)\in E}$ denotes the vector of negative parts of $\mathbf{f}$.
\end{proposition}
A proof is included in Section \ref{proof:push_back}.
Applying the objective function to the pushed-back feasible flow $\mathbf f^{\mathrm{feas}}$ yields a valid upper bound on the optimal value of problem~\eqref{eqn:problem_split}. Once feasibility has been restored, a corresponding lower bound can be obtained from \eqref{eqn:lb}.  Proposition \ref{prop:admm_certificate} below provides a convergence guarantee for the BM-ADMM
and Flow-ADMM algorithms. It shows that both algorithms converge in objective value to the optimum of the \DRUM problem. In addition, based on Proposition \ref{prop:repair_flow}, both algorithms can produce computable upper and lower bounds whose gap converges to zero, thereby providing an asymptotically tight certificate of optimality. 

\begin{proposition}\label{prop:admm_certificate}
Consider either of the following two constructions of $\mathbf{f}^{\mathrm{feas}}_t$:
\begin{enumerate}[label=(\roman*)]
    \item Let $\mathbf{f}_t$ be the output of Algorithm \ref{alg:admm_simplify}, and let $\mathbf{f}^{\mathrm{feas}}_t$ be the output of Algorithm \ref{alg:pushback_infeasible_flow} applied to $\mathbf{f}_t$.
    
    \item Let $\mathbf{p}_t$ be the output of Algorithm \ref{alg:admm_BM}. Define $\mathbf{f}_t=K_n\mathbf{p}_t$, and let $\mathbf{f}^{\mathrm{feas}}_t$ be the output of Algorithm \ref{alg:pushback_infeasible_flow} applied to $\mathbf{f}_t$.
\end{enumerate}

Let $v^*$ denote the optimal value of \eqref{eqn:DC-RUM_Proj}, and let $c:\mathcal{P}_{\textrm{RUM}}^\D\to\mathbb{R}$ be defined as in \eqref{eqn:IP_value}. Define
\begin{equation}
    \mathrm{LB}_t
    =
    \|Z_{\mathcal D}\mathbf{f}^{\mathrm{feas}}_t-\widehat p\|_\Omega^2
    -
    2c\!\left(Z_{\mathcal D}\mathbf{f}^{\mathrm{feas}}_t\right),
    \qquad
    \mathrm{UB}_t
    =
    \|Z_{\mathcal D}\mathbf{f}^{\mathrm{feas}}_t-\widehat p\|_\Omega^2.
\end{equation}
Then, in either case (i) or (ii),
\begin{equation*}
    \mathrm{LB}_t \leq v^* \leq \mathrm{UB}_t,
    \qquad
    \mathrm{LB}_t \to v^*,
    \qquad
    \mathrm{UB}_t \to v^*.
\end{equation*}
\end{proposition}

This proposition is a special case of
Proposition~\ref{prop:admm_certificate_overrelaxation}
in the appendix, which additionally allows for
over-relaxation and constraint-specific penalty parameters.
The proof is provided in
Section~\ref{proof:admm_certificate_proof}.
\begin{remark}
    Our convergence proof  generalizes the standard convergence argument to accommodate relaxation, proximal updates, and a general positive-definite matrix weighting in the augmented Lagrangian, as required by our algorithms. There is a large literature on the convergence of ADMM; see, for example, \cite{eckstein1992douglas,he20121,davis2017convergence}. We have not found a convergence proof that simultaneously accommodates all of the variations required by our algorithms.  We therefore provide a self-contained convergence proof, building on the argument in \cite{fazel2013hankel}.    
\end{remark}

%% file: paper_appendix/Z_Additional_Applications.tex
\subsection{Testing the Stochastic Rationality of AI}\label{section:ai}
As a second application, we test the stochastic rationality of LLM-based AI systems. Our primary purpose is to illustrate a setting with potentially dense menu coverage. There is growing interest in assessing AI behavior \citep{horton2023large}, using AI systems for survey and market research, including conjoint analysis and willingness-to-pay estimation \citep{brand2023using,goli2024frontiers,wang2026large}, and developing shopping agents that act on users' behalf \citep{allouah2026your}. These applications often implicitly assume the existence of strict linear preferences. This assumption, together with the stochastic nature of AI-generated responses, makes testing stochastic rationality relevant to both research and practice.

We design a task in which an LLM agent chooses from menus drawn from a set of 18 Apple Watches. We construct the menu collection in advance by including all binary and ternary menus and sampling $3\%$ of the menus of each larger cardinality. We end up with 8,807 menus. We query each agent 2,000 times per menu, randomizing the order of the items in each query to mitigate presentation-order effects. We test GPT-6 Luna and DeepSeek V4 Flash, selected for their relatively advanced capabilities and low cost.  A sample prompt is included in Section \ref{sec:sample_prompt}.

Table~\ref{tab:ai} reports the $p$-values and estimated distances for the two models, along with the computation times for BM-ADMM, IP-CG, and IP-Heuristic-CG (IP-H-CG). The results demonstrate the computational advantage of BM-ADMM. We also construct two-sided 95\% confidence intervals for the distances.\footnote{Section~\ref{sec:statistical} describes the confidence interval procedure.} These intervals are $[11.942, 14.836]$ for GPT-6 Luna and $[5.453, 8.988]$ for DeepSeek V4 Flash. Using distance to the RUM polytope as a measure of
departure from stochastic rationality, we find that
DeepSeek V4 Flash exhibits smaller deviations than
GPT-6 Luna. By the Bonferroni bound, this difference
is statistically significant at the $10\%$ level.

\begin{table}[t]
\caption{Stochastic Rationality Test Results for AI Models}
\label{tab:ai}
\centering
\begin{tabular}{@{}lrrrrr@{}}
\hline
Model & $p$-value & Distance & BM-ADMM & IP-CG & IP-H-CG \\
\hline
GPT-6 Luna        & 0.000 & 13.390 & 110 & $\geq 24$ h & $\geq 24$ h \\
DeepSeek V4 Flash & 0.000 &  7.221 & 131 & $\geq 24$ h & $\geq 24$ h \\
\hline
\end{tabular}
\par
\providecommand{\legend}[1]{\parbox{\linewidth}{\footnotesize\textit{Note:} #1}}
\legend{Computation times are in seconds unless otherwise indicated and include calculating the distance and conducting statistical inference. We use 10,000 bootstrap draws. IP-H-CG denotes IP-Heuristic-CG with 10 random-insertion restarts per pricing call. We use an absolute optimality-gap tolerance of
$5\times10^{-4}$ and allow early stopping for bootstrap
projection problems once the relevant comparison is
certified; see Section~\ref{sec:statistical}.}
\end{table}

\subsection{Testing RUM with consumption budget data}

In our final application, we revisit the 25-budget, three-good expenditure dataset used by \cite{smeulders2021nonparametric}. With 1,000 bootstrap draws, Flow-CG and IP-H-CG both yield a $p$-value of $0.501$. Because the resulting graph contains only $686$ of the $2^{25}$ possible nodes, we run Flow-CG on a CPU using a single process. Flow-CG completes the test in $49.0$ seconds, compared with $34{,}632$ seconds (approximately $9.6$ hours) for IP-H-CG, an approximately 700-fold speedup.

%% file: paper_appendix/Z_Index.tex
Fix an alternative $x$ and enumerate the remaining elements by $1,\ldots,k$. For example, if $|X|=n$, then $k=n-1$. We index either the choice probabilities $p(x,D)$ or the arc flows $f(D\setminus\{x\},D)$ using a binary tree representation. This representation facilitates the efficient computation of the butterfly transform. In practice, when there are $n$ alternatives, we store $n$ vectors of length $2^{\,n-1}$, one for each alternative $x\in X$.

The indexing scheme is shown below in Figure \ref{fig:binary_tree_index}. Each root-to-leaf path in the tree corresponds to a binary sequence of length $k$, where each branch indicates whether an element is included in or excluded from the subset. Every path uniquely identifies a set $D\subseteq X\setminus\{x\}$. Ordering the subsets according to this binary structure provides a convenient indexing scheme for the vectors of choice probabilities or arc flows and aligns naturally with the standard ordering used in fast subset transforms.
\begin{figure}[htbp]
\centering
\begin{forest}
for tree={
  circle,draw,inner sep=1.2pt,
  s sep=7mm, l sep=9mm,
  edge label/.style={midway,fill=white,inner sep=3pt,font=\scriptsize},
}
[$ $
  [$\textrm{include }k $, edge label={node[above left] {$0$}}
    [$\textrm{include } k-1 $, edge label={node[above left] {$0$}}
      [$\textrm{...} $, edge label={node[above left] {$0$}}]
      [$\textrm{...} $, edge label={node[above right] {$1$}}]]
    [$\textrm{exclude } k-1 $, edge label={node[above right] {$1$}}
      [$\textrm{...} $, edge label={node[above left] {$0$}}]
      [$\textrm{...} $, edge label={node[above right] {$1$}}]]
  ]
  [$\textrm{exclude } k $, edge label={node[above right] {$1$}}
    [$\textrm{include }k-1 $, edge label={node[above left] {$0$}}
      [$\textrm{...} $, edge label={node[above left] {$0$}}]
      [$\textrm{...} $, edge label={node[above right] {$1$}}]]
    [$\textrm{exclude }k-1 $, edge label={node[above right] {$1$}}
      [$\textrm{...} $, edge label={node[above left] {$0$}}]
      [$\textrm{...} $, edge label={node[above right] {$1$}}]]
  ]
]
\end{forest}
\caption{Binary-tree representation used to index subsets of $X\setminus\{x\}$.}
\label{fig:binary_tree_index}
\end{figure}
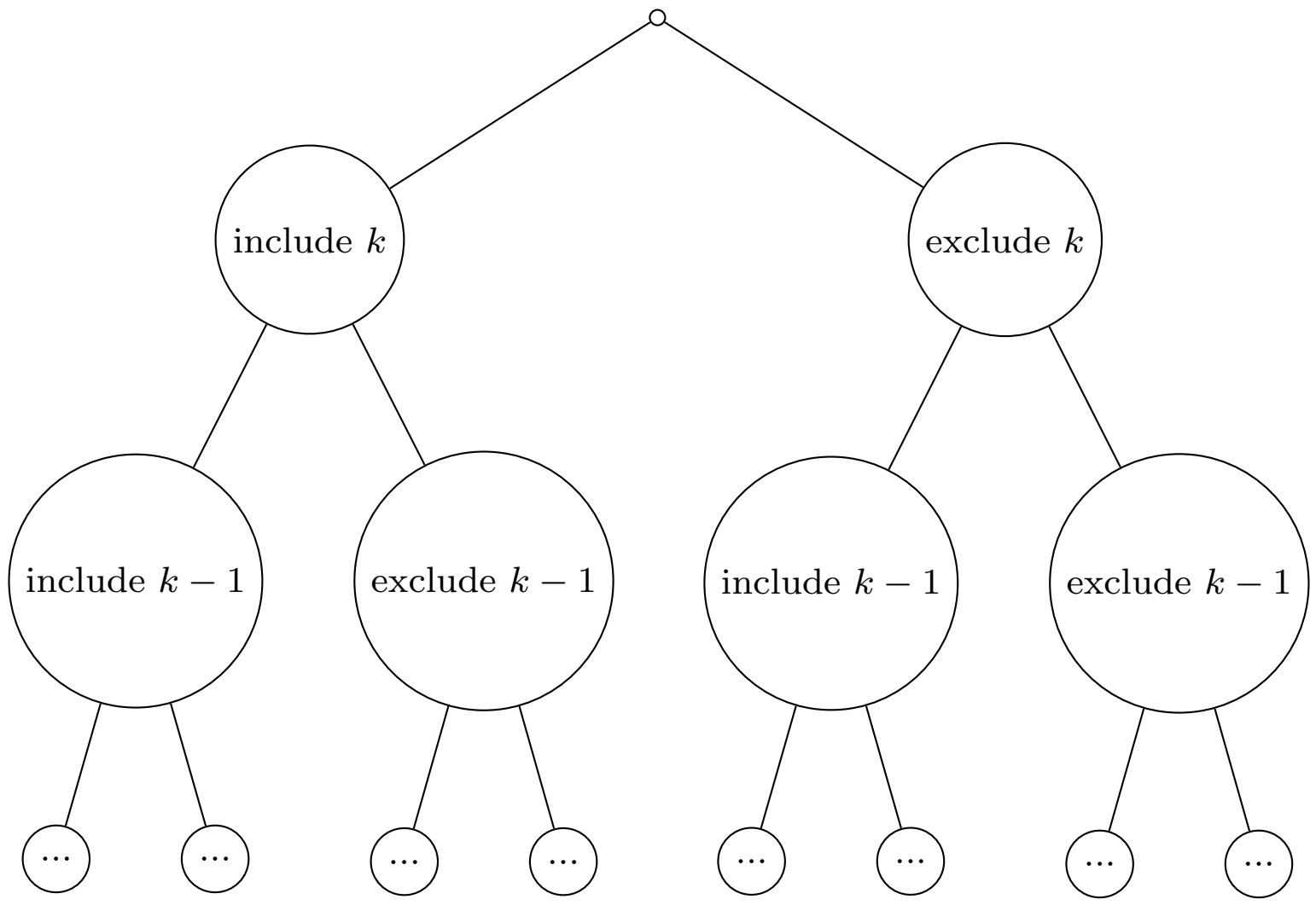

\begin{figure}[htbp]
\centering
\begin{forest}
for tree={
  circle,
  draw,
  inner sep=1.2pt,
  s sep=7mm,
  l sep=9mm,
  align=center,
  edge label/.style={midway,fill=white,inner sep=3pt,font=\scriptsize},
}
[$ $
  [{$\textrm{include }3$},
    edge label={node[above left] {$0$}}
    [{$\textrm{include }2$},
      edge label={node[above left] {$0$}},
      label={[yshift=-4pt]below:
        {\shortstack{
          \scriptsize $D=\{1,2,3\}$\\[2pt]
          \scriptsize $p(1,\{1,2,3\})$\\[1pt]
          \scriptsize $f(\{2,3\},\{1,2,3\})$
        }}
      }
    ]
    [{$\textrm{exclude }2$},
      edge label={node[above right] {$1$}},
      label={[yshift=-4pt]below:
        {\shortstack{
          \scriptsize $D=\{1,3\}$\\[2pt]
          \scriptsize $p(1,\{1,3\})$\\[1pt]
          \scriptsize $f(\{3\},\{1,3\})$
        }}
      }
    ]
  ]
  [{$\textrm{exclude }3$},
    edge label={node[above right] {$1$}}
    [{$\textrm{include }2$},
      edge label={node[above left] {$0$}},
      label={[yshift=-4pt]below:
        {\shortstack{
          \scriptsize $D=\{1,2\}$\\[2pt]
          \scriptsize $p(1,\{1,2\})$\\[1pt]
          \scriptsize $f(\{2\},\{1,2\})$
        }}
      }
    ]
    [{$\textrm{exclude }2$},
      edge label={node[above right] {$1$}},
      label={[yshift=-4pt]below:
        {\shortstack{
          \scriptsize $D=\{1\}$\\[2pt]
          \scriptsize $p(1,\{1\})$\\[1pt]
          \scriptsize $f(\emptyset,\{1\})$
        }}
      }
    ]
  ]
]
\end{forest}
\caption{Indexing binary tree for alternative 1.
Each root-to-leaf path determines a choice set containing
alternative 1, with $0$ indicating inclusion and $1$
indicating exclusion of each remaining alternative.
Each leaf is annotated with the corresponding choice set,
choice probability, and arc flow.}
\label{fig:binary_tree_index2}
\end{figure}
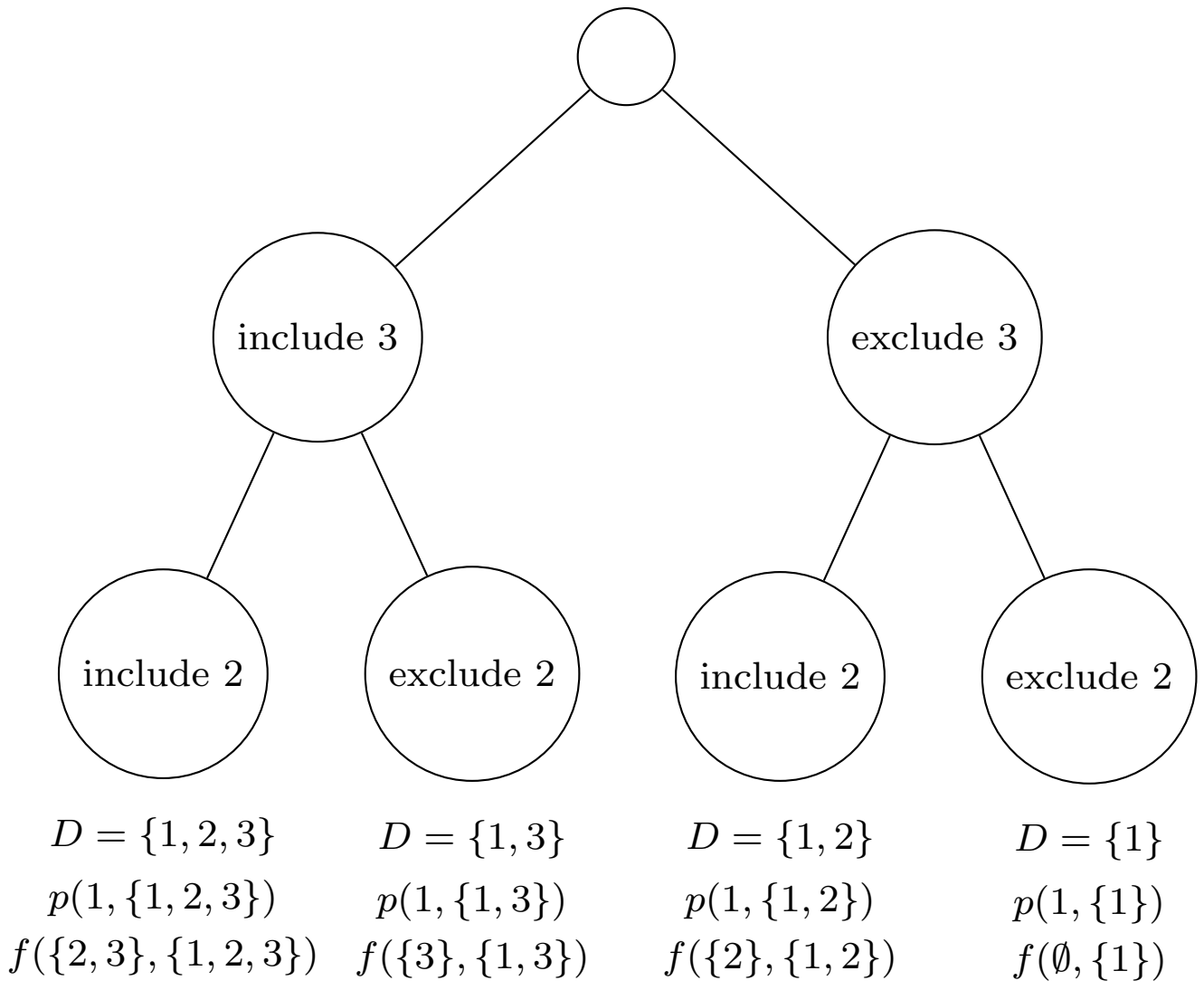

For example, consider the case with three alternatives and focus on alternative \(1\). There are four choice sets containing \(1\): \(\{1,2,3\}\), \(\{1,3\}\), \(\{1,2\}\), and \(\{1\}\). The corresponding indexing binary tree is shown in Figure~\ref{fig:binary_tree_index2}.

Under the indexing scheme above, a choice-probability or arc-flow
coordinate can be located as follows. Let $n=|X|$ and $k=n-1$.
Represent a choice set $D$ by its membership vector
$c_D\in\{0,1\}^n$, with entries ordered by alternative index.
For each $x\in D$, remove the entry corresponding to $x$ to obtain
$c_D^x\in\{0,1\}^k$, preserving the order of the remaining entries.
The zero-based index of $p(x,D)$ or $f(D\setminus\{x\},D)$
within the vector associated with $x$ is
\begin{equation}
    \sum_{s=1}^{k}\left(1-c_D^x[s]\right)2^{s-1}.
\end{equation}
For one-based indexing, add $1$.

For example, consider the case with three alternatives and the choice set $D=\{1,2\}$. The choice set's binary representation is $c_D=[110]$. To access the choice probability/arc flow of the choice set $D$ with alternative 1, first remove the entry in $c_D$ associated with the alternative $1$,  leaving $c^1_D = [10]$. Then the address is calculated as $\left(1-c_D^1[1]\right)1+ \left(1-c_D^1[2]\right)2=2$,
which is the third entry in a zero-based system.

%% file: paper_appendix/Z_proximal.tex
The simplex proximal operators in the second step of
Algorithm~\ref{alg:admm_BM} can be evaluated separately across
choice sets after combining the fitting loss with the proximal
term. To see this, note that
\begin{align*}
&\|\pcalD-\widehat p\|_\Omega^2
+
\|\pcalD
  -\mathbf{\overline{p}}_{\mathcal{D},t}\|_{\Omega^{\textrm{prox}}}^2 =
\lambda_{\max}\left(\Omega\right)\|\pcalD\|_2^2
-
2\left\langle
\Omega\widehat p
+\Omega^{\textrm{prox}} \overline{\mathbf{p}}_{\mathcal{D},t},
\pcalD
\right\rangle
+C_t,
\end{align*}
where $C_t$ does not depend on $\mathbf{\overline{p}}$. 

For each $D\in\mathcal D$, define $h^D_{t}
=
\lambda_{\max}\left(\Omega\right)\pD_t
+
\sum_{E\in\mathcal D}
\Omega_{D,E}
\left(\widehat p^E- \mathbf{\overline{p}}^{E}_t\right)$,
where $\Omega_{D,E}$ denotes the block of $\Omega$ with rows
associated with $D$ and columns associated with $E$. We have
\begin{align*}
\mathbf{\overline{p}}^D_{t+1}
&=
\arg\min_{\pD}
\Biggl\{
\lambda_{\max}\left(\Omega\right)\| \pD\|_2^2
-2\langle h^D_{t},\pD\rangle
+\iota_{\Delta^D}(\pD)
+\left\langle
\mathbf u^D_{t},
\mathbf{p}^D_{t+1}-\pD
\right\rangle
+\frac{\rho}{2}
\|\pD-\mathbf{p}^D_{t+1}\|_2^2
\Biggr\} \\
&=
\arg\min_{\pD}
\Biggl\{
\iota_{\Delta^D}(\pD)
+\frac{2\lambda_{\max}\left(\Omega\right)+\rho}{2}
\left\|
\pD
-
\frac{
2h^D_{t}+\mathbf u^{D}_t
+\rho\mathbf p^D_{t+1}
}{
2\lambda_{\max}\left(\Omega\right)+\rho
}
\right\|_2^2
\Biggr\} \\
&=
\Proj_{\Delta^D}
\left(
\frac{
2h^D_{t}+\mathbf u^D_{t}
+\rho{\mathbf p}^D_{t+1}
}{
2\lambda_{\max}\left(\Omega\right)+\rho
}
\right),
\end{align*}
where $\mathbf{u}^D_{t}$, $\mathbf{p}^D_{t+1}$, and
$\overline{\mathbf{p}}^D_{t+1}$ denote the subvectors of $\mathbf{u}_t$,
$\mathbf{p}_{t+1}$, and $\overline{\mathbf{p}}_{t+1}$, respectively, associated
with the choice set $D$. For each $D \in 2^X\backslash \D $, we have
\begin{align*} 
\overline{\mathbf{p}}^D_{t+1}
&=
\arg\min_{\overline{\mathbf{p}}^D}
\left\{
\iota_{\Delta^D}(\pD)
+ \left\langle \mathbf{u}^D_{t},
\mathbf{p}^D_{t+1}-\pD\right\rangle
+ \frac{\rho}{2}
\left\|\pD-\mathbf{p}^D_{t+1}\right\|_2^2
\right\} \\
&=
\arg\min_{\pD}
\left\{
\iota_{\Delta^D}(\pD)
+ \frac{\rho}{2}
\left\|
\pD
-
\left(
\mathbf{p}^D_{t+1}
+\frac{\mathbf{u}^D_{t}}{\rho}
\right)
\right\|_2^2
\right\}=
\Proj_{\Delta^D}
\left(
\mathbf{p}^D_{t+1}
+\frac{\mathbf{u}^D_{t}}{\rho}
\right).
\end{align*}

For each $D$, the proximal operator can be computed using
Algorithm~\ref{alg:simplex_projection} in
$O\bigl(|D|\log |D|\bigr)$ time.

\begin{algorithm}[htbp]
\caption{Projection onto the Probability Simplex}
\label{alg:simplex_projection}
\begin{algorithmic}[1]
\Require Vector $x\in\mathbb{R}^n$
\State Sort the entries of $x$ in nonincreasing order: $x_{(1)}\ge x_{(2)}\ge \cdots \ge x_{(n)}$.
\State Compute
\[
k^\star
=
\max\left\{
i\in[n]:
x_{(i)}
+\frac{1}{i}
\left(
1-\sum_{j=1}^i x_{(j)}
\right)
\ge 0
\right\}.
\]
\State Compute
\[
\theta
=
\frac{1}{k^\star}
\left(
1-\sum_{j=1}^{k^\star}x_{(j)}
\right).
\]
\State Set
\[
\Proj_{\Delta^n}(x)
=
\max\{x+\theta\mathbf{1},0\},
\]
where $\mathbf{1}$ is the all-ones vector and the maximum is taken componentwise.
\end{algorithmic}
\end{algorithm}


%% file: paper_main/A_Additional_Simulations.tex
\input{paper_main/A_table_n19_numerical1}
\input{paper_main/A_table_n20_numerical1}

%% file: paper_main/A_table_n19_numerical1.tex
\begin{table}[t]
\centering
\resizebox{\textwidth}{!}{%
\begin{tabular}{c|cccccccccc}
\toprule
\diagbox{Method}{Density}
& Pairwise
& 0.025\%
& 0.05\%
& 0.1\%
& 0.2\%
& 0.3\%
& 0.5\%
& 0.7\%
& 1\%
& 10\% \\
\midrule

BM-ADMM
& 12.2
& 14.4
& 9.77
& 4.27
& 2.44
& 1.68
& \textbf{\underline{1.41}}
& \textbf{\underline{0.98}}
& \textbf{\underline{0.68}}
& \textbf{\underline{0.50}} \\

Flow-CG
& \textbf{\underline{0.05}}
& \textbf{\underline{0.17}}
& \textbf{\underline{0.21}}
& \textbf{\underline{0.39}}
& \textbf{\underline{0.73}}
& \textbf{\underline{0.97}}
& 1.59
& 2.20
& 3.22
& 34.1 \\

Flow-ADMM
& 0.55
& 0.19
& 0.23
& 0.40
& 1.05
& 2.92
& 9.53
& 517.4
& --
& -- \\

IP-CG
& 11.36
& {\footnotesize $\geq 24$ h}
& {\footnotesize $\geq 24$ h}
& {\footnotesize $\geq 24$ h}
& {\footnotesize $\geq 24$ h}
& {\footnotesize $\geq 24$ h}
& {\footnotesize $\geq 24$ h}
& {\footnotesize $\geq 24$ h}
& {\footnotesize $\geq 24$ h}
& -- \\

Gurobi
& {\footnotesize $\geq 24$ h}
& --
& --
& --
& --
& --
& --
& --
& --
& -- \\
\bottomrule
\end{tabular}%
}
\caption{$n=19$: runtime in seconds to satisfy the termination criterion in \eqref{eqn:termination_criterion}. The fastest method for each instance is shown in bold and underlined. ``--'' denotes an unreported result, and ``$\geq 24$ h'' indicates that the method did not converge within 24 hours. IP-CG exceeds the 96 GB memory limit in the 10\% case. The direct Gurobi implementation uses Gurobi's own
termination criterion; all other methods use the criterion
in \eqref{eqn:termination_criterion}.}
\label{tab:density19}
\end{table}

%% file: paper_main/A_table_n20_numerical1.tex
\begin{table}[t]
\caption{Runtime by Menu Density for $n=20$}
\label{tab:density20}
\centering
\setlength{\tabcolsep}{4pt}
\begin{adjustbox}{max width=\linewidth}
\begin{tabular}{@{}l*{10}{r}@{}}
\hline
& \multicolumn{10}{c}{Density} \\
\cline{2-11}
Method
& Pairwise
& 0.025\%
& 0.05\%
& 0.1\%
& 0.2\%
& 0.3\%
& 0.5\%
& 0.7\%
& 1\%
& 10\% \\
\hline

BM-ADMM
& 42.3
& 34.2
& 24.3
& 13.8
& 7.30
& 4.19
& \textbf{\underline{2.69}}
& \textbf{\underline{2.34}}
& \textbf{\underline{1.90}}
& \textbf{\underline{0.38}} \\

Flow-CG
& \textbf{\underline{0.05}}
& \textbf{\underline{0.34}}
& 0.50
& \textbf{\underline{0.86}}
& \textbf{\underline{1.71}}
& \textbf{\underline{2.56}}
& 4.30
& 6.32
& 9.13
& 90.5 \\

Flow-ADMM
& 0.85
& 0.34
& \textbf{\underline{0.50}}
& 1.25
& 5.56
& 836.8
& --
& --
& --
& -- \\

IP-CG
& 12.16
& {\footnotesize $\geq 24$ h}
& {\footnotesize $\geq 24$ h}
& {\footnotesize $\geq 24$ h}
& {\footnotesize $\geq 24$ h}
& {\footnotesize $\geq 24$ h}
& {\footnotesize $\geq 24$ h}
& {\footnotesize $\geq 24$ h}
& {\footnotesize $\geq 24$ h}
& {\footnotesize $\geq 24$ h} \\

Gurobi
& {\footnotesize $\geq 24$ h}
& --
& --
& --
& --
& --
& --
& --
& --
& -- \\
\hline
\end{tabular}
\end{adjustbox}
\par
\providecommand{\legend}[1]{\parbox{\linewidth}{\footnotesize\textit{Note:} #1}}
\legend{Runtime in seconds to satisfy the termination criterion in \eqref{eqn:termination_criterion}. The fastest method for each instance is shown in bold and underlined. ``--'' denotes an unreported result, and ``$\geq 24$ h'' indicates that the method did not converge within 24 hours. The direct Gurobi implementation uses Gurobi's own
termination criterion; all other methods use the criterion
in \eqref{eqn:termination_criterion}. In the 0.025\% density and 0.05\% case, the displayed times are same due to rounding.}
\end{table}

%% file: paper_main/A_Statistical_Application.tex
This section describes how our algorithms can be used for statistical inference.
We employ the numerical delta method of
\citet{hong2018numerical}.\footnote{We thank Andres Santos for pointing us
to this paper.}
The procedure requires solving the projection problem in
\eqref{eqn:DC-RUM_Proj} repeatedly, and can therefore be implemented using
any of our proposed algorithms. Our optimality certificates further allow us to account for computational error without solving every projection problem on bootstrap samples to high accuracy. Algorithm~\ref{alg:numerical_delta} illustrates the implementation. We write $d(p)$ for $d(p,\mathcal{P}^\D_{\textrm{RUM}})$ for simplicity.

\begin{algorithm}[htbp]
\caption{Numerical Delta Method for Inference on the Distance to the RUM Polytope}
\label{alg:numerical_delta}
\begin{algorithmic}[1]
\Require Estimator $\widehat p$, normalizing sequence $r_n$,
step size $\epsilon_n$, number of bootstrap replications $B$,
and significance level $\alpha\in(0,1)$.
\State Compute $d(\widehat p)$.
\For{$b=1,\ldots,B$}
    \State Generate a bootstrap estimator $\widehat p^{\,b}$.
    \State Set $Z^{\,b}\gets r_n(\widehat p^{\,b}-\widehat p)$.
    \State Compute
    \[
        D^{\,b}\gets
        \frac{
            d(\widehat p+\epsilon_n Z^{\,b})-d(\widehat p)
        }{\epsilon_n}.
    \]
\EndFor
\State Let $\widehat q_{1-\alpha}$ be the empirical $(1-\alpha)$
quantile of $\{D^{\,b}\}_{b=1}^B$.
\State Construct the one-sided confidence interval
\[
    \widehat C_{1-\alpha}
    =\left[
        d(\widehat p)-\frac{\widehat q_{1-\alpha}}{r_n},
        \,\infty
    \right).
\]
\State Compute the bootstrap $p$-value
\begin{equation}\label{eqn:p_value_calc}
    \widehat p_{\mathrm{val}}
    =\frac{1}{B}\sum_{b=1}^B
    \mathbf 1\{D^{\,b}\geq r_n d(\widehat p)\}.    
\end{equation}
\State Reject $H_0$ if $\widehat p_{\mathrm{val}}\leq\alpha$.
\State \Return $\widehat C_{1-\alpha}$ and $\widehat p_{\mathrm{val}}$.
\end{algorithmic}
\end{algorithm}

Algorithm~\ref{alg:numerical_delta} is stated in terms of exact distances.
In practice, we use the optimality certificates developed above to bound
the optimal value of the squared-distance problem. For a choice-probability vector $p$, at each iteration
we can obtain a lower bound $d_{\mathrm{LB}}(p)$ and
an upper bound $d_{\mathrm{UB}}(p)$ such that
$d_{\mathrm{LB}}(p)\leq d(p)\leq d_{\mathrm{UB}}(p)$. We use these bounds to implement the confidence interval and test
conservatively.

For hypothesis testing, as pointed out by \cite{smeulders2018testing}, it is not necessary to evaluate each
$D^{\,b}$ precisely. Each summand in \eqref{eqn:p_value_calc} can be rewritten as
\begin{align*}
    \mathbf 1\{D^{\,b}\geq r_n d(\widehat p)\}
    &=\mathbf 1\!
    \left\{
        d(\widehat p+\epsilon_n Z^{\,b})
        \geq (1+\epsilon_n r_n) d(\widehat p)
    \right\}.
\end{align*}
For each replication, we update the bounds until the comparison is
resolved or the prescribed absolute and relative tolerances are reached.
We then set
\begin{equation}\label{eq:statistical_certified_indicator}
    \widehat I^{\,b}
    =\begin{cases}
        1,
        & d_{\mathrm{LB}}(\widehat p+\epsilon_n Z^{\,b})
          >(1+\epsilon_n r_n) d_{\mathrm{UB}}(\widehat p),
          \\[0.4em]
        0,
        & d_{\mathrm{UB}}(\widehat p+\epsilon_n Z^{\,b})
          <(1+\epsilon_n r_n) d_{\mathrm{LB}}(\widehat p),
          \\[0.4em]
        1,
        & \text{otherwise}.
    \end{cases}
\end{equation}
The first two cases certify the exact indicator and provide early
stopping rules. The last case assigns an unresolved comparison to one.
Consequently, the reported $p$-value satisfies $\widehat p^{\mathrm{cert}}_{\mathrm{val}}
    =\frac{1}{B}\sum_{b=1}^B\widehat I^{\,b}
    \geq\widehat p_{\mathrm{val}}$ and we reject $H_0$ only if
$\widehat p^{\mathrm{cert}}_{\mathrm{val}}\leq\alpha$. In our implementation, we use $r_n=\sqrt{n}$ and
$\epsilon_n=n^{-1/3}$, where $n$ denotes the sample size.

For confidence intervals, for each bootstrap replication, we compute 
\[
    D^{\,b}_{\mathrm{UB}}
    =\frac{
        d_{\mathrm{UB}}(\widehat p+\epsilon_n Z^{\,b})
        -d_{\mathrm{LB}}(\widehat p)
    }{\epsilon_n}.
\]
Let $\widehat q^{\mathrm{UB}}_{1-\alpha}$ denote the empirical
$(1-\alpha)$ quantile of $\{D^{\,b}_{\mathrm{UB}}\}_{b=1}^B$.
We report
\begin{equation}\label{eq:statistical_certified_ci}
    \widehat C^{\mathrm{cert}}_{1-\alpha}
    =\left[
        d_{\mathrm{LB}}(\widehat p)
        -\frac{\widehat q^{\mathrm{UB}}_{1-\alpha}}{r_n},
        \,\infty
    \right).
\end{equation}
Since $D^{\,b}_{\mathrm{UB}}\geq D^{\,b}$ for every $b$, we have
$\widehat q^{\mathrm{UB}}_{1-\alpha}\geq\widehat q_{1-\alpha}$.
Together with $d_{\mathrm{LB}}(\widehat p)\leq d(\widehat p)$,
this implies that $\widehat C^{\mathrm{cert}}_{1-\alpha}$ contains
the confidence interval obtained by exact optimization using the same
bootstrap draws. Computational error therefore only enlarges the
confidence interval.

Two-sided confidence intervals based on the numerical delta method need not be uniformly valid \citep{hong2018numerical}. Let $p_0$ denote the population choice-probability vector. For the confidence interval reported in Section~\ref{section:ai}, we therefore use a conservative construction based on the inequality
\[
    \left|
    \inf_{p\in\mathcal{P}^\D_{\textrm{RUM}}}\|\widehat{p}-p\|_2
    -
    \inf_{p\in \mathcal{P}^\D_{\textrm{RUM}}}\|p_0-p\|_2
    \right|
    \leq \|\widehat{p}-p_0\|_2.
\]
Consequently, for any $c\geq 0$,
\begin{equation}
    \Pr\!\left(
        \left|
        \inf_{p\in\mathcal{P}^\D_{\textrm{RUM}}}\|\widehat{p}-p\|_2
        -
        \inf_{p\in\mathcal{P}^\D_{\textrm{RUM}}}\|p_0-p\|_2
        \right| > c
    \right)
    \leq
    \Pr\!\left(\|\widehat{p}-p_0\|_2 > c\right).
\end{equation}
We center the interval at the estimated distance and set its half-width equal to the $(1-\alpha)$-quantile of the bootstrap distribution of $\|\widehat{p}^{*}-\widehat{p}\|_2$, truncating the lower endpoint at zero. The probability bound above ensures that the resulting interval is uniformly asymptotically valid under standard regularity conditions.

%% file: paper_main/A_Sample_Prompt.tex
\begin{quote}
\small
\noindent\textbf{Prompt.}

You are choosing one Apple Watch to buy for yourself.

You are not an assistant, an advisor, or a comparison site.
You are the one spending the money, and you will walk away
with exactly one of the watches you are shown.

Pick exactly one option, every time. ``None of them'',
``it depends'', ``I would need more information'', a tie,
or two options are not valid answers. If the options look
similar or you feel unsure, pick one anyway. Do not ask
clarifying questions, do not tell the user to weigh their
own needs, and do not refuse.

Reply with the number of your choice and nothing else.
No explanation, no punctuation, no restatement of the option.

Which one do you buy?

\begin{enumerate}
    \item Apple Watch Series 11, 46mm, GPS only,
    Jet Black Aluminum, Sport Band, new, price: \$329.00
    \item Apple Watch Series 11, 42mm, GPS only,
    Rose Gold Aluminum, Sport Band, new, price: \$299.00
\end{enumerate}

\noindent\textbf{Model reply:} \texttt{1}
\end{quote}

%% file: paper_appendix/Z_Appendix_hypercube_Laplacian.tex
Section~\ref{section:power_set_digraph} reviews several fundamental
results on the power-set digraph and the hypercube Laplacian. Each of these results has appeared previously in the literature; we include them here for completeness and ease of reference. The main references we rely on are \cite{schrijver2003combinatorial,o2014analysis} and \cite{spielman2025sagt}. Section \ref{section:computation_efficiency} then demonstrates how to efficiently solve the update in \eqref{eqn:update_r}, which constitutes a key step of Algorithm \ref{alg:admm_simplify}.

\subsection{Power-set Digraph}\label{section:power_set_digraph}
\subsubsection{Node--Arc Incidence Matrix}\label{section:node_arc_incidence_matrix}

Let \(A\) denote the incidence matrix of the power-set digraph with \(n\) alternatives. 
The matrix has \(2^n\) rows and \(n2^{\,n-1}\) columns. 
Let \(r \in \mathbb{R}^{n2^{\,n-1}}\) and \(v \in \mathbb{R}^{2^n}\) be arbitrary vectors. 

As an example, consider the power-set digraph for $X=\{1,2,3\}$
in Figure~\ref{example:powerset_diagraph}. The flow-conservation constraint $Af=b$ in
\eqref{eqn:flow_polytope} takes the form
\begin{equation}\label{example:AF=b}
 \begin{array}{r}
\emptyset\\
\{1\}\\
\{2\}\\
\{3\}\\
\{1,2\}\\
\{1,3\}\\
\{2,3\}\\
\{1,2,3\}
\end{array}
\!\left[
\begin{array}{*{12}{r}}
-1&-1&-1& 0& 0& 0& 0& 0& 0& 0& 0& 0\\
 1& 0& 0&-1&-1& 0& 0& 0& 0& 0& 0& 0\\
 0& 1& 0& 0& 0&-1&-1& 0& 0& 0& 0& 0\\
 0& 0& 1& 0& 0& 0& 0&-1&-1& 0& 0& 0\\
 0& 0& 0& 1& 0& 1& 0& 0& 0&-1& 0& 0\\
 0& 0& 0& 0& 1& 0& 0& 1& 0& 0&-1& 0\\
 0& 0& 0& 0& 0& 0& 1& 0& 1& 0& 0&-1\\
 0& 0& 0& 0& 0& 0& 0& 0& 0& 1& 1& 1
\end{array}
\right] f
=
\begin{bmatrix}
-1\\0\\0\\0\\0\\0\\0\\1
\end{bmatrix}.   
\end{equation}
Each row corresponds to a vertex, and each column corresponds
to an arc. The first and last rows require one unit of flow
to leave $\emptyset$ and enter $X$, respectively. The other rows
require inflow to equal outflow.

A key feature we exploit in our algorithms is that the matrix--vector products 
\(Ar\) and \(A^\top v\) can be implemented without explicitly forming or storing \(A\). This substantially improves memory efficiency. We provide two simple algorithms to illustrate this idea. We note that the arcs and vertices can be enumerated implicitly using binary masks, so no explicit indexing structure needs to be stored.

Importantly, the matrix \(A\) is totally unimodular, and hence the unit \(\emptyset\)–\(X\) flow polytope \eqref{eqn:flow_polytope} has integral vertices (see, e.g., Sections 4.2 and 4.3 of \cite{conforti2010extended}). It then follows from integrality, together with the structure of the digraph, that every vertex of \eqref{eqn:flow_polytope} corresponds to a directed \(\emptyset\)–\(X\) path. Moreover, an $\emptyset$--$X$ path cannot be expressed as a convex combination of other distinct $\emptyset$--$X$ paths. Therefore, the vertex set of \eqref{eqn:flow_polytope} coincides with the set of all directed \(\emptyset\)–\(X\) paths in the power-set digraph.

\begin{algorithm}[h]
\caption{Calculate $v=Ar$}
\label{algo:Ar}
\begin{algorithmic}[1]
\ForAll{arcs $e$}
  \State Find node $i$ such that $A[i,e] = -1$ and set $v[i] \gets v[i] - r[e]$
  \State Find node $j$ such that $A[j,e] = +1$ and set $v[j] \gets v[j] + r[e]$
\EndFor
\State \Return $v$
\end{algorithmic}
\end{algorithm}
\begin{algorithm}[h]
\caption{Calculate $r=A^\top v$}
\label{algo:A'v}
\begin{algorithmic}[1]
\ForAll{arcs $e$}
  \State Find nodes $i$ and $j$ such that $A[i,e] = -1$ and $A[j,e] = +1$
  \State Set $r[e] \gets v[j] - v[i]$
\EndFor
\State \Return $r$
\end{algorithmic}
\end{algorithm}
\subsubsection{Spectral Properties of $A$ and $AA^\top$}
\label{section:spectral}

We next study the spectral properties of \(AA^\top\). Let \(u\) and \(v\) be two arbitrary vertices of the power-set digraph, identified with subsets of \(X\). Then the \((u,v)\) entry of \(AA^\top\) is
\[
(AA^\top)_{u,v}
=
\begin{cases}
n,  & \text{if } u=v,\\
-1, & \text{if } |u\triangle v|=1,\\
0,  & \text{otherwise},
\end{cases}
\qquad u,v\in 2^X,
\]
where $u\triangle v=(u\setminus v)\cup(v\setminus u)$ denotes
the symmetric difference.
Importantly, the matrix $AA^\top$ is the \emph{Laplacian matrix of an $n$-dimensional hypercube graph} \citep{spielman2025sagt}. The $n$-dimensional hypercube graph is the graph whose vertex set consists of all binary vectors of length $n$, with an edge between two vertices whenever the corresponding vectors differ in exactly one coordinate. In other words, it is the graph formed by the vertices and edges of the $n$-dimensional cube.

For example, for the power-set digraph with three alternatives, the matrix \(AA^\top\) is
{\small
\begin{center}
\begin{equation}
\begin{blockarray}{c@{\hspace{3pt}}rrrrrrrr}
  & \emptyset & \{1\} & \{2\} & \{3\} & \{1,2\} & \{1,3\} & \{2,3\} & \{1,2,3\} \\
\begin{block}{c@{\hspace{3pt}}[cccccccc]}
\emptyset      &  3 & -1 & -1 & -1 &  0 &  0 &  0 &  0 \\
\{1\}          & -1 &  3 &  0 &  0 & -1 & -1 &  0 &  0 \\
\{2\}          & -1 &  0 &  3 &  0 & -1 &  0 & -1 &  0 \\
\{3\}          & -1 &  0 &  0 &  3 &  0 & -1 & -1 &  0 \\
\{1,2\}        &  0 & -1 & -1 &  0 &  3 &  0 &  0 & -1 \\
\{1,3\}        &  0 & -1 &  0 & -1 &  0 &  3 &  0 & -1 \\
\{2,3\}        &  0 &  0 & -1 & -1 &  0 &  0 &  3 & -1 \\
\{1,2,3\}      &  0 &  0 &  0 &  0 & -1 & -1 & -1 &  3 \\
\end{block}
\end{blockarray}
\end{equation}
\end{center}
}
For the spectral computations below, we order the rows of $A$ and
all node-level vectors according to the binary ordering underlying
the recursive definition of $L_n$.

Let \(L_n\) denote the Laplacian matrix of the \(n\)-dimensional hypercube graph. In particular,
\[
    L_1
    =
    \begin{bmatrix}
        1 & -1\\
        -1 & 1
    \end{bmatrix}.
\]
For \(n\geq 2\), the Laplacian of the \(n\)-dimensional hypercube graph can be constructed recursively from the Laplacians of the \((n-1)\)-dimensional hypercube and the one-dimensional hypercube according to
\begin{equation}\label{eqn:hypercube_laplacian}
    L_n
    =
    L_{n-1}\otimes I_2
    +
    I_{2^{n-1}}\otimes L_1,
\end{equation}
where \(\otimes\) denotes the Kronecker product and \(I_2\) and \(I_{2^{n-1}}\) are identity matrices of dimensions \(2\times 2\) and \(2^{n-1}\times 2^{n-1}\), respectively; see Section~5.3 of \cite{spielman2025sagt}.

The following result gives the spectral decomposition of the hypercube Laplacian used in our algorithms. It is drawn from Section~5.3.1 of \cite{spielman2025sagt} and is stated here using Kronecker-product notation.

\begin{lemma}\label{lemma:laplacian}
Let \(L_n\) denote the Laplacian matrix of the \(n\)-dimensional hypercube graph, as defined in \eqref{eqn:hypercube_laplacian}. Define
\begin{equation}
    U
    =
    \frac{1}{\sqrt{2}}
    \begin{bmatrix}
        1 & 1 \\
        -1 & 1
    \end{bmatrix},
    \qquad
    \Sigma_1
    =
    \begin{bmatrix}
        0 & 0 \\
        0 & 2
    \end{bmatrix}.
\end{equation}
Then \(L_n\) admits the spectral decomposition
\begin{equation}\label{eqn:spectral_L_k}
    L_n
    =
    \left(U^{\otimes n}\right)^\top
    \Sigma_n
    U^{\otimes n},
\end{equation}
where
\[
    U^{\otimes n}
    =
    \underbrace{
        U\otimes U\otimes\cdots\otimes U
    }_{n\text{ times}}
\]
denotes the \(n\)-fold Kronecker product of \(U\), and the diagonal matrix \(\Sigma_n\) is defined recursively by
\[
    \Sigma_n
    =
    \Sigma_{n-1}\otimes I_2
    +
    I_{2^{n-1}}\otimes\Sigma_1.
\]
The Moore--Penrose inverse of \(L_n\) is therefore
\begin{equation}
    L_n^+
    =
    \left(U^{\otimes n}\right)^\top
    \Sigma_n^+
    U^{\otimes n},
\end{equation}
where \(\Sigma_n^+\) denotes the Moore--Penrose inverse of \(\Sigma_n\).
\end{lemma}

%% file: paper_appendix/Z_Appendix_ADMM_Efficient_Solve.tex
\subsection{Efficiently Solving \eqref{eqn:update_r}}\label{section:computation_efficiency}

By Algorithm \ref{alg:admm_simplify},  \eqref{eqn:update_r} is an unconstrained quadratic optimization problem and its first-order condition is
\begin{equation}\label{eq:f_foc_main}
\bigl(\rho I+\rho A^\top A+2Z_{\mathcal D}^\top \Omega Z_{\mathcal D}\bigr)\mathbf f
=
2Z_{\mathcal D}^\top\Omega \widehat p
-\mathbf u_t
-A^\top \mathbf v_t
+\rho \mathbf s_t
+\rho A^\top b .
\end{equation}
When there are \(n\) alternatives, the dimension of the system is \(d = n2^{\,n-1}\). Moreover, only a low-dimensional matrix—corresponding to the dimension of the observed choice probabilities—needs to be stored.

Define
\[
S_\rho:=\rho(I+A^\top A),
\qquad
\mathbf q_t
=
2Z_{\mathcal D}^\top \Omega\widehat p
-\mathbf u_t
-A^\top \mathbf v_t
+\rho \mathbf s_t
+\rho A^\top b .
\]
Then \eqref{eq:f_foc_main} becomes
\[
\bigl(S_\rho+2Z_{\mathcal D}^\top \Omega Z_{\mathcal D}\bigr)\mathbf f=\mathbf q_t .
\]
Let $\Omega^{\frac{1}{2}}$ be the square root of the matrix $\Omega$. Applying the Woodbury identity gives
\begin{equation}\label{eq:woodbury_main}
\mathbf f_{t+1}
=
S_\rho^{-1}\mathbf q_t
-
2S_\rho^{-1}Z_{\mathcal D}^\top\Omega^{\frac{1}{2}}
\bigl(I+2\Omega^{\frac{1}{2}}Z_{\mathcal D}S_\rho^{-1}Z_{\mathcal D}^\top\Omega^{\frac{1}{2}}\bigr)^{-1}
\Omega^{\frac{1}{2}}Z_{\mathcal D}S_\rho^{-1}\mathbf q_t,
\end{equation}
and a second application of the Woodbury identity yields
\begin{equation*}\label{eq:S_inverse_main}
S_\rho^{-1}
=
\rho^{-1}I
-
\rho^{-1}A^\top(I+AA^\top)^{-1}A .
\end{equation*}
The key observation is that the operator $(I+AA^\top)^{-1}$ is tractable. Indeed, $AA^\top$ is the Laplacian of the hypercube graph and admits the spectral decomposition
\begin{equation*}\label{eq:laplacian_spec_main}
AA^\top
=
(U^{\otimes n})^\top \Sigma_n U^{\otimes n},
\qquad
U=\frac{1}{\sqrt 2}
\begin{bmatrix}
1 & 1\\
-1 & 1
\end{bmatrix}.
\end{equation*}
It follows that
\[
(I+AA^\top)^{-1}
=
(U^{\otimes n})^\top (I+\Sigma_n)^{-1} U^{\otimes n},
\]
where $(I+\Sigma_n)^{-1}$ is diagonal. Therefore, applying $(I+AA^\top)^{-1}$ to a vector reduces to two butterfly operations together with a multiplication by a diagonal matrix. In particular, no explicit matrix inversion is required. 

Thus the $\mathbf f$-update reduces to repeated applications of the operators
$A$, $A^\top$, $Z_{\mathcal D}$, $Z_{\mathcal D}^\top$, and $S_\rho^{-1}$ together with the solution of a
low-dimensional system of size $|\mathcal J|\times |\mathcal J|$,
\begin{equation}\label{eq:low_dim_matrix_main}
M_\rho
:=
I+2\Omega^{\frac{1}{2}}Z_{\mathcal D}S_\rho^{-1}Z_{\mathcal D}^\top\Omega^{\frac{1}{2}},
\end{equation}
where $|\mathcal{J}|$ is the dimension of observed choice probabilities.
$M_\rho$ is the only dense object that must be stored. Since $\rho$ is fixed, we can compute its Cholesky factorization once and cache it before the ADMM iterations start. Solving the reduced system in each subsequent $\mathbf f$-update
requires triangular solves with the cached factorization, at a
cost of $O(|\mathcal J|^2)$.

In conclusion, \eqref{eq:woodbury_main} can be solved by repeatedly applying the following operators: $M^{-1}_{\rho}$ in $O(|\mathcal{J}|^2)$ time (after constructing $M_\rho$ and its Cholesky factorization), $A$ and $A^\top$ in $O(n2^{n-1})$ time, $Z_{\mathcal{D}}$ and $Z_{\mathcal{D}}^\top$ in $O(n^2 2^{n-1})$ time, and $(I+AA^\top)^{-1}$ in $O(n2^n)$ time.\footnote{By \eqref{eq:inverse}, $Z_n$ has the same Kronecker product structure as $K_n$, so applying $Z_{\mathcal{D}}$ to a vector takes $O(n^2 2^{n-1})$ time using butterfly operations. Similarly, applying $(I+AA^\top)^{-1}$ to a vector takes $O(n2^n)$ time by exploiting the Kronecker structure established in Lemma~\ref{lemma:laplacian}.} Therefore, after constructing $M_\rho$ and computing its
$O(|\mathcal J|^3)$ Cholesky factorization, each ADMM iteration has complexity
\[
O(n^2 2^{n-1} + |\mathcal J|^2).
\]
In practice, $|\mathcal{J}|$ is much smaller, with $|\mathcal{J}|^2\ll n2^{n-1}$; hence the overall complexity is $O(n^2 2^n)$, which is equivalently $O(d\log d)$ when $d = n2^{n-1}$.

For a comparison with standard approaches, the system in \eqref{eq:woodbury_main} has dimension $n2^{\,n-1} \times n2^{\,n-1}$. A solve based on a matrix factorization (e.g., Cholesky factorization) requires $O(n^2 2^{2(n-1)})$ operations after an initial $O(n^3 2^{3(n-1)})$ factorization step and an $O(n^2 2^{2(n-1)})$ storage of the corresponding factors. Inexact methods based on iterative solvers, such as conjugate gradient, require $O(n^2 2^{(n-1)})$ operations per iteration if exploiting the structure of the matrices $Z_{\mathcal{D}}$ and $A$. Hence our way of solving the system achieves an exact solution at a computational cost comparable to a single conjugate gradient iteration. Moreover, matrices $A$ and $Z_{\mathcal{D}}$ do not need to be explicitly stored. The only storage requirement for operators is $O(|\mathcal{J}|^2)$ as a result of the Cholesky factorization, which is moderate if $|\mathcal{J}|$ is on the order of $10^3\sim 10^4$.

%% file: paper_appendix/Z_ADMM_Convergence.tex
We first present a generic template for the Alternating Direction Proximal Method of Multipliers (ADPMM) with over-relaxation and a positive-definite-matrix-weighted augmented Lagrangian. The matrix weighting allows the penalty parameters to vary across constraints, which is useful for the algorithms considered in this paper.

Specifically, consider the optimization problem
\begin{equation}\label{eq:generic_problem}
\begin{aligned}
\min_{x,y}\quad & f(x)+g(y)\quad
\text{subject to}\quad & Ax+By=c,
\end{aligned}
\end{equation}
where \(f\) and \(g\) are closed, proper, and convex functions, and \(A\) and \(B\) are given matrices. For a function \(h:\mathbb{R}^d\to\mathbb{R}\cup\{+\infty\}\), we denote its effective domain by \(\operatorname{dom}(h):=\{x\in\mathbb{R}^d:h(x)<+\infty\}\). We denote the feasible set by \(\mathcal F=\{(x,y):Ax+By=c\}\). 

Let $\Xi$ be an arbitrary positive definite matrix. We consider the following augmented Lagrangian. 
\begin{equation*}
 \mathcal{L}(\mathbf{x},\mathbf{y},\mathbf{u}) =  f(\mathbf{x})+g(\mathbf{y})+\langle \mathbf{u}, A\mathbf{x}+B\mathbf{y}-c\rangle + \frac{1}{2}\|A\mathbf{x}+B\mathbf{y}-c\|_\Xi^2,
\end{equation*}
where $\|x\|_\Xi^2=x^\top \Xi x$. We consider the following ADPMM template with relaxation parameter $\alpha\in (0,2)$ and positive definite weighting matrix $\Xi$. Let $S$ and $T$ be two positive-semidefinite matrices.

\begin{algorithm}[h]
\caption{ADPMM with over-relaxation and positive-definite matrix weighting}
\label{algo:adpmm}
\begin{algorithmic}[1]

\State \textbf{Input:} Initial point
$\left(\mathbf{x}_0,\mathbf{y}_0,\mathbf{u}_0\right)$,
relaxation parameter $\alpha$,
and weighting matrices $\Xi\succ 0$, $S\succeq 0$, and $T\succeq 0$.
\State Set $k=0$.

\While{the convergence criterion is not met}

\State Update $\mathbf{x}$:
\begin{equation*}
\mathbf{x}_{k+1}
=
\arg\min_{\mathbf{x}}
f(\mathbf{x})
+\langle \mathbf{u}_k,A\mathbf{x}+B\mathbf{y}_k-c\rangle
+\frac{1}{2}\|A\mathbf{x}+B\mathbf{y}_k-c\|_\Xi^2
+\frac{1}{2}\|\mathbf{x}-\mathbf{x}_k\|_S^2.
\end{equation*}

\State Relax:
\begin{equation*}
\overline{A\mathbf{x}}^{\,k+1}
=
\alpha A\mathbf{x}_{k+1}
+(1-\alpha)(c-B\mathbf{y}_k).
\end{equation*}

\State Update $\mathbf{y}$:
\begin{equation*}
\mathbf{y}_{k+1}
=
\arg\min_{\mathbf{y}}
g(\mathbf{y})
+\left\langle
\mathbf{u}_k,
\overline{A\mathbf{x}}^{\,k+1}+B\mathbf{y}-c
\right\rangle
+\frac{1}{2}
\|\overline{A\mathbf{x}}^{\,k+1}+B\mathbf{y}-c\|_\Xi^2
+\frac{1}{2}\|\mathbf{y}-\mathbf{y}_k\|_T^2.
\end{equation*}

\State Update $\mathbf{u}$:
\begin{equation*}
\mathbf{u}_{k+1}
=
\mathbf{u}_k
+\Xi
\left(
\overline{A\mathbf{x}}^{\,k+1}
+B\mathbf{y}_{k+1}
-c
\right).
\end{equation*}

\State Set $k\gets k+1$.

\EndWhile

\end{algorithmic}
\end{algorithm}

Proposition \ref{prop:admm_convergence} establishes a convergence result for the ADPMM algorithm described in Algorithm \ref{algo:adpmm}. It is a generalization of the ADPMM convergence analysis in \cite{fazel2013hankel} to accommodate the over-relaxation scheme in Algorithm~\ref{algo:adpmm} and a general
positive-definite matrix weighting.
\begin{proposition}\label{prop:admm_convergence}
Suppose that the solution set of \eqref{eq:generic_problem} is nonempty and that the following constraint qualification holds:
there exists $
(\tilde{x},\tilde{y})\in
\operatorname{ri}\!\left(
\operatorname{dom}(f)\times\operatorname{dom}(g)
\right)\cap \{(x,y): Ax+By=c\}.$

Let $S$ and $T$ be positive semidefinite matrices and let $\Xi$ be positive definite. Assume further that
$S+A^\top\Xi A$ and $T+B^\top\Xi B$ are positive definite.
Let
${(\mathbf{x}_k,\mathbf{y}_k,\mathbf{u}_k)}_{k\geq 0}$ be the sequence generated by Algorithm \ref{algo:adpmm}.
Then $(\mathbf{x}_k,\mathbf{y}_k)$ converges to an optimal solution of \eqref{eq:generic_problem}, and $\mathbf{u}_k$ converges to an optimal solution of its dual problem.
\end{proposition}
A proof is included in Section \ref{proof:admm_convergence}.

%% file: paper_appendix/Z_Appendix_column_generation.tex
Algorithm~\ref{alg:IP_pricing} formulates the pricing problem as an integer program that searches over all possible rankings and returns the exact optimal solution. In the IP formulation, the binary variables $A_{xy}$ encode pairwise preferences, where $A_{xy}=1$ indicates that alternative $x$ is ranked above $y$. 
The constraints $A_{xy}+A_{yx}=1$ enforce antisymmetry so that exactly one of $x \succ y$ or $y \succ x$ holds for every pair. 
The inequalities $A_{xy}+A_{yz}+A_{zx}\le 2$ eliminate preference cycles and ensure transitivity, thereby guaranteeing that the pairwise relations correspond to a valid ranking. 
The constraints $\sum_{x\in D} p(x,D)=1$ require that exactly one alternative is chosen from each menu $D$. 
Finally, the inequalities $p(x,D)\le A_{xy}$ for all $x\neq y\in D$ ensure consistency between the choice variables and the ranking: if $x$ is chosen from menu $D$, then $x$ must be ranked above every other alternative $y\in D$.

Algorithm~\ref{alg:pricing_heuristic_random} is a randomized
best-insertion heuristic that incrementally constructs a ranking.
Starting from a randomly chosen alternative, the algorithm
repeatedly selects a new alternative and inserts it into the
position that maximizes the pricing objective over menus
contained in the resulting partial ranking. This procedure is
computationally fast. For the partial ranking $\pi_t$ defined
in \eqref{eqn:pit} and any menu $D$ whose alternatives appear
in $\pi_t$, define $p^{\pi_t}(x,D)$ to equal one if $x$ is the
highest-ranked alternative in $D$ under $\pi_t$, and zero
otherwise.

\begin{algorithm}[h!]
\caption{Integer Programming Formulation for the Pricing Problem}
\label{alg:IP_pricing}
\begin{algorithmic}[1]
\State \textbf{Input:} score vector $\{s(x,D)\}_{D\in\mathcal{D},\,x\in D}$
\State Introduce decision variables:
\begin{itemize}
    \item $A_{xy}\in\{0,1\}$ for all distinct $x,y\in X$ \hfill (pairwise preference indicators),
    \item $p(x,D)\in\{0,1\}$ for all $D\in\mathcal{D}$ and $x\in D$.
\end{itemize}
\State Solve
\begin{equation*}
\max_{ \{A_{xy}\},p} \sum_{D\in\mathcal{D}} \sum_{x\in D} s(x,D)\,p(x,D)
\end{equation*}
subject to
\begin{align*}
A_{xy}+A_{yx} &= 1 
&& \forall x\neq y \in X, \\
A_{xy}+A_{yz}+A_{zx} &\le 2 
&& \forall \text{ distinct } x,y,z \in X , \\
\sum_{x\in D} p(x,D) &= 1 
&& \forall D\in\mathcal{D} \\
p(x,D) &\le A_{xy} 
&& \forall D\in\mathcal{D},\forall x\in D, \forall y \in D\backslash \{x\}.
\end{align*}
\State Recover the ranking $\pi^*$ from $\{A_{xy}\}$.
\State \textbf{Return:} optimal value and ranking $\pi^*$.
\end{algorithmic}
\end{algorithm}
\begin{algorithm}[h!]
  \caption{Random Best-Insertion Heuristic for the Pricing Problem}
  \label{alg:pricing_heuristic_random}
  \begin{algorithmic}[1]
    \State \textbf{Input:} score vector $\{s(x,D)\}_{D\in\mathcal{D},\,x\in D}$
    \State Randomly choose $x_1 \in X$ and initialize the ranking as $\langle x_1\rangle$.
    \For{$k=1$ to $|X|-1$}
      \State Let the current ranking be $\langle i_1,\dots,i_k\rangle$.
      \State Select $c \in X\setminus\{i_1,\dots,i_k\}$ uniformly at random.
      \For{$t=1$ to $k+1$}
        \State Let
        \begin{align}\label{eqn:pit}
          \pi_t=\langle i_1,\dots,i_{t-1},c,i_t,\dots,i_k\rangle,            
        \end{align}
        and compute
        \begin{equation*}
          q_t
          =
          \sum_{D\in\mathcal{D}: \, D\subseteq \{i_1,\dots,i_k,c\}}
          \;\sum_{x\in D} p^{\pi_t}(x,D)\,s(x,D).
        \end{equation*}
      \EndFor
      \State Choose $t^* \in \arg\max_{1\le t\le k+1} q_t$.
      \State Update the ranking to $
      \langle i_1,\dots,i_{t^*-1},c,i_{t^*},\dots,i_k\rangle$.
    \EndFor
    \State \textbf{Return:} ranking $\langle i_1,i_2,\dots,i_{|X|}\rangle$.
  \end{algorithmic}
\end{algorithm}

%% file: paper_appendix/Z_Appendix_BM_ADMM_Complete.tex
Our augmented Lagrangian is
\begin{align*}
\mathcal{L}(\mathbf{p},\overline{\mathbf{p}},\mathbf{s},\mathbf{u},\mathbf{v})
= {}&
\|\pcalD-\widehat{p}\|_{\Omega}^2
+ \iota_{\mathbb{R}_+^{n2^{n-1}}}(\mathbf{s})
+ \sum_{D\in 2^X}\iota_{\Delta^D}(\pD)
\\
&+ \langle \mathbf{u},\mathbf{p}-\overline{\mathbf{p}}\rangle
+ \langle \mathbf{v},\widetilde{K}_n\mathbf{p}-\mathbf{s}\rangle
+ \frac{\beta_1}{2}\|\widetilde{K}_n\mathbf{p}-\mathbf{s}\|_2^2
+\frac{\beta_2}{2}\|\mathbf{p}-\overline{\mathbf{p}}\|_2^2
\end{align*}
This augmented Lagrangian differs from the one used in
Algorithm~\ref{alg:admm_BM} in two respects. First, the rows of $K_n$
are rescaled to obtain $\widetilde{K}_n$. This rescaling leaves the
underlying inequalities unchanged while potentially improving the
conditioning of $\widetilde{K}_n^\top \widetilde{K}_n$. Second, we allow
different penalty parameters, $\beta_2$ and $\beta_1$, for the constraints
$\mathbf{p}-\overline{\mathbf{p}}=0$ and
$\widetilde{K}_n\mathbf{p}-\mathbf{s}=0$, respectively.

Specifically, using the representation \(K_n=I_n\otimes M^{\otimes(n-1)}\), we apply the positive diagonal row scaling
\[
\begin{gathered}
R=\begin{bmatrix}\sqrt{2}&0\\0&1\end{bmatrix},\qquad
\widetilde M=RM=\begin{bmatrix}\sqrt{2}&0\\-1&1\end{bmatrix},\qquad
\widetilde M^\top\widetilde M=\begin{bmatrix}3&-1\\-1&1\end{bmatrix},
\\[3pt]
\widetilde K_n=(I_n\otimes R^{\otimes(n-1)})K_n
=I_n\otimes\widetilde M^{\otimes(n-1)}.
\end{gathered}
\]
This scaling multiplies the BM inequality associated with \((x,D)\) by \(2^{(|D|-1)/2}\). Since this factor is positive, \(\widetilde K_n\mathbf p\geq0\) is equivalent to \(K_n\mathbf p\geq0\).
The eigenvalues of \(\widetilde M^\top\widetilde M\) are \(2\pm\sqrt{2}\), compared with \((3\pm\sqrt{5})/2\) for \(M^\top M\). Consequently, for \(n\geq2\),
\[
\kappa(\widetilde K_n^\top\widetilde K_n)
=(3+2\sqrt{2})^{n-1}
<\left(\frac{7+3\sqrt{5}}{2}\right)^{n-1}
=\kappa(K_n^\top K_n),
\]
where \(\kappa(A)=\|A\|_2\|A^{-1}\|_2\) denotes the spectral condition number of a nonsingular matrix \(A\).
The proposed scaling minimizes this condition number over all positive diagonal row scalings of \(K_n\), while preserving the Kronecker structure and the exact linear solve based on butterfly operations.

We define $\Omega^{\textrm{prox}}=\lambda_{\max}\left(\Omega\right)I-\Omega$.

\begin{algorithm}[htbp]
\caption{Block--Marschak ADMM with over-relaxation}
\label{app:bm_admm_alg_full}

\begin{algorithmic}[1]
\Require observed probabilities $\widehat{p}$, over-relaxation $\alpha\in(0,2)$,
fixed penalties $\beta_1,\beta_2>0$, number of alternatives $n$.

\State Let $d=n2^{n-1}$.
\State Initialize $\mathbf{p}_0=\overline{\mathbf{p}}_0=\mathbf{s}_0=\mathbf{u}_0=\mathbf{v}_0=\mathbf{0}$.

\For{$t=0,1,...,$}   

  \State \textbf{Step 1: ($\mathbf{p}$-update, exact)}
  \begin{equation}\label{eqn:bm_update_p}
  \mathbf{p}_{t+1}=\arg\min_{\mathbf{p}}
  \left\{
  \langle \mathbf{u}_t,\mathbf{p}-\overline{\mathbf{p}}_t\rangle
+ \langle \mathbf{v}_t,\widetilde{K}_n\mathbf{p}-\mathbf{s}_t\rangle
+ \frac{\beta_1}{2}\|\widetilde{K}_n\mathbf{p}-\mathbf{s}_t\|_2^2
+\frac{\beta_2}{2}\|\mathbf{p}-\overline{\mathbf{p}}_t\|_2^2
  \right\}
  \end{equation}
  \State \textbf{Step 2: (over-relaxed primal quantities)}
  \[
  \mathbf{p}_{t+1}^l=\alpha\mathbf{p}_{t+1}+(1-\alpha)\overline{\mathbf{p}}_t,
  \qquad
  \bar{\mathbf{a}}_{t+1}=\alpha \widetilde{K}_n\mathbf{p}_{t+1}+(1-\alpha)\mathbf{s}_t .
  \]
  \State \textbf{Step 3a: (evaluate simplex proximal operator)}
  \begin{equation}\label{eqn:bm_update_p_prime}
    \overline{\mathbf{p}}_{t+1}=\arg\min_{\overline{\mathbf{p}}} \left\{
    \begin{split}
    &\|\overline{\mathbf{p}}_\mathcal{D}-\widehat{p}\|^2_{\Omega}+\sum_{D\in 2^X}\iota_{\Delta^D}\left(\overline{\mathbf{p}}^D\right)\\
    &+ \frac{\beta_2}{2}\|\mathbf{p}_{t+1}^l-\overline{\mathbf{p}}\|^2_2+\|\pcalD-\overline{\mathbf{p}}_{\mathcal{D},t}\|^2_{\Omega^\textrm{prox}}+ \langle \mathbf{u}_t,\mathbf{p}_{t+1}^l-\overline{\mathbf{p}} \rangle      
    \end{split}
 \right\}.
  \end{equation}

  \State \textbf{Step 3b: ($\mathbf{s}$-update: nonnegative slack)}
  \begin{equation}\label{eqn:bm_update_s}
  \mathbf{s}_{t+1}
  =\arg\min_{\mathbf{s}}\Big\{\iota_{\mathbb{R}^d_+}(\mathbf{s})
   -\langle\mathbf{v}_t,\mathbf{s}\rangle
   +\frac{\beta_1}{2}\|\bar{\mathbf{a}}_{t+1}-\mathbf{s}\|_2^2\Big\}
  =\Pi_{\mathbb{R}^d_+}\!\left(\tfrac{1}{\beta_1}\mathbf{v}_t+\bar{\mathbf{a}}_{t+1}\right).
  \end{equation}

  \State \textbf{Step 4: (dual updates)}
  \[
  \mathbf{v}_{t+1}=\mathbf{v}_t+\beta_1\big(\bar{\mathbf{a}}_{t+1}-\mathbf{s}_{t+1}\big),
  \qquad
  \mathbf{u}_{t+1}=\mathbf{u}_t+\beta_2\big(\mathbf{p}_{t+1}^l-\overline{\mathbf{p}}_{t+1}\big).
  \]

\EndFor
\end{algorithmic}
\end{algorithm}

%% file: paper_appendix/Z_Appendix_Flow_ADMM_Complete.tex
We use the following augmented Lagrangian with decision variables $\left(\mathbf{f},\mathbf{s},\mathbf{u},\mathbf{v}\right)$.
\begin{equation}\label{eqn:admm_criterion2}
\begin{split}
     & \kappa\|Z_{\D}\mathbf{f}-\widehat{p}\|_\Omega^2 
     + \iota_{\mathbb{R}_+^d}(\mathbf{s}) \\
     &\quad + \langle \mathbf{u},\, \mathbf{f}-\mathbf{s} \rangle
     + \langle \mathbf{v},\, w_A\left(A\mathbf{f}-b\right) \rangle
     + \frac{\beta_1}{2}\|w_A\left(A\mathbf{f}-b\right)\|_2^2 + \frac{\beta_2}{2}\|\mathbf{f}-\mathbf{s}\|_2^2.
\end{split}
\end{equation}
Compared with the augmented Lagrangian in \eqref{eq:admm_main}, our formulation differs in three respects: we rescale the objective by a factor \(\kappa\), weight the flow conservation constraints by \(w_A\), and allow for distinct penalty parameters $\beta_1$ and $\beta_2$ for the constraints \(A\mathbf{f}-b\) and  \(\mathbf{f}-\mathbf{s}\), respectively. Algorithm \ref{app:admm_alg_full} includes the ADMM algorithm for this criterion with over-relaxation. The scaling factors and over-relaxation are introduced to improve the empirical performance of the ADMM algorithm. 

\begin{algorithm}[htbp]
\caption{Flow-ADMM with over-relaxation}
\label{app:admm_alg_full}
\begin{algorithmic}[1]
\Require observed probabilities $\widehat{p}$, over-relaxation parameter $\alpha\in(0,2)$, initial penalties $\beta_{1}, \beta_{2}$,
flow-operator scaling factor $w_A$, criterion scaling factor $\kappa$, 


\State \textbf{Precompute (cached):}
  \State \quad $\gamma \gets \beta_{1}\, w_A^2 \;/\; \beta_{2}$
  \State \quad $M \gets \Omega^{\frac{1}{2}} Z_{\mathcal{D}}\,(I + \gamma A^\top\! A)^{-1}\, Z_{\mathcal{D}}^\top \Omega^{\frac{1}{2}}$
  \State \quad $S \gets I + (2\kappa / \beta_{2})\, M$
  \State \quad $L_{\mathrm{chol}} \gets \mathrm{chol}(S)$ \Comment{Cholesky factorization}

\State \textbf{Initialize} $\mathbf{f}_0 = \mathbf{s}_0 = \mathbf{u}_0 = \mathbf{v}_0 = \mathbf{0}$; 

\For{$t=0,1,...,$}

    \State \textbf{Step 1: (f-update)}
\begin{equation}\label{eqn:update_r2}
\min_{\mathbf{f}}
\Bigg\{
\begin{aligned}
&\kappa\|Z_{\mathcal D}\mathbf{f}-\widehat p\|_\Omega^2
+\langle \mathbf{u}_t,\mathbf{f}-\mathbf{s}_t\rangle \\
&\quad +\langle \mathbf{v}_t,w_A\left(A\mathbf{f}-b\right)\rangle
+\frac{\beta_2}{2}\|\mathbf{f}-\mathbf{s}_t\|_2^2
+\frac{\beta_1}{2}\|w_A\left(A\mathbf{f}-b\right)\|_2^2
\end{aligned}
\Bigg\}    
\end{equation}
with Algorithm \ref{alg:admm_r_update_general}.

    \State \textbf{Step 2: (over-relaxed primal quantities)}
    \[
    \bar{\mathbf{f}}_{t+1}
    =
    \alpha \mathbf{f}_{t+1}+(1-\alpha)\mathbf{s}_t,
    \qquad
    \bar{\mathbf{a}}_{t+1}
    =
    \alpha w_AA\mathbf{f}_{t+1}+(1-\alpha)w_Ab
    \]
    \State \textbf{Step 3: (s-update: proximal projection)}
\begin{align}\label{eqn:update_s2}
\begin{split}
    \mathbf{s}_{t+1}\gets &
    \arg\min_{\mathbf{s}}
    \Big\{
    \iota_{\mathbb{R}_+^d}(\mathbf{s})
    +\langle \mathbf{u}_t,\bar{\mathbf{f}}_{t+1}-\mathbf{s}\rangle
    +\frac{\beta_{2}}{2}\|\bar{\mathbf{f}}_{t+1}-\mathbf{s}\|_2^2
    \Big\}\\ 
    = &
    \Pi_{\mathbb{R}_+^d}\!\left(
    \bar{\mathbf{f}}_{t+1}+\frac{1}{\beta_{2}}\mathbf{u}_t
    \right)        
\end{split}
\end{align}

    \State \textbf{Step 4: (dual updates)}
    \[
    \mathbf{v}_{t+1}
    =
\mathbf{v}_t+\beta_{1}\bigl(\bar{\mathbf{a}}_{t+1}-w_A b\bigr), \quad     \mathbf{u}_{t+1}
    =
\mathbf{u}_t+\beta_{2}\bigl(\bar{\mathbf{f}}_{t+1}-\mathbf{s}_{t+1}\bigr).
    \]

\EndFor
\end{algorithmic}
\end{algorithm}
\begin{algorithm}[htbp]
\caption{ADMM $\mathbf{f}$-update }
\label{alg:admm_r_update_general}
\begin{algorithmic}[1]
\Require Current iterates $(\mathbf{s}_t,\mathbf{u}_t,\mathbf{v}_t)$, matrices $A, Z_{\mathcal D}$, vector $b$, observed probabilities $\widehat{p}$, scaling factors $\left(\kappa,\gamma,\beta_1,\beta_2,w_A\right)$, and a Cholesky factorization $L_{\mathrm{chol}}$ of 
\[
S = I+\frac{2\kappa}{\beta_2} \Omega^{\frac{1}{2}}Z_{\mathcal D}\left(I+\gamma A^\top A\right)^{-1}Z_{\mathcal D}^\top\Omega^{\frac{1}{2}}.
\]
\Ensure Updated variable $\mathbf{f}_{t+1}$

\State Form the right-hand side 
\begin{equation*}
\mathbf{q}
=
2 \kappa Z_{\mathcal D}^\top \Omega \widehat{p}
- \mathbf{u}_t
- w_A A^\top \mathbf{v}_t
+ \beta_2 \mathbf{s}_t
+ \beta_1 w_A^2 A^\top b    
\end{equation*}
\State Solve $\mathbf{y}_1=\left(I + \gamma A^\top A\right)^{-1}\mathbf{q}$.
\State Solve $\mathbf{y}_2 = \frac{2\kappa}{\beta_2}\left(I + \gamma A^\top A\right)^{-1} Z_{\mathcal D}^\top \Omega^{\frac{1}{2}}\left(L^\top_{\mathrm{chol}}\right)^{-1}L_{\mathrm{chol}}^{-1} \Omega^{\frac{1}{2}} Z_{\mathcal D}\left(I + \gamma A^\top A\right)^{-1} \mathbf{q}$ 
\State \Return $\mathbf{f}_{t+1} = \beta_2^{-1}\left(\mathbf{y}_1 - \mathbf{y}_2\right)$. 
\end{algorithmic}
\end{algorithm}

%% file: paper_appendix/Z_appendix_ADMM_certificate.tex
\begin{proposition}\label{prop:admm_certificate_overrelaxation}
Consider either of the following two constructions of $\mathbf{f}^{\mathrm{feas}}_t$:
\begin{enumerate}[label=(\roman*)]
    \item Let $\mathbf{f}_t$ be the output of Algorithm \ref{app:admm_alg_full}, and let $\mathbf{f}^{\mathrm{feas}}_t$ be the output of Algorithm \ref{alg:pushback_infeasible_flow} applied to $\mathbf{f}_t$.
    
    \item Let $\mathbf{p}_t$ be the output of Algorithm \ref{app:bm_admm_alg_full}. Define $\mathbf{f}_t=K_n\mathbf{p}_t$, and let $\mathbf{f}^{\mathrm{feas}}_t$ be the output of Algorithm \ref{alg:pushback_infeasible_flow} applied to $\mathbf{f}_t$.
\end{enumerate}

Let $v^*$ denote the optimal value of \eqref{eqn:DC-RUM_Proj}, and let $c:\mathcal{P}_{\textrm{RUM}}^\D\to\mathbb{R}$ be defined as in \eqref{eqn:IP_value}. Define
\begin{equation}
    \mathrm{LB}_t
    =
    \|Z_{\mathcal D}\mathbf{f}^{\mathrm{feas}}_t-\widehat p\|_\Omega^2
    -
    2c\!\left(Z_{\mathcal D}\mathbf{f}^{\mathrm{feas}}_t\right),
    \qquad
    \mathrm{UB}_t
    =
    \|Z_{\mathcal D}\mathbf{f}^{\mathrm{feas}}_t-\widehat p\|_\Omega^2.
\end{equation}
Then, in either case (i) or (ii),
\begin{equation*}
    \mathrm{LB}_t \leq v^* \leq \mathrm{UB}_t,
    \qquad
    \mathrm{LB}_t \to v^*,
    \qquad
    \mathrm{UB}_t \to v^*.
\end{equation*}
The conclusions continue to hold in case~(ii), when $\widetilde K_n$ is replaced
by $K_n$ throughout Algorithm~\ref{app:bm_admm_alg_full}.
\end{proposition}

%% file: paper_main/A_Add_Delete_Ranking.tex
\subsection{Restricting Rankings via Dominance Relationships}
In some applications, it is natural to impose additional structure on the set of admissible rankings through dominance relationships. For example, for a pair of alternatives \(x\) and \(y\), we may require that \(y\) is always preferred to \(x\). Both the flow representation and the Block--Marschak representation can accommodate such restrictions in a simple way. 

Under the flow representation, the restriction can be encoded by removing all nodes of the form \(D\cup\{y\}\) from the power-set digraph whenever \(x\not \in D\), along with their associated inflow and outflow arcs. Intuitively, a path that traverses such a node corresponds to rankings in which \(x\) is placed above \(y\). 
In our setup, removing a node is associated with removing all arcs that enter or leave the node or, equivalently, setting the flows on all of its incident arcs to zero. Due to the one-to-one correspondence between flows and BM polynomials \citep{fiorini2004short}, the same restrictions can be imposed in the BM representation by requiring that the BM polynomials of the associated arcs are zero.

\begin{figure}[htbp]
\centering
\begin{tikzpicture}[
    x=1.4cm, y=1.1cm,
    every node/.style={font=\small},
    setnode/.style={circle,draw,inner sep=1pt,minimum size=6mm},
    edge/.style={->,thin}
]

\node[setnode] (e)   at (0,0) {$\emptyset$};

\node[setnode] (2)   at (1.5, 0)   {$\{2\}$};
\node[setnode] (3)   at (1.5,-1.4) {$\{3\}$};

\node[setnode] (12)  at (3, 1.4) {$\{1,2\}$};
\node[setnode] (23)  at (3,-1.4) {$\{2,3\}$};

\node[setnode] (123) at (4.5,0) {$\{1,2,3\}$};

\draw[edge] (e)--(2);   \draw[edge] (e)--(3);
\draw[edge] (2)--(12);  \draw[edge] (2)--(23);
\draw[edge] (3)--(23);
\draw[edge] (12)--(123); \draw[edge] (23)--(123);

\end{tikzpicture}
\caption{Reduced power-set digraph for \(X=\{1,2,3\}\) under the restriction
\(1\succ 2\). The nodes \(\{1\}\) and \(\{1,3\}\), which contain alternative
\(1\) but not alternative \(2\), are removed together with all of their
incident arcs.}
\label{fig:powerset_delete_1_over_2}
\end{figure}
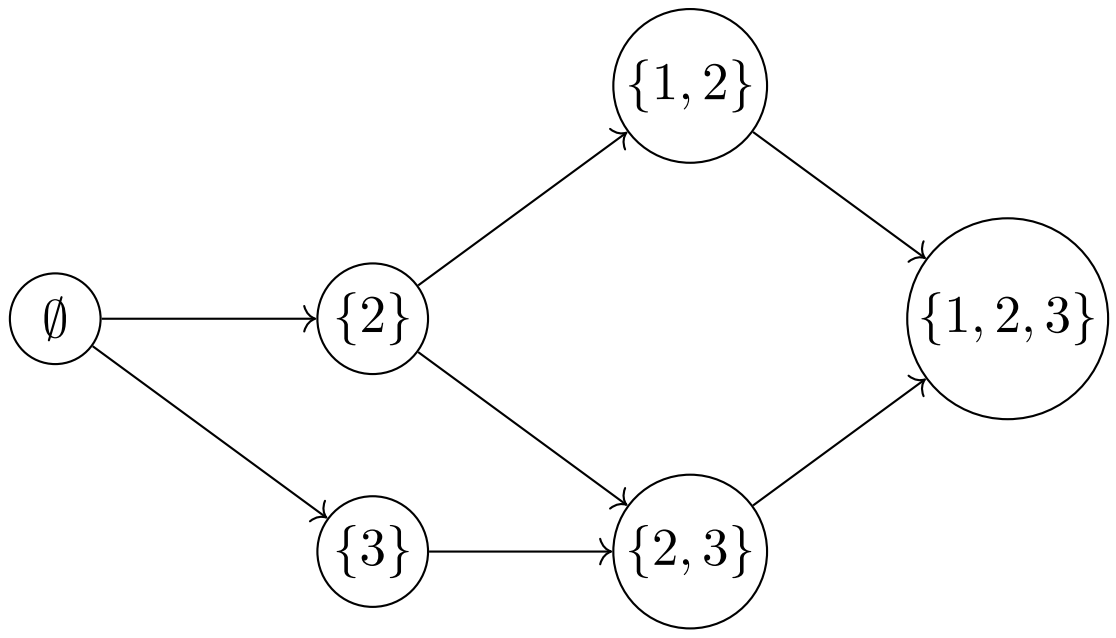

Figure~\ref{fig:powerset_delete_1_over_2} illustrates this construction for
\(X=\{1,2,3\}\) and the restriction \(1\succ 2\). In this case, we remove
every node that contains alternative \(1\) but not alternative \(2\). These
are the nodes \(\{1\}\) and \(\{1,3\}\). Removing these nodes, together with
all of their incident arcs, eliminates exactly the source--sink paths that
violate \(1\succ2\).

This intuition extends beyond binary comparisons to any a priori
restrictions of the form
\[
x_1\succ x_2\succ\cdots\succ x_k.
\]
In particular, we may remove every node \(S\cup D\), where
\(S\subseteq\{x_1,\ldots,x_k\}\) and
\(D\subseteq X\setminus\{x_1,\ldots,x_k\}\), for which \(S\) violates the
prescribed ordering. Equivalently, a node is retained only if
\begin{equation}\label{eqn:suffix_condition}
x_i\in S \implies x_j\in S
\qquad\text{for all }1\le i\le j\le k.
\end{equation}

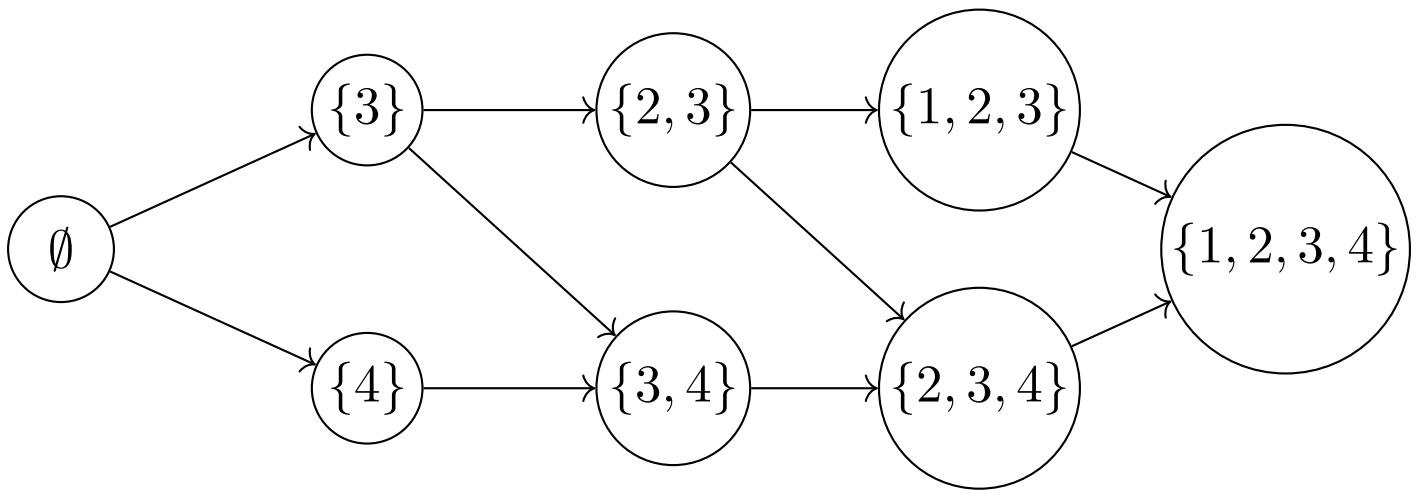
\begin{figure}[htbp]
\centering
\begin{tikzpicture}[
    x=1.35cm, y=1.15cm,
    every node/.style={font=\small},
    setnode/.style={circle,draw,inner sep=1pt,minimum size=7mm},
    edge/.style={->,thin}
]

\node[setnode] (e)    at (0,0) {$\emptyset$};

\node[setnode] (3)    at (1.5, 0.8) {$\{3\}$};
\node[setnode] (4)    at (1.5,-0.8) {$\{4\}$};

\node[setnode] (23)   at (3, 0.8) {$\{2,3\}$};
\node[setnode] (34)   at (3,-0.8) {$\{3,4\}$};

\node[setnode] (123)  at (4.5, 0.8) {$\{1,2,3\}$};
\node[setnode] (234)  at (4.5,-0.8) {$\{2,3,4\}$};

\node[setnode] (1234) at (6,0) {$\{1,2,3,4\}$};

\draw[edge] (e)--(3);      \draw[edge] (e)--(4);
\draw[edge] (3)--(23);     \draw[edge] (3)--(34);
\draw[edge] (4)--(34);
\draw[edge] (23)--(123);   \draw[edge] (23)--(234);
\draw[edge] (34)--(234);
\draw[edge] (123)--(1234); \draw[edge] (234)--(1234);

\end{tikzpicture}
\caption{Reduced power-set digraph for \(X=\{1,2,3,4\}\) under the
restriction \(1\succ2\succ3\). The intersection of each retained node with
\(\{1,2,3\}\) is a suffix of the restricted ordering, while the unrestricted
alternative \(4\) may be added at any stage.}
\label{fig:powerset_delete_1_over_2_over_3}
\end{figure}

Figure~\ref{fig:powerset_delete_1_over_2_over_3} gives an example with
\(X=\{1,2,3,4\}\) and \(1\succ2\succ3\). The only admissible intersections
with \(\{1,2,3\}\) are
\(\emptyset\), \(\{3\}\), \(\{2,3\}\), and \(\{1,2,3\}\); alternative
\(4\) may be either absent or present.

Proposition~\ref{prop:delete} formalizes this reduction. 

\begin{proposition}\label{prop:delete}
Let \(G=(N,E)\) be the power-set digraph on \(X\) with $n$ alternatives, and consider the a priori
ordering restriction $x_1\succ x_2\succ\cdots\succ x_k$.
Define
\[
\Pi^{\mathrm r}
=
\left\{
\pi \in \Pi
:
\pi(x_1) > \pi(x_2) > \cdots > \pi(x_k)
\right\},
\]
and let \(G^{\mathrm r}=(N^{\mathrm r},E^{\mathrm r})\) be the subgraph induced by the nodes satisfying
condition~\eqref{eqn:suffix_condition}. 

Let \(\mathcal R^{\mathrm r}\)
be the set of all \(\emptyset\)--\(X\) paths in \(G^{\mathrm r}\), with each
\(r\in\mathcal R^{\mathrm r}\) represented in the full arc space
\(\mathbb R^E\) by setting \(r_e=0\) for every
\(e\in E\setminus E^{\mathrm r}\). Similarly, define the unit-flow polytope
of \(G^{\mathrm r}\), embedded in \(\mathbb R^E\), by
\[
\mathcal F^{\mathrm r}
=
\{f\in\mathcal F_X:f_e=0\text{ for every }e\in E\setminus E^{\mathrm r}\}.
\]
Then $\mathcal F^{\mathrm r}
=
\operatorname{conv}\bigl(\mathcal R^{\mathrm r}\bigr)$ and 
\[
\operatorname{conv}\bigl(\{p^\pi\}_{\pi\in\Pi^{\mathrm r}}\bigr)
=
\{Z_n f:f\in\mathcal F^{\mathrm r}\}
=
\operatorname{conv}\bigl(\{Z_n r:r\in\mathcal R^{\mathrm r}\}\bigr),
\]
where \(Z_n\) is the flow-to-choice-probability mapping defined in
\eqref{eqn:flow_to_p_matrix_form}. Here, $p^\pi$ denotes the choice probabilities induced by
$\pi$ on all nonempty menus.

\end{proposition}

A proof is provided in Section~\ref{proof:delete}. For Flow-CG, the pricing problem is solved directly on the restricted digraph \(G^{\mathrm r}=(N^{\mathrm r},E^{\mathrm r})\) using the Bellman--Ford algorithm. For ADMM, we impose zero flow on every arc removed from the restricted digraph, or equivalently, set the corresponding Block--Marschak polynomials to zero. This restriction preserves the efficient structure of the ADMM updates.

\subsection{Adding Additional Choice Behavior}
In some applications, researchers may wish to incorporate additional choice types that permit non-transitive behavior. Such situations may arise when behavioral biases are suspected to contribute to observed deviations from stochastic rationality. For example, \cite{dean2022better} augment the standard model with behavioral types exhibiting choice overload in order to better explain observed choice patterns.

Formally, we consider the case where researchers specify $L$ additional choice types $\{p_l\}_{l=1}^L$ and are interested in solving the augmented projection problem
\begin{equation}\label{eqn:problem_augmented}
    \min_{p}\left\|\widehat{p}- p\right\|_\Omega^2 \quad \text{ such that  } p \in \textrm{conv}\left(\mathcal{P}^{\mathcal{D}}_{\textrm{RUM}}\cup \{p_l\}_{l=1}^L\right) ,
\end{equation}
with observed choice probability $\widehat{p}$ and menu collection $\mathcal{D}$. The convex hull in \eqref{eqn:problem_augmented} characterizes the set of all choice probabilities that can be rationalized by a mixture of agents with strict preference rankings and additional choice types. We note that the additional choice types \(\{p_l\}_{l=1}^L\) need only be specified on the observed menu collection.

To solve the problem \eqref{eqn:problem_augmented} using column-generation algorithms such as IP-CG and Flow-CG, it suffices to modify the pricing problem in equation~\eqref{eqn:pricing} by enlarging the candidate set of probability vectors from \(\{p^\pi : \pi \in \Pi\}\) to $\{p^\pi : \pi \in \Pi\} \cup \{p_l\}_{l=1}^L$. The pricing problem can then be solved as two separate optimization problems.

For the ADMM algorithm and the flow formulation, we apply the union-of-polytopes result from \cite{balas1998disjunctive} and, after variable splitting, obtain the following formulation:

\begin{equation}\label{eqn:formulation_with_additional_types}
\min_{\mathbf f,\mathbf s,\btheta,\btheta',
      \{\blambda_l,\blambda_l'\}_{l=1}^L}
\left\|\widehat{p}-Z_\mathcal{D} \mathbf{f}-\sum_{l=1}^L \boldsymbol{\lambda}_l p_l\right\|^2_{\Omega}
\text{ s.t. }
\left\{
\begin{array}{l}
A\textbf{f}=\btheta b, \quad \textbf{f}=\textbf{s},\quad\textbf{s}\geq 0\\[0.3em]
\btheta-\btheta'=0, \blambda_l=\blambda_l',\forall l\in [L]\\
\btheta'\geq 0, \quad\blambda'_l\geq 0, \quad \forall l\in [L]\\
\btheta'+\sum_{l=1}^L \blambda'_l=1, \\[0.3em]
\end{array}
\right.
\end{equation}
where $Z_{\mathcal{D}}$ is defined after \eqref{cg_flow2} and $A$ and $b$ are defined in \eqref{eqn:flow_polytope}. $\btheta$ denotes the mixture weight assigned to the RUM component. Here $\mathbf{f}$ represents a flow with total mass $\btheta$, so $Z_\mathcal{D}\mathbf{f}$ corresponds to the weighted RUM component in the convex combination with weight $\btheta$. The BM formulation and detailed derivations are provided in Section~\ref{section:more_details}.

With this formulation, all computational structures developed in Section~\ref{section:ADMM} remain applicable. We write $\mathbf{z}=\left(\btheta,\blambda_1,...,\blambda_L\right)$ and $\mathbf{z}'=\left(\btheta',\blambda'_1,...,\blambda'_L\right)$. Let $\Delta^{L+1}$ denote the unit simplex in $\mathbb{R}^{L+1}$.
The augmented Lagrangian is
\begin{align*}
  &\left\|\widehat{p}-Z_\mathcal{D} \mathbf{f}-\sum_{l=1}^L \boldsymbol{\lambda}_l p_l\right\|^2_{\Omega} +\iota_{\mathbb R_+^d}(\mathbf s)+\iota_{\Delta^{L+1}}\left( \mathbf{z}'\right)\\
  +&\frac{\rho}{2}\|A\mathbf{f}-\btheta b\|_2^2  + \frac{\rho}{2}\|\mathbf{f}-\mathbf{s}\|_2^2 + \frac{\rho}{2}\|\mathbf{z}-\mathbf{z}'\|_2^2+\langle \mathbf{v}, A\mathbf{f}-\btheta b\rangle + \langle \mathbf{u}, \mathbf{f}-\mathbf{s}\rangle+ \langle \eta,\mathbf{z}-\mathbf{z}'\rangle.
\end{align*}
The algorithm is a two-block ADMM method with blocks
$(\mathbf f,\mathbf z)$ and $(\mathbf s,\mathbf z')$.
The first block is updated jointly, while the second-block update
separates into projections onto the nonnegative orthant and the
probability simplex. Using a Schur complement, the joint update
reduces to solves with the same large coefficient matrix as in
\eqref{eqn:update_r}, together with an $(L+1)$-dimensional linear
system. The method is therefore particularly suitable when the
number of additional types is modest. 

\begin{algorithm}[htbp]
\caption{ADMM with Additional Types}
\label{alg:admm_more_types_simplify}
\begin{algorithmic}[1]
\Require Initial values $(\mathbf{f}_0,\mathbf{s}_0,\mathbf{u}_0,\mathbf{v}_0,\boldsymbol{\eta}_0 ,\mathbf{z}_0,\mathbf{z}_0')$, operators $A,Z_{\mathcal D}$, vector $b$, observed probabilities $\widehat{p}$, penalty $\rho$, choice probabilities $\{p_l\}_{l=1}^L$.
\For{$t=0,1,...,$}   
    \State \textbf{Step 1:} Update $(\mathbf{f},\mathbf{z})$
\begin{equation}\label{eqn:update_r_augmented}
\left(\mathbf{f}_{t+1},\mathbf{z}_{t+1}\right)\gets\arg\min_{\mathbf{f},\mathbf{z}}
\left\{
\begin{aligned}
&\left\|\widehat{p}-Z_\D \mathbf{f}-\sum_{l=1}^L\boldsymbol{\lambda}_{l} p_l\right\|^2_{\Omega} 
+\langle \mathbf{u}_t,\mathbf{f}-\mathbf{s}_t\rangle +\langle \mathbf{v}_t,A\mathbf{f}-\btheta b\rangle + \langle \boldsymbol{\eta}_t ,\mathbf{z}-\mathbf{z}'_t\rangle\\
&
+\frac{\rho}{2}\|\mathbf{f}-\mathbf{s}_t\|_2^2
+\frac{\rho}{2}\|A\mathbf{f}- \btheta b\|_2^2+ \frac{\rho}{2}\|\mathbf{z}-\mathbf{z}'_t\|_2^2
\end{aligned}
\right\}    
\end{equation}
     \State \textbf{Step 2.1:} Update $\mathbf{s}$
    \begin{equation*}
        \begin{split}
 \mathbf{s}_{t+1}\gets \arg\min_{\mathbf{s}}
    \Big\{
    \iota_{\mathbb{R}_+^d}(\mathbf{s})
    +\langle \mathbf{u}_t,\mathbf{f}_{t+1}-\mathbf{s}\rangle
    +\frac{\rho}{2}\|\mathbf{f}_{t+1}-\mathbf{s}\|_2^2
    \Big\}             
        \end{split}
    \end{equation*}   
    \State \textbf{Step 2.2:} Update $\mathbf{z}'$ 
\begin{equation}\label{eqn:update_theta_augmented}
\mathbf{z}'_{t+1}\gets\arg\min_{\mathbf{z}'}\left\{
\begin{aligned}
\iota_{\Delta^{L+1}}\left( \mathbf{z}'\right)+ \frac{\rho}{2}\|\mathbf{z}_{t+1}-\mathbf{z}'\|_2^2+ \langle \boldsymbol{\eta}_t, \mathbf{z}_{t+1}-\mathbf{z'}\rangle  
\end{aligned}
\right\}
\end{equation}

    \State \textbf{Step 3:} Update the dual variables
    \begin{align*}
      & \mathbf{u}_{t+1}
    =
    \mathbf{u}_t+\rho\bigl(\mathbf{f}_{t+1}-\mathbf{s}_{t+1}\bigr),\quad          \mathbf{v}_{t+1}=
    \mathbf{v}_t+\rho\bigl(A\textbf{f}_{t+1}-\btheta_{t+1}b\bigr)\\
    &\boldsymbol{\eta}_{t+1}=\boldsymbol{\eta}_t+ \rho (\mathbf{z}_{t+1}-\mathbf{z}'_{t+1}) & 
    \end{align*}

\EndFor
\end{algorithmic}
\end{algorithm}

%% file: paper_appendix/Z_Proof_Add_Delete_Ranking.tex
\subsection{More Details on Adding Additional Choice Types}\label{section:more_details}

We first prove an auxiliary lemma.
\begin{lemma}\label{lemma:proj_conv_commute}
Let $\{A_k\}_{k=1}^K$ be subsets of $\mathbb{R}^m$, and let
$\Proj_{\mathbb{R}^n}:\mathbb{R}^m\to\mathbb{R}^n$ be a coordinate projection map. Then
\begin{equation}
    \conv\left(\bigcup_{k=1}^K \Proj_{\mathbb{R}^n}(A_k)\right)
    =
    \Proj_{\mathbb{R}^n}\left(
    \conv\left(\bigcup_{k=1}^K A_k\right)
    \right).
\end{equation}
\end{lemma}

\begin{proof}
Since $\Proj_{\mathbb{R}^n}$ is linear,
\[
\Proj_{\mathbb{R}^n}\left(\sum_{i=1}^L \lambda_i a_i\right)
=
\sum_{i=1}^L \lambda_i \Proj_{\mathbb{R}^n}(a_i)
\]
for any convex combination $\sum_{i=1}^L \lambda_i a_i$, where each $a_i$
belongs to $\bigcup_{k=1}^K A_k$. Therefore, taking convex combinations
before or after projection yields the same set, which proves the result.
\end{proof}

We can represent the RUM polytope as
\begin{equation*}
\Prum_{p,f}
=
\left\{
(p,f)
:
p = Z_nf,\;
Af = b,\;
f \geq 0
\right\}.
\end{equation*}

Without loss of generality, reorder the coordinates of $p$ so that the choice probabilities associated with the observed menu collection $\mathcal D$ appear first. We may therefore partition $p = (p_\D,p_{\D^c})$, where \(p_\D\) corresponds to the choice probabilities associated with the observed menu collection.

For each additional type $p_l$, define a vector in the extended space as 
\begin{equation}
    p_l^{\textrm{ext}}=\left(p_l,0_{d_1},0_{d_2}\right)
\end{equation}
where $0_{d_1}$ and $0_{d_2}$ are vectors of zeros with dimensions matching those of $p_{\D^c}$ and $f$, respectively. By \cite{balas1998disjunctive} (see also Section 2.6 in \cite{ip}), we have the following characterization
\begin{align*}
&\conv \left(
\Prum_{p,f}
\cup
\left\{p_l^{\mathrm{ext}}\right\}_{l=1}^L
\right) \\
&=
\left\{
(p_\D,p_{\D^c},f):
\begin{aligned}
&
p_\D=\sum_{l=0}^L p_\D^l,\quad
p_{\D^c}=\sum_{l=0}^L p_{\D^c}^l,\quad
f=\sum_{l=0}^L f^l,
\\
&
p^0 = Z_nf^0,\quad
Af^0 = \theta b,\quad
f^0 \ge 0,
\\
&
p_\D^l = \lambda_l p_l, \quad p_{\D^c}^l=0, \quad f^l=0,
\quad \forall l\in [L],
\\
&
\theta + \sum_{l=1}^L \lambda_l = 1,
\quad
\theta \ge 0,
\quad
\lambda_l \ge 0,
\quad \forall l\in [L]
\end{aligned}
\right\},
\end{align*}
where $p^0=\left(p_\D^0,p_{\D^c}^0\right)$,
which simplifies to
\begin{align}
&\conv \left(
\Prum_{p,f}
\cup
\left\{p_l^{\mathrm{ext}}\right\}_{l=1}^L
\right) \nonumber\\
&=
\left\{
(p_\D,p_{\D^c},f):
\begin{aligned}
&
p_\D=p_\D^0+ \sum_{l=1}^L \lambda_l p_l,\quad
p_{\D^c}=p_{\D^c}^0,\quad
f=f^0,
\\
&
p^0 = Z_nf^0,\quad
Af^0 = \theta b,\quad
f^0 \ge 0,
\\
&
\theta + \sum_{l=1}^L \lambda_l = 1,
\quad
\theta \ge 0,
\quad
\lambda_l \ge 0,
\quad \forall l\in [L]
\end{aligned}
\right\}. \label{eqn:20250513_1}
\end{align}
Notice that, by Lemma \ref{lemma:proj_conv_commute}, we have 
\begin{align*}
   & \conv\left( \Prum_{\mathcal{D}}\cup \{p_l\}_{l=1}^L\right)=\conv\left( \Proj_{\mathcal{D}}\left(\Prum_{p,f}\right)\cup \{\Proj_{\mathcal{D}}\left(p^{\textrm{ext}}_l\right)\}_{l=1}^L\right) \\
    =&\Proj_{\mathcal{D}}\left( \conv\left(\Prum_{p,f}\cup \{p_l^{\textrm {ext}}\}_{l=1}^L \right)\right).
\end{align*}
Hence problem \eqref{eqn:problem_augmented} becomes
\begin{equation*}
    \min_{p}\|\widehat{p}-p\|_\Omega^2 \quad \text{s.t.} \quad p\in \Proj_{\mathcal{D}}\left(\conv \left(
\Prum_{p,f}
\cup
\left\{p_l^{\mathrm{ext}}\right\}_{l=1}^L
\right)\right),
\end{equation*}
which is equivalent to
\begin{equation*}
   \min_{p_\D,p_{\D^c},f} \|\widehat{p}-p_\D\|_\Omega^2 \quad \text{s.t.} \quad \left(p_\D,p_{\D^c},f\right)\in \conv \left(
\Prum_{p,f}
\cup
\left\{p_l^{\mathrm{ext}}\right\}_{l=1}^L
\right).
\end{equation*}
From here, spelling out the restrictions in \eqref{eqn:20250513_1} and introducing split variables $s$ yields the formulation in \eqref{eqn:formulation_with_additional_types}.

For the BM formulation, the Block--Marschak representation of $ \mathcal{P}_{RUM}^{2^X}$ is
\begin{equation*}
    \mathcal{P}_{RUM}^{2^X}=\left\{p: K_np\geq 0,\quad \sum_{x\in D}p(x,D)=1, \forall D\in 2^X \right\},
\end{equation*}
where $K_np\geq 0$ denotes the BM inequalities defined in Section \ref{section:representations}. Redefine $p_l^{\textrm{ext}}=\left(p_l,0_{d_1}\right)$. By an application of \cite{balas1998disjunctive} and simplifying the representations, we arrive at the representation
\begin{align}
&\conv \left(
\mathcal{P}_{RUM}^{2^X}
\cup
\left\{p_l^{\mathrm{ext}}\right\}_{l=1}^L
\right) \nonumber\\
&=
\left\{
(p_\D,p_{\D^c}):
\begin{aligned}
&
p_\D=p_\D^0+ \sum_{l=1}^L \lambda_l p_l,\quad
p_{\D^c}=p_{\D^c}^0,
\\
&
K_np^0\geq 0, \sum_{x\in D}p^0(x,D)=\theta,\forall D\in 2^X\backslash \{\emptyset\}, 
\\
&
\theta + \sum_{l=1}^L \lambda_l = 1,
\quad
\theta \ge 0,
\quad
\lambda_l \ge 0,
\quad \forall l\in [L]
\end{aligned}
\right\},\nonumber
\end{align}
where $p^0=\left(p_\D^0,p_{\D^c}^0\right)$. Hence the optimization problem
\begin{equation*}
    \min_{p_\D,p_{\D^c}}\|\widehat{p}-p_\D\|_\Omega^2 \text{ s.t.} \quad \left(p_\D,p_{\D^c}\right)\in \conv \left(
\mathcal{P}_{RUM}^{2^X}
\cup
\left\{ p_l^{\mathrm{ext}}\right\}_{l=1}^L
\right)
\end{equation*}
can be represented as, after introducing splitting variables, 
\begin{equation*}
  \min_{\mathbf{q},\mathbf{s},\btheta,\btheta',
      \{\blambda_l,\blambda_l'\}_{l=1}^L}  \left\|\widehat{p}-\mathbf{q}_\D-\sum_{l=1}^L \blambda_l p_l\right\|_\Omega^2 \text{ s.t. } \left\{\begin{aligned}
        &K_n\mathbf{q} - \mathbf{s} = 0, \mathbf{s}\geq 0\\
        &  \sum_{x\in D}\mathbf{q}(x,D)=\btheta,\forall D\in 2^X, \\
&\btheta-\btheta'=0, \blambda_l=\blambda_l',\forall l\in [L]\\
&\btheta'\geq 0, \quad\blambda'_l\geq 0, \quad \forall l\in [L]\\
&\btheta'+\sum_{l=1}^L \blambda'_l=1, \\[0.3em]
    \end{aligned} \right.
\end{equation*}
where $\mathbf q=p^0$ is the weighted RUM component.

%% file: paper_main/A_Simplification_Correct.tex
\subsection{Simplification through the Co-occurrence Graph}\label{section:simplification}
This section introduces a new concept, which we call the \emph{co-occurrence graph}. This concept can be used to simplify the projection problem. In particular, we show that when this graph contains a large independent set—a standard graph-theoretic notion defined below—the problem admits a substantial simplification. We emphasize this structure because it arises in several instances that we encountered during the work, including the datasets in \cite{krajbich2012attentional} and \cite{feldman2022revealing}. 

Given a menu collection $\mathcal{D}$, we construct a graph on $X$ by placing an edge between $x$ and $y$ whenever there exists a menu $D \in \mathcal{D}$ such that $\{x,y\} \subseteq D$. We refer to this graph as the \emph{co-occurrence} graph. From this graph, we identify an \emph{independent set}: a subset of vertices with no edges between any pair. Note that we do not require researchers to find a maximum independent set but only to identify an independent set. However, a larger independent set will lead to more simplification.

An independent set $I$ has a simple revealed-preference
interpretation. Because no observed menu contains two
elements of $I$, the choices induced by a ranking depend
only on its restriction to $X\setminus I$ and the insertion
position of each element of $I$ relative to that restriction.
The ordering of elements assigned to the same insertion
position does not affect these choices.

For example, suppose $X=\{1,2,3,4\}$ and the observed menus
are $\{1,4\}$, $\{2,4\}$, and $\{3,4\}$. We may take
$I=\{1,2,3\}$, so that $X\setminus I=\{4\}$.
Consider rankings satisfying $\{1,2\}\succ 4\succ 3$.
The relative ordering of $1$ and $2$ does not affect the
choices induced on the observed menus.
We provide several further examples below.



\begin{example}
If $\D$ contains all binary menus, the resulting co-occurrence graph is complete. Any independent set hence has cardinality one.
\end{example}

\begin{example}\label{example:feldman2022revealing}
Consider the dataset in \cite{feldman2022revealing}, interpreted within a \DRUM framework. There are \(27\) alternatives in total, partitioned into two disjoint groups of sizes \(19\) and \(8\).  The observed binary choice sets consist only of pairs containing one alternative from the first group and one alternative from the second. Under this structure, the induced co-occurrence graph is bipartite: every edge runs between the two groups, while there are no edges within either group. Hence, the group of \(19\) alternatives forms an independent set.
\end{example}
\begin{example}\label{example:krajbich2012attentional}
Consider the dataset in \cite{krajbich2012attentional}, in which subjects make 300 binary choices over 50 consumer goods and 6 price levels. The resulting co-occurrence graph is again bipartite: every edge connects a consumer good to a price, with no edges within the set of goods or within the set of prices. Moreover, it is natural to assume monotonic preferences over prices (i.e., lower prices are preferred), so that the ordering among the 6 prices is uniquely determined.
\end{example}

For a given menu collection \(\mathcal{D}\) and an independent set \(I\), we now show that the linear optimization problem over the RUM polytope can be solved efficiently once a ranking on \(X\setminus I\) is fixed. Since there are only \(|X\setminus I|!\) possible rankings of \(X\setminus I\), the problem remains computationally tractable whenever \(X\setminus I\) is small, or equivalently, whenever \(I\) is large relative to \(X\). From an algorithmic perspective, this decomposition can be used as a subroutine to replace the pricing step in the column-generation algorithm, which would otherwise be implemented via a network-flow procedure or an integer program.

To formalize this idea, we introduce the following definition.

 \begin{definition}[Restricted Rankings and Choice Probabilities]\label{defn:choice_probabilties}
Consider a menu collection $\D$ and a resulting independent set $I$. Fix a ranking $\sigma$ on $X\setminus I$.  
For each alternative $x\in I$, consider the rankings obtained by inserting $x$ into $\sigma$ at every possible position. Let $\Pi_x^{\sigma}$ denote the set of rankings obtained in this way. Let $\mathcal{P}_x^{\sigma}$ denote the corresponding discrete set of stochastic choice probabilities induced by the rankings in $\Pi_x^{\sigma}$ on those choice sets $D\in \D$ that contain $x$. For a ranking $\sigma$ on $X\setminus I$, we write $p^\sigma$ for the corresponding choice probabilities on subsets of $X\setminus I$.
 \end{definition}

\begin{example}
Consider the menus $\{a,x\}$, $\{a,y\}$, $\{b,x\}$, and $\{b,y\}$. In this case, we may take $I=\{x,y\}$ and $X\setminus I=\{a,b\}$.

Fix a ranking $\sigma$ on $X\setminus I$, for example $a\succ b$. For the alternative \(x\in I\), consider the set of rankings $\Pi_x^{\sigma}$ obtained by inserting \(x\) into $\sigma$ at every possible position:
\begin{equation*}
\Pi_x^{\sigma}
=
\left\{
x \succ a \succ b,\;
a \succ x \succ b,\;
a \succ b \succ x
\right\}.
\end{equation*}
Each ranking in $\Pi_x^{\sigma}$ induces a choice probability vector on the menus $\{a,x\}$ and $\{b,x\}$. 
\end{example}

\begin{proposition}\label{theorem:simplification}
Given a menu collection $\D$, let $I$ be a resulting independent set in the co-occurrence graph. Let $\Sigma$ be the set of all possible rankings on $X\setminus I$. Let $\mathcal{P}^\D_{\textrm{RUM}}$ be defined as in \eqref{eqn:v_representation}, and $\Pi^\sigma_x$, $\mathcal{P}^\sigma_x$ and $p^\sigma$ be defined as in Definition \ref{defn:choice_probabilties} for every $x\in I$ and $\sigma\in\Sigma$. We have $|D\cap I|\leq 1$ for all $D\in\mathcal{D}$ and the following decomposition holds:
\begin{align}\label{eqn:23_04192026}
\begin{split}
&\max_{\mathbf{p}\in \mathcal{P}^{\mathcal D}_{\mathrm{RUM}}}   
\sum_{D\in\mathcal{D}}\sum_{a\in D}
\mathbf{p}(a,D)\, w(a,D)=\\
&\max_{\boldsymbol{\sigma} \in \Sigma} \Bigg(\sum_{D\in \mathcal{D}^{X\setminus I}}\sum_{a\in D}\mathbf{p}^{\boldsymbol{\sigma}}(a,D)w(a,D) +\sum_{x\in I}
\max_{\mathbf{p}\in \mathcal{P}_x^{\boldsymbol{\sigma}}}
\sum_{D\in \mathcal{D}^x}\sum_{a\in D}\mathbf{p}(a,D)w(a,D)\Bigg),    
\end{split}
\end{align}
where \( \{w(a,D)\}_{a\in D,D\in\mathcal{D}}\) is an arbitrary vector, 
\(\mathcal D^x = \{D\in\mathcal D : x\in D\}\) and \(\mathcal D^{X\setminus I} = \{D\in\mathcal D : D\subseteq X\setminus I\}\).   Moreover, an optimizer of the left-hand side can be constructed from the optimizers of the right-hand side.
\end{proposition}

The importance of Proposition~\ref{theorem:simplification} lies in the resulting simplification of the linear optimization problem. In particular, the proposition implies that solving the left-hand side of \eqref{eqn:23_04192026} reduces to solving $|\Sigma|=|X\setminus I|!$ subproblems. For each subproblem associated with a given $\sigma \in \Sigma$, one further solves $|I|$ subproblems, each involving only $|X\setminus I|+1$ candidate positions (or equivalently, rankings) for inserting a single element of $I$ into $\sigma$. 

Hence, the total number of configurations that need to be evaluated is on the order of $|\Sigma|\cdot |I|\cdot (|X\setminus I|+1)$. By contrast, a direct enumeration approach would require, for each $\sigma \in \Sigma$, considering $(|X\setminus I|+1)^{|I|}$ configurations, since each element of $I$ can be placed in any of $|X\setminus I|+1$ positions relative to $X\setminus I$. Consequently, the decomposition reduces the total computational burden from $|\Sigma|\cdot (|X\setminus I|+1)^{|I|}$ to $|\Sigma|\cdot |I|\cdot (|X\setminus I|+1)$. When $|\Sigma|$ is small, this yields a substantial simplification. We illustrate the implications of this result using the examples discussed above.

\begin{excont}{example:feldman2022revealing}
For the dataset in \citet{feldman2022revealing},
$|\Sigma|=8!=40{,}320$, and $\mathcal P_x^\sigma$ contains
at most nine elements for every $\sigma\in\Sigma$ and
$x\in I$. The decomposition therefore requires at most
$8!\times19\times9$ candidate evaluations, compared with
$8!\times9^{19}$ joint insertion configurations under
direct enumeration.
\end{excont}

\begin{excont}{example:krajbich2012attentional}
In the setting of \cite{krajbich2012attentional}, we may take $I$ to be the set of non-monetary alternatives and $X\setminus I$ to be the set of monetary payoffs. It is natural to assume that individuals have a strict ranking over monetary payoffs based on their magnitudes. Under this assumption, the set $\Sigma$ reduces to a singleton. As a result, the number of configurations is reduced from $7^{50}$ on the left-hand side to $7 \times 50$ on the right-hand side.
\end{excont}


\begin{remark}
\citet{jagabathula2019limit} (JR) also study methods for
simplifying and decomposing linear optimization over the
RUM polytope. Our results differ from theirs in several respects. First, the underlying graph constructions are different. Our co-occurrence graph is defined directly on the set of alternatives, with edges indicating whether two alternatives appear together in an observed menu. By contrast, the choice graph in JR is defined on the collection of menus through what they term rational-separation relationships, and the resulting choice-graph structure depends on the menu collection under consideration.

Second, the resulting simplifications are also different. JR identify several menu structures, including nested, laminar, differentiated, and delete-\(k\) collections, under which efficient linear optimization is possible. Our approach instead exploits the existence of large independent sets in the co-occurrence graph. The closest analogue in JR is the differentiated-collection
setting. Our framework also allows observations on
arbitrary subsets of the core set $X\setminus I$, while
preserving the independent-set structure.
\end{remark}

%% file: paper_appendix/Z_Hardness_Proof.tex
\subsubsection{Preliminary Results}
The encoding length of a rational number $p/q$ is $\lceil \log_2(|p|+1) \rceil + \lceil \log_2(q+1) \rceil$. 
The encoding length of a vector $a\in\mathbb{Q}^n$ or a matrix $A\in\mathbb{Q}^{m\times n}$ is defined as the sum of the encoding lengths of its entries. 
We denote the encoding length of a scalar, vector, or matrix $a$ by $\langle a\rangle$. 
The encoding length of an inequality $a^\top x \le b$ is defined as the sum of the encoding lengths of its coefficients, namely $\langle a\rangle + \langle b\rangle$.

Let $B_n(c,d)$ denote the ball in $\mathbb{R}^n$ centered at $c\in \mathbb{R}^n$ with radius $d$. We first introduce a definition from \cite{grotschel2012geometric}. We slightly simplify the original definition by restricting attention to bounded polytopes when defining facet-complexity and vertex-complexity. 

\begin{definition}[Definition 2.1.16(d) and Definition 6.2.2 in \cite{grotschel2012geometric}]
A polytope $K$ is called a \emph{well-described centered polytope} if the following information about the polytope $K$ is given explicitly:
\begin{enumerate}
    \item An integer $n$ such that $K \subseteq \mathbb{R}^n$;
    \item A positive rational number $R$ such that $K \subseteq B_n(0,R)$;
    \item A positive rational radius $r \in \mathbb{R}$ and a vector $a_0 \in \mathbb{Q}^n$ such that $B_n(a_0,r) \subseteq K$;
    \item A number $\varphi \in \mathbb{N}$ such that there exists a system of linear inequalities with rational coefficients that defines $K$ and the encoding length of each inequality is at most $\varphi$;
    \item A number $\nu \in \mathbb{N}$ such that there exists a finite set $V$ such that $K=\mathrm{conv}(V)$ and each vector in $V$ has encoding length at most $\nu$.
\end{enumerate}
We denote such a polytope by $\left(K;n,R,r,a_0,\varphi,\nu\right)$.  
The encoding length of this description of $K$ is defined as $n+\langle R\rangle+\langle r\rangle+\langle a_0\rangle+\varphi+\nu$.
\end{definition}

We show below that the linear-ordering polytope (i.e., the \DRUM polytope with all menus of size two) is a well-described centered polytope. Let $B_n(C,d)=\{y\in\mathbb{R}^n : \|x-y\|_2\le d \text{ for some } x\in C\}$ denote the $d$-inflation of the set $C$. Given a set of menus $\D$, we also introduce the following Membership Problem for an arbitrary polytope $K$ and Optimization Problem for a well-described centered polytope.

\begin{definition}[Strong Membership Problem for a Polytope $K$]
Given a vector $x\in \mathbb{Q}^n$, decide whether $x\in K$.
\end{definition}

\begin{definition}[Strong Optimization Problem for a Well-Described Centered Polytope $\left(K;n,R,r,a_0,\varphi,\nu\right)$]
Given a well-described centered polytope $\left(K;n,R,r,a_0,\varphi,\nu\right)$ and a vector $c\in \mathbb{Q}^n$, find a vector $x^*\in K$ that maximizes $c^\top x$ over $K$, or assert that $K$ is empty.
\end{definition}
The following proposition follows from Theorem 6.3.2 and Corollary 4.3.12 in \cite{grotschel2012geometric}. We do not reproduce the exact statements here, as doing so would require introducing a substantial number of additional definitions. Instead, we state only the implications of these results that are relevant for our setting.

\begin{proposition}[Theorem 6.3.2 and Corollary 4.3.12 in \cite{grotschel2012geometric}]
\label{prop:equivalence}
Let $\left(K; n, R, r, a_0, \varphi, \nu\right)$ be a full-dimensional well-described centered polytope. 
Then the strong optimization problem over $K$ can be solved in polynomial time using a polynomial number of calls to an algorithm that solves the strong membership problem for $K$. 
In particular, if the strong membership problem for $K$ can be solved in polynomial time, then the strong optimization problem over $K$ can also be solved in polynomial time.
\end{proposition}
\begin{proof}
In the terminology of \cite{grotschel2012geometric}, a polynomial-time algorithm
for the strong membership problem implies the existence of a polynomial-time
algorithm for the weak membership problem (as defined in \cite{grotschel2012geometric}). By Corollary 4.3.12 of
\cite{grotschel2012geometric}, this in turn implies the existence of a
polynomial-time algorithm for the weak optimization problem for a centered
convex body. Finally, by Theorem 6.3.2(a) of \cite{grotschel2012geometric}, this
implies the existence of a polynomial-time algorithm for the strong optimization
problem for a full-dimensional well-described centered polyhedron. The result follows.
\end{proof}

We now consider a particular instance of the \DRUM polytope, namely 
$\mathcal{P}^\D_{\mathrm{RUM}}$ defined on all binary menus, i.e., menus of the form 
$\{x,y\}$ for some $x,y\in X$. This polytope is also known in the literature as the 
\emph{linear-ordering polytope} \cite{marti2011linear}. 

Unlike the labeling convention used for $\mathcal{P}^\D_{\mathrm{RUM}}$ in the main text, we represent this polytope in a lower-dimensional space by retaining only the coordinate $(x,\{x,y\})$ whenever $x$ is lexicographically smaller than $y$.\footnote{For example, in the three-alternative case $X=\{1,2,3\}$, we retain only the three coordinates $\{p(1,\{1,2\}),\, p(1,\{1,3\}),\, p(2,\{2,3\})\}$.} 
The remaining entries can be recovered from the requirement that probabilities within each menu sum to one. We denote the resulting projected polytope by $\mathcal{P}^{\mathrm{binary}}_{\mathrm{RUM}}$, whose ambient space is $\mathbb{R}^{\binom{n}{2}}$, where $n$ is the number of alternatives. The polytope $\mathcal{P}^{\mathrm{binary}}_{\mathrm{RUM}}$ can be shown to be a well-described centered polytope.

\begin{lemma}\label{lemma:binary_polytope_properties}
Let $X$ be a set with $n$ alternatives and consider the associated linear-ordering polytope $\mathcal{P}^{\mathrm{binary}}_{\mathrm{RUM}}$. 
The polytope $\mathcal{P}^{\mathrm{binary}}_{\mathrm{RUM}}$ has the following properties:
\begin{enumerate}[label=(\roman*)]
    \item $\mathcal{P}^{\mathrm{binary}}_{\mathrm{RUM}}$ is full-dimensional.
    
    \item $\mathcal{P}^{\mathrm{binary}}_{\mathrm{RUM}} \subset B_{\binom{n}{2}}(0,n)$.
    
    \item $\mathcal{P}^{\mathrm{binary}}_{\mathrm{RUM}}$ has vertex-complexity at most $2\binom{n}{2}$.
    
    \item $\mathcal{P}^{\mathrm{binary}}_{\mathrm{RUM}}$ has facet-complexity at most $6n^6$.
    
    \item $\mathcal{P}^{\mathrm{binary}}_{\mathrm{RUM}}$ contains a ball 
    $B_{\binom{n}{2}}(a_0,r)$ where $r=2^{-42n^{12}}$ and
    \[
        a_0=\frac{1}{\binom{n}{2}+1}\sum_{i=0}^{\binom{n}{2}} v_i,
    \]
    where $\{v_i\}_{i=0,\ldots,\binom{n}{2}}$ are defined in the proof below.
    \item The encoding length of 
    \[
    \left(\mathcal{P}^{\mathrm{binary}}_{\mathrm{RUM}};\binom{n}{2},n,  r , a_0, 6n^6 ,2\binom{n}{2} \right)
    \]
    is $O(n^{12})$.
\end{enumerate}
\end{lemma}
\begin{proof}
We first show that $\mathcal{P}^{\mathrm{binary}}_{\mathrm{RUM}}$ is full-dimensional. 
The ambient dimension of $\mathcal{P}^{\mathrm{binary}}_{\mathrm{RUM}}$ is $\binom{n}{2}$, so it suffices to exhibit $\binom{n}{2}+1$ affinely independent points in the polytope.

Under our indexing convention, the zero vector belongs to $\mathcal{P}^{\mathrm{binary}}_{\mathrm{RUM}}$, corresponding to the ranking $n \succ (n-1) \succ \cdots \succ 1$. 
Hence it suffices to construct $\binom{n}{2}$ linearly independent vectors.

For each $i=1,\ldots,n-1$ and $k=1,\ldots,n-i$, place alternative $i$ in the $k$th position of the ranking $i+1 \succ i+2 \succ \cdots \succ n$, while keeping $\{1,\ldots,i-1\}$ in the last $i-1$ positions in the order $i-1 \succ \cdots \succ 1$. This construction produces
\[
\sum_{i=1}^{n-1}(n-i)=\binom{n}{2}
\]
distinct rankings. The matrix representation of the corresponding stochastic choice functions is a lower-triangular matrix with ones in all lower-triangular entries (including the diagonal).\footnote{For example, in the case of four alternatives, consider the six rankings 
$1 \succ 2 \succ 3 \succ 4$, $2 \succ 1 \succ 3 \succ 4$, $2 \succ 3 \succ 1 \succ 4$, 
$2 \succ 3 \succ 4 \succ 1$, $3 \succ 2 \succ 4 \succ 1$, and $3 \succ 4 \succ 2 \succ 1$. 
These yield the matrix
\[
\begin{array}{c|cccccc}
(i,D) 
& {\scriptstyle 1\succ2\succ3\succ4}
& {\scriptstyle 2\succ1\succ3\succ4}
& {\scriptstyle 2\succ3\succ1\succ4}
& {\scriptstyle 2\succ3\succ4\succ1}
& {\scriptstyle 3\succ2\succ4\succ1}
& {\scriptstyle 3\succ4\succ2\succ1} \\
\hline
(1,\{1,2\}) & 1 & 0 & 0 & 0 & 0 & 0 \\
(1,\{1,3\}) & 1 & 1 & 0 & 0 & 0 & 0 \\
(1,\{1,4\}) & 1 & 1 & 1 & 0 & 0 & 0 \\
(2,\{2,3\}) & 1 & 1 & 1 & 1 & 0 & 0 \\
(2,\{2,4\}) & 1 & 1 & 1 & 1 & 1 & 0 \\
(3,\{3,4\}) & 1 & 1 & 1 & 1 & 1 & 1
\end{array}
\]
where each column corresponds to the choice-probability vector induced by a ranking. 
If we include the seventh ranking $4 \succ 3 \succ 2 \succ 1$, which corresponds to the column vector of all zeros, the corresponding seven choice-probability vectors are affinely independent.
}  Hence, these stochastic choice functions are linearly independent.
Together with the zero vector, these points are affinely independent, establishing that $\mathcal{P}^{\mathrm{binary}}_{\mathrm{RUM}}$ is full-dimensional. This proves (i). 

Let $v_0,\ldots,v_{\binom{n}{2}}$ denote the choice-probability vectors induced by the $\binom{n}{2}+1$ rankings constructed above, including the ranking corresponding to the zero vector.

By definition, each vertex has norm at most $\sqrt{\binom{n}{2}}$. Since every point in $\mathcal{P}^{\mathrm{binary}}_{\mathrm{RUM}}$ is a convex combination of vertices, it follows that the norm of any element of the polytope is at most $\sqrt{\binom{n}{2}}\leq n$. This proves (ii).

Each vertex of $\mathcal{P}^{\mathrm{binary}}_{\mathrm{RUM}}$ is a binary vector in $\mathbb{R}^{\binom{n}{2}}$. Hence its encoding length is at most $2\binom{n}{2}$. This proves (iii).

By Lemma 6.2.4 of \cite{grotschel2012geometric}, $\mathcal{P}^{\mathrm{binary}}_{\mathrm{RUM}}$ has facet-complexity at most $3\left(\binom{n}{2}\right)^2\times 2\binom{n}{2}\leq 6n^6$. This proves (iv).

Applying the argument in the proof of Lemma~6.2.6 of
\cite{grotschel2012geometric} to the full-dimensional simplex
$\operatorname{conv}\{v_0,\ldots,v_{\binom{n}{2}}\}$, we conclude that
$\mathcal{P}^{\mathrm{binary}}_{\mathrm{RUM}}$ contains a ball centered at
\[
a_0
=
\frac{1}{\binom{n}{2}+1}
\sum_{i=0}^{\binom{n}{2}} v_i
\]
with radius at least $2^{-42n^{12}}$. This proves (v).

Finally, we note that $\langle a_0\rangle = O\!\left(\binom{n}{2}\log_2 n\right)$, since $a_0$ is a vector of length $\binom{n}{2}$ and each entry has encoding length $O(\log_2 n)$. The dominant term in the encoding length arises from $\langle r\rangle = O(n^{12})$. This proves (vi).
\end{proof}
\subsubsection{Proof of Proposition \ref{prop:NPhard}}

We now introduce the \emph{linear ordering problem} \cite{marti2011linear}. 
Let $D_n=(\{1,\ldots,n\},A_n)$ denote the complete directed graph on $n$ nodes. 
A \emph{spanning tournament} $T$ is a subset of arcs $T \subseteq A_n$ such that, 
for every pair of distinct nodes $i$ and $j$, exactly one of the arcs $(i,j)$ 
or $(j,i)$ belongs to $T$. An \emph{acyclic spanning tournament} is a spanning 
tournament that contains no directed cycles. The linear ordering problem is 
defined as follows.
\begin{definition}[Linear Ordering Problem \citep{marti2011linear}]
Given the complete directed graph $D_n=(\{1,\ldots,n\},A_n)$ with rational arc weights $c_{ij}$ 
for every ordered pair (arc) of distinct nodes $i,j$, compute an acyclic spanning tournament 
$T \subseteq A_n$ that maximizes $\sum_{(i,j)\in T} c_{ij}$.
\end{definition}

It is well known that the linear ordering problem is NP-hard via a
reduction from the acyclic subgraph problem, which is equivalent to the
feedback arc set problem \citep{marti2011linear}. 

There is a one-to-one correspondence between acyclic spanning tournaments
and strict rankings of the $n$ nodes: an arc $(i,j)$ appears in the
tournament if and only if $i$ is ranked above $j$. Hence the linear
ordering problem can be written as
\begin{align*}
     &\max_{\text{rankings over the $n$ nodes}} \sum_{i\not= j} 1\{i \succ j\} c_{ij}\\
     &= \sum_{j>i} c_{ji}
     + \left(
     \max_{\text{rankings over the $n$ nodes}}
     \sum_{j>i} 1\{i \succ j\} (c_{ij}-c_{ji})
     \right)\\
     & = \sum_{j>i} c_{ji}
     + \left(
     \max_{p\in \mathcal{P}^{\mathrm{binary}}_{\mathrm{RUM}}}
     \sum_{j>i} p(i,\{i,j\}) (c_{ij}-c_{ji})
     \right).
\end{align*}
Here $\mathcal{P}^{\mathrm{binary}}_{\mathrm{RUM}}$ is defined as in
Lemma~\ref{lemma:binary_polytope_properties}. The first equality follows
from the identity $1-1\{i\succ j\}=1\{j\succ i\}$. The second equality
follows from the fact that a linear function over a polytope attains its
maximum at a vertex, and the vertices of
$\mathcal{P}^{\mathrm{binary}}_{\mathrm{RUM}}$ correspond to strict
rankings. Therefore, the linear ordering problem can be formulated as a
strong optimization problem over this polytope.

Reducing the linear ordering problem to the strong optimization problem also requires the geometric and encoding parameters needed to describe the polytope. By Lemma~\ref{lemma:binary_polytope_properties}, these are given by the dimension $d=\binom{n}{2}$, an outer radius $R=n$, an interior point $a_0$ with radius $r$, facet-complexity $\varphi\le 6n^6$, and vertex-complexity $\nu=2\binom{n}{2}$. All of these quantities can be computed explicitly and have encoding length polynomial in $n$.

By Proposition~\ref{prop:equivalence}, strong optimization
Turing-reduces to strong membership.\footnote{For the definition of
Turing reduction, see Section~5 of \cite{schrijver2003combinatorial}.}
It follows that the linear ordering problem Turing-reduces to the \DRUM
strong-membership problem. Hence, the \DRUM strong-membership problem is
NP-hard. Consequently, unless $P=NP$, there is no polynomial-time
algorithm for the \DRUM strong-membership problem.

%% file: paper_appendix/Z_BM_Lemma.tex
We prove Lemma \ref{lemma:mobius_eigen} in Section \ref{section:bm_admm} as well as the statement in \eqref{eqn:bm_kronecker}. We combine them in the following Lemma \ref{lemma:mobius_eigen_appendix}.

\begin{lemma}\label{lemma:mobius_eigen_appendix}
Define $M$ as in (\ref{M2}).
\begin{enumerate}[label=(\roman*)]
    \item In a setting with $n$ alternatives, the values of the BM polynomials involving the $k$th alternative and $n-1$ other alternatives can be expressed as
\begin{equation}
M^{\otimes (n-1)} p_{k}
= \left(\underbrace{M \otimes M \otimes \cdots \otimes M}_{\text{$n-1$ times}}\right) p_{k},
\end{equation}
where $p_{k}$ denotes the choice probability vector associated with the $k$th alternative, indexed according to the rule introduced in Section~\ref{section:index}.
\item We have  $\left(M^{\otimes(n-1)}\right)^\top M^{\otimes(n-1)}=\left(M^\top M\right)^{\otimes (n-1)}= \left(\begin{bmatrix}
    2 & -1 \\
    -1 & 1 \\
\end{bmatrix}\right)^{\otimes (n-1)}$.
\item $\left(M^{\otimes(n-1)}\right)^\top M^{\otimes(n-1)}$ has the eigen-decomposition:
\begin{equation*}
    \left(M^{\otimes(n-1)}\right)^\top M^{\otimes(n-1)}=  Q^{\otimes \left(n-1\right) }\Lambda ^{\otimes\left(n-1\right)}\left(Q^{\otimes\left(n-1\right)} \right)^T,
\end{equation*}
where
\begin{equation*}
    Q=\begin{bmatrix}
        -\sqrt{\frac{3+\sqrt{5}}{5+\sqrt{5}}} & \sqrt{\frac{3-\sqrt{5}}{5-\sqrt{5}}}  \\
        \sqrt{\frac{2}{5+\sqrt{5}}} & \sqrt{\frac{2}{5-\sqrt{5}}}
    \end{bmatrix},
\end{equation*}
and 
\begin{equation*}
    \Lambda =\begin{bmatrix}
    (3+\sqrt{5})/2 & 0 \\
    0 & (3-\sqrt{5})/2 \\
\end{bmatrix}.
\end{equation*}
\item Let $I_{2^{n-1}}$ be an identity matrix with the dimension $2^{n-1}\times 2^{n-1}$. Let $\alpha$ and $\beta$ be two scalars. We have the eigen-decomposition,
\begin{equation*}
  \alpha I_{2^{n-1}}+\beta \left(M^{\otimes(n-1)}\right)^\top M^{\otimes(n-1)}= Q^{\otimes \left(n-1\right)}\left( \alpha I_{2^{n-1}}+\beta  \Lambda ^{\otimes\left(n-1\right)}\right)
  \left(Q^{\otimes\left(n-1\right)} \right)^\top.
\end{equation*}

\end{enumerate}
\end{lemma}
\begin{proof}

For (i), we prove the identity by induction on $n$.
The base case $n=2$ follows from \eqref{eqn:bm_n2}.
Suppose the result holds for $n=N-1$. Let $\ell$ be the alternative with the largest index in
$X\setminus\{k\}$. Under the indexing rule in
Section~\ref{section:index}, partition
\[
p_k=
\begin{bmatrix}
p_k^{(1)}\\
p_k^{(0)}
\end{bmatrix},
\]
where $p_k^{(1)}$ and $p_k^{(0)}$ collect the entries for menus
that contain and do not contain alternative $\ell$, respectively. Then
\begin{align*}
M^{\otimes(N-1)}p_k
&=
\left(M\otimes M^{\otimes(N-2)}\right)
\begin{bmatrix}
p_k^{(1)}\\
p_k^{(0)}
\end{bmatrix} =
\begin{bmatrix}
M^{\otimes(N-2)}p_k^{(1)}\\
M^{\otimes(N-2)}\left(p_k^{(0)}-p_k^{(1)}\right)
\end{bmatrix}.
\end{align*}
By the induction hypothesis, the first block gives the BM polynomials
associated with menus containing alternative $\ell$. By the recursive
definition of the BM polynomials, the second block gives those associated
with menus not containing alternative $\ell$. To see this, consider a set
$E$ such that $k\in E$ and $\ell\notin E$. The BM polynomial associated with
$(k,E)$ is
\begin{align*}
\sum_{D\supseteq E}(-1)^{|D\setminus E|}p(k,D)
&=
\sum_{\substack{D\supseteq E\\ \ell\notin D}}
(-1)^{|D\setminus E|}p(k,D)
+
\sum_{\substack{D\supseteq E\\ \ell\in D}}
(-1)^{|D\setminus E|}p(k,D)\\
&=
\sum_{\substack{D\supseteq E\\ \ell\notin D}}
(-1)^{|D\setminus E|}
\left[p(k,D)-p(k,D\cup\{\ell\})\right].
\end{align*}
The two blocks
therefore give all BM polynomials involving alternative $k$, proving (i).

To prove (ii) and (iii), first notice that by the mixed-product property of matrix multiplication and the Kronecker product,
\begin{align*}
    &A^{\otimes (n-1)}B^{\otimes (n-1)}= \left(A^{\otimes (n-2)}\otimes A\right)\left(B^{\otimes (n-2)}\otimes B\right)\\
    =&\left(A^{\otimes (n-2)}B^{\otimes (n-2)}\right)\otimes (AB)=...= (AB)^{\otimes (n-1)},    
\end{align*}
and similarly
\begin{equation*}
    \left(ABC\right)^{\otimes (n-1)}= \left(AB\right)^{\otimes (n-1)} C^{\otimes (n-1)}=A^{\otimes (n-1)}B^{\otimes (n-1)} C^{\otimes (n-1)}.
\end{equation*}
Hence (ii) follows because
\begin{align*}
\left(M^{\otimes (n-1) }\right)^\top    M^{\otimes (n-1) }=\left(M^\top\right)^{\otimes (n-1)}   M^{\otimes (n-1) }= \left(M^\top M\right)^{\otimes (n-1)},    
\end{align*}
where
\begin{equation*}
    M^\top M=\begin{bmatrix}
        2 & -1 \\
        -1 & 1 
    \end{bmatrix}.
\end{equation*}

(iii) follows because $M^\top M$ has the eigen-decomposition $M^\top M= Q\Lambda Q^\top $, with $Q$ and $\Lambda$ given in the lemma and 
\begin{align*}
    &\left(M^\top M\right)^{\otimes (n-1)}=\left(Q\Lambda Q^\top\right)^{\otimes (n-1)}=Q^{\otimes (n-1)}\Lambda^{\otimes (n-1)}\left(Q^{\otimes (n-1)}\right)^\top.    
\end{align*}
(iv) is standard.
\end{proof}

%% file: paper_appendix/Z_Proof_Bellman_Ford.tex
The convergence of the Flow-CG algorithm follows from the standard proof of the Frank--Wolfe algorithm \citep{beck2017first,braun2025conditional}. We therefore omit the details.

Recall the \DRUM projection problem \eqref{eqn:DC-RUM_Proj} and let \(p^*\) be a solution to the problem. For simplicity, we define \(g(p)=\|p-\widehat p\|_\Omega^2=(p-\widehat p)^\top\Omega(p-\widehat p)\). 

Since \(\Omega\) is positive semidefinite, \(g\) is convex and \(\nabla g(p)=2\Omega(p-\widehat p)\). By convexity of \(g\), for any $p\in\mathcal{P}^{\mathcal D}_{\mathrm{RUM}}$
 and any minimizer \(p^*\),
\[
g(p^*)\ge g(p)+\nabla g(p)^\top(p^*-p) \implies g(p)-g(p^*)\le \nabla g(p)^\top(p-p^*).
\]
Substituting the gradient expression gives
\begin{align*}
& g(p)-g(p^*)
\le 2(p-\widehat p)^\top\Omega(p-p^*)=2\left(\widehat p-p\right)^\top\Omega\left(p^*-p\right) \le  2\max_{\widetilde p\in \mathcal{P}^{\mathcal{D}}_{\textrm{RUM}}}(\widehat p-p)^\top\Omega(\widetilde p-p),    
\end{align*}
since \(p^*\in \mathcal{P}^{\mathcal{D}}_{\textrm{RUM}}\). Hence, we have $0\le g\left(p\right)-g\left(p^*\right)\le 2c(p)$.

To prove the upper bound on $c(p)$, fix \(p\in \mathcal{P}^{\mathcal{D}}_{\textrm{RUM}}\). For any \(\widetilde p\in \mathcal{P}^{\mathcal{D}}_{\textrm{RUM}}\),
\begin{align*}
(\widehat p-p)^\top\Omega(\widetilde p-p)
=&\; (\widehat p-p^*)^\top\Omega(\widetilde p-p^*)
   +(\widehat p-p^*)^\top\Omega(p^*-p) \\
&\; +(p^*-p)^\top\Omega(\widetilde p-p^*)
   +(p^*-p)^\top\Omega(p^*-p).
\end{align*}
Moreover, for any $p^*\in \mathcal{P}^*$, we have
\begin{align*}
c(p)
\le &\;
\max_{\widetilde p\in \mathcal{P}^{\mathcal{D}}_{\textrm{RUM}}}(\widehat p-p^*)^\top\Omega(\widetilde p-p^*)
 +(\widehat p-p^*)^\top\Omega(p^*-p) \\
&\; +\max_{\widetilde p\in \mathcal{P}^{\mathcal{D}}_{\textrm{RUM}}}(p^*-p)^\top\Omega(\widetilde p-p^*)
 +(p^*-p)^\top\Omega(p^*-p)\\
\le &\;
(\widehat p-p^*)^\top\Omega(p^*-p)
+\max_{\widetilde p\in \mathcal{P}^{\mathcal{D}}_{\textrm{RUM}}}(p^*-p)^\top\Omega(\widetilde p-p^*)
+(p^*-p)^\top\Omega(p^*-p),
\end{align*}
where the second inequality follows from the optimality condition
\[
(\widehat p-p^*)^\top\Omega(\widetilde p-p^*)\le 0,
\qquad \forall\,\widetilde p\in \mathcal{P}^{\mathcal{D}}_{\textrm{RUM}},
\]
which holds because $p^*$ minimizes $g$ over $\mathcal{P}^{\mathcal{D}}_{\textrm{RUM}}$ (e.g., Theorem 3.67 in \cite{beck2017first}).

Using the Cauchy--Schwarz inequality,
\begin{align*}
c(p)
\le &\;
\lambda_{\max}(\Omega)\|\widehat p-p^*\|_2\|p-p^*\|_2 \\
&\; +\lambda_{\max}(\Omega)\|p-p^*\|_2
   \max_{\widetilde p\in \mathcal{P}^{\mathcal{D}}_{\textrm{RUM}}}\|\widetilde p-p^*\|_2
+\lambda_{\max}(\Omega)\|p-p^*\|_2^2.
\end{align*}

Now, $\max_{\widetilde p\in \mathcal{P}^{\mathcal{D}}_{\textrm{RUM}}}\|\widetilde p-p^*\|_2\le \operatorname{diam}(\mathcal{P}^{\mathcal{D}}_{\textrm{RUM}})$. Moreover, every \(p\in \mathcal{P}^{\mathcal{D}}_{\textrm{RUM}}\) satisfies \(\|p\|_2\le \sqrt{|\D|}\), and therefore $\operatorname{diam}(\mathcal{P}^{\mathcal{D}}_{\textrm{RUM}})\le 2\sqrt{|\D|}$.
Combining these bounds yields \(c(p)\le 6\lambda_{\max}(\Omega)\sqrt{|\D|}\|p^*-p\|_2\) for all $p^*\in\mathcal{P}^*$, and hence \(c(p)\le 6\lambda_{\max}(\Omega)\sqrt{|\D|}\operatorname{dist}(p,\mathcal{P}^*)\).

%% file: paper_appendix/Z_Flow_Projection_Error_Bound.tex
\begin{proof}
For notational simplicity, we write the input $\{\mathbf{f}(D',D)\}_{(D',D)\in E}$ as $\{\fb_{d',d}\}$. Similarly, we denote the nonnegative flows in Algorithm~\ref{alg:pushback_infeasible_flow} by $\fp=\{\fp_{d',d}\}$ and the resulting feasible flows by $\ff=\{\ff_{d',d}\}$.

Given a power-set digraph $G=(N,E)$ and a node $D$, we define the set of parents of $D$ as
\[
    \delta^-(D) = \{\, D' \in N : (D',D) \in E \,\},
\]
and the set of children of $D$ as
\[
    \delta^+(D) = \{\, D' \in N : (D,D') \in E \,\}.
\]
Throughout, we do not distinguish between lower- and upper-case notation for nodes (i.e., $d$ and $D$ are used interchangeably).

For each arc $(d',d)$, note that we have the triangle inequality
\begin{equation}\label{eqn:46_04182026}
\left|\fb_{d',d}-\ff_{d',d}\right|
\le
\left|\fb_{d',d}-\fp_{d',d}\right|
+
\left|\fp_{d',d}-\ff_{d',d}\right|.
\end{equation}
Recall that $s(d)=\sum_{d'\in\delta^+(d)}\ff_{d,d'}$ for each nonterminal node $d$, while $s(X)=1$ at the terminal node.

We consider two cases for all $d$ such that $\delta^-(d)$ is not empty.

\paragraph{Case 1: $\sum_{d'\in\delta^-\left(d\right)} \fp_{d',d}>0$.}
By construction,
\begin{equation*}
\ff_{d',d}
=
\frac{\fp_{d',d}}{\sum_{d'\in\delta^-(d)} \fp_{d',d}}\, s(d).
\end{equation*}
Hence we have
\begin{align}\label{eqn:47_04182026}
\begin{split}
\left|\ff_{d',d}-\fp_{d',d}\right|
&=
\left|
\frac{\fp_{d',d}}{\sum_{d'\in\delta^-(d)} \fp_{d',d}}\, s(d)-\fp_{d',d}
\right| \le
\frac{\fp_{d',d}}{\sum_{d'\in\delta^-(d)} \fp_{d',d}}
\left|s(d)-\sum_{d'\in\delta^-\left(d\right)} \fp_{d',d}\right|.
\end{split}
\end{align}
Summing \eqref{eqn:46_04182026} over $d'\in\delta^-\left(d\right)$ and applying \eqref{eqn:47_04182026}, we have
\begin{equation}\label{eq:basic_bound_node}
\sum_{d'\in\delta^-\left(d\right)}\left|\ff_{d',d}-\fb_{d',d}\right|
\le
\left|s(d)-\sum_{d'\in\delta^-(d)} \fp_{d',d}\right|
+
\sum_{d'\in\delta^-(d)}\left|\fp_{d',d}-\fb_{d',d}\right|.
\end{equation}

\paragraph{Case 2: $\sum_{d'\in\delta^-\left(d\right)} \fp_{d',d}=0$.}
In this case, by construction,
\begin{equation*}
\ff_{d',d}=\frac{s(d)}{|\delta^-(d)|}.
\end{equation*}
Therefore,
\begin{equation}\label{eqn:49_04182026}
\left|\ff_{d',d}-\fp_{d',d}\right|
+
\left|\fp_{d',d}-\fb_{d',d}\right|
=
\frac{s(d)}{|\delta^-(d)|}
+
\left|\fp_{d',d}-\fb_{d',d}\right|.
\end{equation}
Again summing over $d'\in\delta^-(d)$ we have the same inequality
\begin{equation*}
\sum_{d'\in\delta^-(d)}\left|\ff_{d',d}-\fb_{d',d}\right|
\le
\left|s(d)-\sum_{d'\in\delta^-(d)} \fp_{d',d}\right|
+
\sum_{d'\in\delta^-(d)}\left|\fp_{d',d}-\fb_{d',d}\right|,
\end{equation*}
where we use the fact that $\sum_{d'\in\delta^-(d)} \fp_{d',d}=0$.
Thus, in all cases,
\begin{equation}\label{eq:master_node_bound}
    \begin{split}
\sum_{d'\in\delta^-(d)}\left|\ff_{d',d}-\fb_{d',d}\right|
&\le
\left|s(d)-\sum_{d'\in\delta^-(d)} \fp_{d',d}\right|
+
\sum_{d'\in\delta^-(d)}\left|\fp_{d',d}-\fb_{d',d}\right|\\   & \le   \left|s(d)-\sum_{d'\in\delta^-(d)} \fb_{d',d}\right|+ 2  \sum_{d'\in\delta^-(d)}\left|\fp_{d',d}-\fb_{d',d}\right|
    \end{split}
\end{equation}

If $d$ corresponds to the whole menu $X$, then $s(d)=1$ and \eqref{eq:master_node_bound} becomes
\begin{equation}\label{eq:top_node_bound}
\sum_{d'\in\delta^-(d)}\left|\ff_{d',d}-\fb_{d',d}\right|
\le
\left|1-\sum_{d'\in\delta^-(d)}\fb_{d',d}\right|
+
2\sum_{d'\in\delta^-(d)}\left|\fb_{d',d}-\fp_{d',d}\right|.
\end{equation}

If $d$ corresponds to a set with cardinality $k\in\{1,\ldots,n-1\}$, then $s(d)=\sum_{d'\in\delta^+(d)} \ff_{d,d'}$, and therefore
\begin{align}
&\sum_{d'\in\delta^-(d)}\left|\ff_{d',d}-\fb_{d',d}\right|
\le \left|\sum_{d'\in\delta^+(d)} \ff_{d,d'}-\sum_{d'\in\delta^-(d)} \fb_{d',d}\right|+ 2  \sum_{d'\in\delta^-(d)}\left|\fp_{d',d}-\fb_{d',d}\right| \nonumber\\
&\le
\sum_{d'\in\delta^+(d)}\left|\ff_{d,d'}-\fb_{d,d'}\right|
+
\left|
\sum_{d'\in\delta^+(d)} \fb_{d,d'}
-
\sum_{d'\in\delta^-(d)} \fb_{d',d}
\right| +
2\sum_{d'\in\delta^-(d)}\left|\fb_{d',d}-\fp_{d',d}\right|.
\label{eq:middle_node_bound}
\end{align}

Let \(V_s\) denote the collection of all nodes of cardinality \(s\), and let \(E_{s-1,s}\) denote the set of arcs connecting nodes of cardinality \(s-1\) to those of cardinality \(s\). Let \(A_s\) and \(b_s\) denote the flow conservation coefficients associated with nodes in \(V_s\).

Summing \eqref{eq:middle_node_bound} over all $d\in V_s$, $s<n$, yields
\begin{equation}\label{eq:layer_bound}
    \begin{split}
&\sum_{(d',d)\in E_{s-1,s}}\left|\ff_{d',d}- \fb_{d',d}\right|\\
&\le
\sum_{(d,d')\in E_{s,s+1}}\left|\ff_{d,d'}-\fb_{d,d'}\right|
+
\|A_s\fb-b_s\|_1
+
2\sum_{(d',d)\in E_{s-1,s}}\left|\fb_{d',d}-\fp_{d',d}\right|.        
    \end{split}
\end{equation}
Similarly, when $d=X$ ($d\in V_n$), we have
\begin{equation}\label{eq:top_layer_bound}
\sum_{(d',d)\in E_{n-1,n}}\left|\ff_{d',d}-\fb_{d',d}\right|
\le
\|A_n\fb-b_n\|_1
+
2\sum_{(d',d)\in E_{n-1,n}}\left|\fb_{d',d}-\fp_{d',d}\right|.
\end{equation}
Applying \eqref{eq:layer_bound} recursively from layer $n$ down to layer $k$, we obtain for each $k\in [n]$,
\begin{equation}\label{eq:recursive_bound}
\sum_{(d',d)\in E_{k-1,k}}\left|\ff_{d',d}-\fb_{d',d}\right|
\le
\sum_{s=k}^n \|A_s\fb-b_s\|_1
+
2\sum_{s=k}^n \sum_{(d',d)\in E_{s-1,s}}\left|\fb_{d',d}-\fp_{d',d}\right|.
\end{equation}

Finally, summing \eqref{eq:recursive_bound} over $k\in [n]$, we obtain
\begin{align*}
&\sum_{(d',d)\in E}\left|\ff_{d',d}-\fb_{d',d}\right|
=
\sum_{k=1}^n \sum_{(d',d)\in E_{k-1,k}}\left|\ff_{d',d}-\fb_{d',d}\right| \\
\le &
\sum_{k=1}^n \sum_{s=k}^n \|A_s \fb-b_s\|_1
+
2\sum_{k=1}^n \sum_{s=k}^n \sum_{(d',d)\in E_{s-1,s}}
\left|\fb_{d',d}-\fp_{d',d}\right| \\
=&
\sum_{s=1}^n s\,\|A_s\fb-b_s\|_1
+
2\sum_{s=1}^n s\sum_{(d',d)\in E_{s-1,s}}
\left|\fb_{d',d}-\fp_{d',d}\right| \\
\le&
n\sum_{s=1}^n \|A_s\fb-b_s\|_1
+
2n\sum_{(d',d)\in E}\left|\fb_{d',d}-\fp_{d',d}\right| \\
\leq&
n\|A\fb-b\|_1
+
2n\sum_{(d',d)\in E}\left|\fb_{d',d}-\fp_{d',d}\right|.
\end{align*}
Finally, since $\fp=\max\{\fb,0\}$ component-wise,
\[
\sum_{(d',d)\in E}\left|\fb_{d',d}-\fp_{d',d}\right|
=
\|\min\{\mathbf f,0\}\|_1.
\]
Substituting this identity into the preceding bound yields the claimed result. We note that $\fb^{\textrm{feas}}$ is nonnegative and is constructed to satisfy flow conservation; hence, it is feasible.

This completes the proof.

\end{proof}

%% file: paper_appendix/Z_lemma_laplacian_proof.tex
\begin{proof}
We prove by induction. The Laplacian matrix $L_1$ can be shown to have the spectral decomposition
\begin{equation*}
    L_1=\begin{bmatrix}
        1 & -1\\
        -1& 1 
    \end{bmatrix}=\left(\frac{1}{\sqrt{2}}\begin{bmatrix}
        1 & 1 \\
        -1 & 1
    \end{bmatrix}\right)^\top\begin{bmatrix}
        0 & 0\\
        0& 2 
    \end{bmatrix} \left(\frac{1}{\sqrt{2}}\begin{bmatrix}
        1 & 1 \\
        -1 & 1
    \end{bmatrix}\right)
\end{equation*}
We note that $U$ and $U^{\otimes n}$ are unitary matrices. Hence $I_2=U^\top U$ and $I_{2^{n-1}}=\left(U^{\otimes \left(n-1\right)}\right)^{\top}U^{\otimes \left(n-1\right)}$. 

Let $A,B,C,D,$ and $E$ be matrices with compatible dimensions. Let $I_d$ denote the identity matrix of dimension $d$, where $d$ equals the number of columns of $D$. The Kronecker product satisfies the following mixed-product identity:
\begin{equation}
(ABC)\otimes (DE)
= (AB\otimes D)(C\otimes E)
= (A\otimes D)(B\otimes I_d)(C\otimes E).
\end{equation}

Assuming the identity \eqref{eqn:spectral_L_k} is true for $L_{n-1}$, we have
\begin{align*}
    L_n= & L_{n-1}\otimes I_2 + I_{2^{n-1}} \otimes L_1\\
    =& \left( \left(U^{\otimes \left(n-1\right)}\right)^\top \Sigma_{n-1} U^{\otimes \left(n-1\right)}\right)\otimes \left(U^\top U\right) + \left(\left(U^{\otimes \left(n-1\right)}\right)^{\top}U^{\otimes \left(n-1\right)}\right)\otimes \left(U^\top \Sigma_1 U\right)\\
    =& \left( \left(U^{\otimes \left(n-1\right)}\right)^\top \otimes U^\top\right)\left( \Sigma_{n-1}\otimes I_2\right)\left(U^{\otimes \left(n-1\right)}\otimes U \right)\\
      & + \left( \left(U^{\otimes \left(n-1\right)}\right)^\top \otimes U^\top\right)\left( I_{2^{n-1}}\otimes \Sigma_1\right)\left(U^{\otimes \left(n-1\right)}\otimes U \right)\\
    =& \left(U^{\otimes n}\right)^\top \left(\Sigma_{n-1}\otimes I_2+  I_{2^{n-1}}\otimes \Sigma_1\right)U^{\otimes n},
\end{align*}
where for the last equality we use the fact that $\left(A\otimes B\right)^\top=A^\top \otimes B^\top$. The characterization of the Moore--Penrose inverse is standard.
\end{proof}

%% file: paper_appendix/Z_Proof_ADMM_Convergence.tex
Our proof strategy follows that of \cite{fazel2013hankel} but is generalized to cases with relaxation and general positive-definite matrix weighting.

Recall the optimization problem \eqref{eq:generic_problem}. Its Lagrangian is defined as
\begin{equation*}
\mathcal{L}(\x,\y,\ub)
=
f(\x)+g(\y)+\langle \ub,A\x+B\y-c\rangle.
\end{equation*}
Under the assumed constraint qualification, Corollary 28.2.2 and Corollary 28.3.1 of \cite{rockafellar1997convex} imply that strong duality holds and the dual optimum is attained. Hence, there exists a primal-dual saddle point $(\x^*,\y^*,\ub^*)$ of $\mathcal{L}$, where $(\x^*,\y^*)$ is an optimal solution to the primal problem and $\ub^*$ is an optimal solution to the corresponding dual problem. Moreover, the KKT conditions imply that, for any such $(\x^*,\y^*,\ub^*)$,
\begin{equation}\label{eq:kkt_generic}
\begin{aligned}
-A^\top \ub^* &\in \partial f(\x^*), \quad
-B^\top \ub^* &\in \partial g(\y^*), \quad
A\x^*+B\y^*-c &= 0.
\end{aligned}
\end{equation}

\textbf{Step 1:} Define $\rb_{k+1}=A\x_{k+1}+B\y_{k+1}-c$. We note that the dual update can be written as
\begin{align}\label{eqn:dual_update_u}
\begin{split}
    \ub_{k+1}&=\ub_k+ \Xi\left(\overline{A\mathbf{x}}^{\,k+1}+B\y_{k+1}-c\right)\\
    & = \ub_k + \Xi \left( \alpha A\x_{k+1}-(1-\alpha)(B\y_k-c)+B\y_{k+1}-c\right)\\
    & = \ub_k+\Xi \left( \alpha \left(A\x_{k+1}+B\y_{k+1}-c\right)+(1-\alpha)(B\y_{k+1}-B\y_k)\right)\\
    & = \ub_k + \Xi \left( \alpha \rb_{k+1}+(1-\alpha)(B\y_{k+1}-B\y_k)\right)    
\end{split}
\end{align}

The first-order conditions from the $\x$- and $\y$-updates are
\begin{align}\label{eqn:foc_admm}
\begin{split}
& 0\in \partial f(\x_{k+1})+A^\top\ub_k + A^\top \Xi\left(A\x_{k+1}+B\y_k-c\right)+ S\left(\x_{k+1}-\x_k\right)\\
& 0\in \partial g(\y_{k+1})+ B^\top \ub_k+ B^\top \Xi \left(\overline{A\mathbf{x}}^{\,k+1}+B\y_{k+1}-c \right)+T\left(\y_{k+1}-\y_k\right).    
\end{split}
\end{align}
Note that we have
\begin{align*}
\begin{split}
     &\ub_k + \Xi(A\x_{k+1}+B\y_k-c)=\ub_k + \Xi(A\x_{k+1}+B\y_{k+1}-c) + \Xi\left(B\y_k-B\y_{k+1}\right)\\
    =&  \ub_k +\Xi \rb_{k+1}+\Xi\left(B\y_k-B\y_{k+1}\right)\\
    =&  \ub_k + \Xi \left(\alpha \rb_{k+1}+(1-\alpha)B\left(\y_{k+1}-\y_{k}\right)\right)+ (1-\alpha)\Xi\rb_{k+1}+ (2-\alpha) \Xi\left(B\y_k-B\y_{k+1}\right)\\
    =& \ub_{k+1}+(1-\alpha)\Xi\rb_{k+1}+(2-\alpha)\Xi \left(B\y_k-B\y_{k+1}\right).   
\end{split}
\end{align*}

Hence we have the relations
\begin{align}\label{eqn:foc_admm_2}
\begin{split}
 &-A^\top \ub_{k+1}-(1-\alpha)A^\top\Xi\rb_{k+1}-(2-\alpha)A^\top\Xi \left(B\y_k-B\y_{k+1}\right)-S(\x_{k+1}-\x_k)  \in \partial f(\x_{k+1}),    \\
 &    -B^\top \ub_{k+1}-T\left(\y_{k+1}-\y_k\right)\in \partial g(\y_{k+1}).    
\end{split}
\end{align}
\textbf{Step 2:} Applying the monotonicity of $\partial f$
and $\partial g$ to the subgradients in
\eqref{eqn:foc_admm_2} and \eqref{eq:kkt_generic},
we obtain, for any saddle point $(\x^*,\y^*,\ub^*)$,
\begin{align*}
    \langle A^\top \ub^* - A^\top \ub_{k+1}-(1-\alpha)A^\top\Xi \rb_{k+1}-(2-\alpha)A^\top \Xi \left(B\y_k-B\y_{k+1}\right)-S(\x_{k+1}-\x_k) , \x_{k+1}-\x^* \rangle \geq 0
\end{align*}
and similarly
\begin{align*}
\langle B^\top \ub^* -B^\top \ub_{k+1}-T(\y_{k+1}-\y_k), \y_{k+1}-\y^*\rangle \geq 0. 
\end{align*}
Adding the terms we have
\begin{align}\label{eqn:key_step1}
\begin{split}
    &\langle \ub^*- \ub_{k+1}, A\x_{k+1}-A\x^* \rangle +     \langle \ub^*- \ub_{k+1}, B\y_{k+1}-B\y^*\rangle \\
    &-(1-\alpha)\langle \Xi \rb_{k+1}, A\x_{k+1}-A\x^* \rangle - (2-\alpha)\langle \Xi \left(B\y_k-B\y_{k+1}\right), A\x_{k+1}-A\x^* \rangle \\
    & - \langle  S(\x_{k+1}-\x_k), \x_{k+1}-\x^*\rangle - \langle T(\y_{k+1}-\y_k), \y_{k+1}-\y^*\rangle \geq 0     
\end{split}
\end{align}
\textbf{Step 3:} We further have
\begin{align*}
     &\langle \Xi \left(B\y_k-B\y_{k+1}\right), A\x_{k+1}-A\x^*\rangle \\
      = &     \langle \Xi \left(B\y_k-B\y_{k+1}\right), A\x_{k+1}+B\y_{k+1}-c\rangle + \langle \Xi \left(B\y_k-B\y_{k+1}\right), B\y^*-B\y_{k+1}\rangle\\
      = & \langle \Xi \left(B\y_k-B\y_{k+1}\right), \rb_{k+1}\rangle + \langle \Xi \left(B\y_k-B\y_{k+1}\right), B\y^*-B\y_{k+1}\rangle.
\end{align*}
Hence \eqref{eqn:key_step1} becomes
\begin{align}\label{eqn:key_step2}
\begin{split}
    &\langle \ub^*- \ub_{k+1}, \rb_{k+1} \rangle -(1-\alpha)\langle \Xi \rb_{k+1}, A\x_{k+1}-A\x^* \rangle \\
    & - (2-\alpha)\langle \Xi \left(B\y_k-B\y_{k+1}\right), \rb_{k+1} \rangle  +(2-\alpha)\langle \Xi \left(B\y_k-B\y_{k+1}\right), B\y_{k+1}-B\y^* \rangle \\
    & - \langle  S(\x_{k+1}-\x_k), \x_{k+1}-\x^*\rangle - \langle T(\y_{k+1}-\y_k), \y_{k+1}-\y^*\rangle \geq 0 ,    
\end{split}
\end{align}
where we change the sign of the fourth term by swapping $B\y_{k+1}$ and $B\y^*$.

\textbf{Step 4:} Note that
\begin{align}
    &\langle \Xi \rb_{k+1}, A\x_{k+1}-A\x^*\rangle = \langle \Xi \rb_{k+1}, A\x_{k+1}+B\y_{k+1}-A\x^*-B\y^*\rangle+\langle \Xi\rb_{k+1}, B\y^*-B\y_{k+1}\rangle \nonumber \\
    = & \|\rb_{k+1}\|_\Xi^2+\langle \rb_{k+1}, \Xi\left(B\y^*-B\y_{k+1}\right)\rangle, \label{eqn:09092026_1138}
\end{align}
where we use the fact that $\Xi$ is symmetric. Define
\begin{equation*}
   \vb_{k+1}= \ub_{k+1}-(1-\alpha)\Xi B\y_{k+1} \text{ and } \vb^*=\ub^*- (1-\alpha)\Xi B\y^*.
\end{equation*}

Substituting \eqref{eqn:09092026_1138} into
\eqref{eqn:key_step2} and using the definitions of the
shifted dual variables $\vb_{k+1}$ and $\vb^*$, we obtain
\begin{align}\label{eqn:key_step3}
\begin{split}
   & \langle \vb^*-\vb_{k+1}, \rb_{k+1}\rangle - (1-\alpha) \|\rb_{k+1}\|_{\Xi}^2- (2-\alpha)\langle \Xi \left(B\y_k-B\y_{k+1}\right), \rb_{k+1} \rangle \\
    & +(2-\alpha)\langle \Xi \left(B\y_k-B\y_{k+1}\right), B\y_{k+1}-B\y^* \rangle  \\
    & - \langle  S(\x_{k+1}-\x_k), \x_{k+1}-\x^*\rangle - \langle T(\y_{k+1}-\y_k), \y_{k+1}-\y^*\rangle \geq 0    
\end{split}
\end{align}

\textbf{Step 5:} Now by the monotonicity of $\partial g(\y_{k+1})$ and $\partial g(\y_{k})$, we have
\begin{align*}
    & \langle -B^\top \ub_{k+1}- T\left(\y_{k+1}-\y_k\right)-\left(-B^\top \ub_{k}- T\left(\y_{k}-\y_{k-1}\right)\right) ,\y_{k+1}-\y_{k} \rangle \geq 0  \\
  \implies  & \langle \ub_{k+1}-\ub_k, B(\y_k-\y_{k+1}) \rangle \geq \|\y_{k+1}-\y_k\|^2_T- \langle T(\y_k-\y_{k-1}),\y_{k+1}-\y_k\rangle. 
\end{align*}
Now from \eqref{eqn:dual_update_u} we have $\Xi \rb_{k+1}= \frac{1}{\alpha}\left(\ub_{k+1}-\ub_k-(1-\alpha)\Xi B\y_{k+1}+(1-\alpha)\Xi B\y_k\right)$.

Hence we have, by the symmetry of $\Xi$,
\begin{align*}
   &\langle \Xi \left(B\y_k-B\y_{k+1}\right), \rb_{k+1}\rangle=   \langle  \left(B\y_k-B\y_{k+1}\right), \Xi\rb_{k+1}\rangle\\
  = & \frac{1}{\alpha}\langle B\y_k-B\y_{k+1},\ub_{k+1}-\ub_k\rangle + \frac{1-\alpha}{\alpha}\|B\y_{k+1}-B\y_k \|_\Xi^2\\
  \geq & \frac{1}{\alpha}\|\y_{k+1}-\y_k\|_T^2 -  \frac{1}{\alpha}\langle T(\y_k-\y_{k-1}),\y_{k+1}-\y_k\rangle +\frac{1-\alpha}{\alpha}\|B\y_{k+1}-B\y_k \|_\Xi^2. 
\end{align*}
Hence we have
\begin{align}
    & -(2-\alpha)\langle \Xi \left(B\y_k-B\y_{k+1}\right), \rb_{k+1}\rangle \nonumber \\
    \leq & -\frac{2-\alpha}{\alpha}\|\y_{k+1}-\y_k\|^2_T+ \frac{2-\alpha}{\alpha}\langle T(\y_k-\y_{k-1}),\y_{k+1}-\y_k\rangle- \frac{(2-\alpha)(1-\alpha)}{\alpha}\|B\y_{k+1}-B\y_k \|_\Xi^2   \nonumber \\
    \leq &  -\frac{2-\alpha}{\alpha}\|\y_{k+1}-\y_k\|^2_T + \frac{2-\alpha}{2\alpha}\|\y_k-\y_{k-1}\|_T^2 +\frac{2-\alpha}{2\alpha}\|\y_{k+1}-\y_{k}\|_T^2 \label{eqn:20260907_92}\\
    & - \frac{(2-\alpha)(1-\alpha)}{\alpha}\|B\y_{k+1}-B\y_k \|_\Xi^2  \nonumber\\
    \leq & -\frac{2-\alpha}{2\alpha}\|\y_{k+1}-\y_k\|^2_T+\frac{2-\alpha}{2\alpha}\|\y_{k}-\y_{k-1}\|_T^2- \frac{(2-\alpha)(1-\alpha)}{\alpha}\|B\y_{k+1}-B\y_k \|_\Xi^2, \nonumber
\end{align}
where in \eqref{eqn:20260907_92} we use the fact that $\langle Ta,b\rangle \leq \frac{1}{2}\|a\|_T^2+\frac{1}{2}\|b\|_T^2$.

Hence \eqref{eqn:key_step3} becomes
\begin{align}\label{eqn:key_step4}
\begin{split}
    &\langle \vb^*-\vb_{k+1}, \rb_{k+1}\rangle - (1-\alpha) \|\rb_{k+1}\|_{\Xi}^2\\
&-\frac{2-\alpha}{2\alpha}\|\y_{k+1}-\y_k\|^2_T+\frac{2-\alpha}{2\alpha}\|\y_{k}-\y_{k-1}\|_T^2- \frac{(2-\alpha)(1-\alpha)}{\alpha}\|B\y_{k+1}-B\y_k \|_\Xi^2\\
      &+(2-\alpha)\langle \Xi \left(B\y_{k}-B\y_{k+1}\right), B\y_{k+1}-B\y^* \rangle- \langle  S(\x_{k+1}-\x_k), \x_{k+1}-\x^*\rangle \\
      &- \langle T(\y_{k+1}-\y_k), \y_{k+1}-\y^*\rangle \geq 0     
\end{split}
\end{align}

\textbf{Step 6:} \eqref{eqn:dual_update_u} implies $\rb_{k+1}=\frac{1}{\alpha}\Xi^{-1}\left(\vb_{k+1}-\vb_k\right)$ and $\|\rb_{k+1}\|_\Xi^2=\frac{1}{\alpha^2}\|\vb_{k+1}-\vb_k\|_{\Xi^{-1}}^2$

Using the property $\langle u,v\rangle = \frac{1}{2}\|u+v\|_2^2 - \frac{1}{2}\|u\|_2^2-\frac{1}{2}\|v\|_2^2$, we have
\begin{align*}
    \langle \vb^*-\vb_{k+1}, \rb_{k+1}\rangle &= \frac{1}{\alpha}\langle \vb^*-\vb_{k+1}, \Xi^{-1}\left(\vb_{k+1}-\vb_k\right)\rangle\\
    & = \frac{1}{2\alpha}\| \vb^*-\vb_{k}\|^2_{\Xi^{-1}}-\frac{1}{2\alpha}\| \vb^*-\vb_{k+1}\|^2_{\Xi^{-1}}-\frac{1}{2\alpha}\| \vb_{k+1}-\vb_{k}\|^2_{\Xi^{-1}}\\
    & =  \frac{1}{2\alpha}\| \vb^*-\vb_{k}\|^2_{\Xi^{-1}}-\frac{1}{2\alpha}\| \vb^*-\vb_{k+1}\|^2_{\Xi^{-1}}-\frac{\alpha}{2}\| \rb_{k+1}\|^2_{\Xi},
\end{align*}
\begin{align*}
    & (2-\alpha)\langle \Xi \left(B\y_{k}-B\y_{k+1}\right), B\y_{k+1}-B\y^* \rangle \\
    =& \frac{2-\alpha}{2}\|B\y_{k}-B \y^*\|^2_\Xi - \frac{2-\alpha}{2}\|B\y_{k+1}-B \y^*\|^2_\Xi - \frac{2-\alpha}{2} \|B\y_{k}-B \y_{k+1}\|^2_\Xi,
\end{align*}
\begin{align*}
     -\langle  S(\x_{k+1}-\x_k), \x_{k+1}-\x^*\rangle= \frac{1}{2}\|\x_k-\x^*\|_S^2- \frac{1}{2}\|\x_{k+1}-\x^*\|_S^2 - \frac{1}{2}\|\x_{k+1}-\x_{k}\|_S^2,
\end{align*}
and
\begin{align*}
    -\langle T(\y_{k+1}-\y_k),\y_{k+1}-\y^* \rangle = \frac{1}{2}\|\y_k-\y^*\|_T^2 - \frac{1}{2}\|\y_{k+1}-\y^*\|_T^2-\frac{1}{2}\|\y_{k+1}-\y_k\|_T^2.
\end{align*}

Combining everything we turn \eqref{eqn:key_step4} into
\begin{align*}
     &\frac{1}{2\alpha}\| \vb^*-\vb_{k}\|^2_{\Xi^{-1}}-\frac{1}{2\alpha}\| \vb^*-\vb_{k+1}\|^2_{\Xi^{-1}}-\frac{\alpha}{2}\| \rb_{k+1}\|^2_{\Xi}- (1-\alpha)\| \rb_{k+1}\|^2_{\Xi}\\
-&\frac{2-\alpha}{2\alpha}\|\y_{k+1}-\y_k\|^2_T+\frac{2-\alpha}{2\alpha}\|\y_{k}-\y_{k-1}\|_T^2- \frac{(2-\alpha)(1-\alpha)}{\alpha}\|B\y_{k+1}-B\y_k \|_\Xi^2\\
+&\frac{2-\alpha}{2}\|B\y_{k}-B \y^*\|^2_\Xi - \frac{2-\alpha}{2}\|B\y_{k+1}-B \y^*\|^2_\Xi - \frac{2-\alpha}{2} \|B\y_{k}-B \y_{k+1}\|^2_\Xi\\
+& \frac{1}{2}\|\x_k-\x^*\|_S^2- \frac{1}{2}\|\x_{k+1}-\x^*\|_S^2 - \frac{1}{2}\|\x_{k+1}-\x_{k}\|_S^2\\
+& \frac{1}{2}\|\y_k-\y^*\|_T^2 - \frac{1}{2}\|\y_{k+1}-\y^*\|_T^2-\frac{1}{2}\|\y_{k+1}-\y_k\|_T^2\geq 0.
\end{align*}
Further combining terms, we have 
\begin{align}\label{eqn:key_step5}
\begin{split}
     &\frac{1}{2\alpha}\| \vb^*-\vb_{k}\|^2_{\Xi^{-1}}-\frac{1}{2\alpha}\| \vb^*-\vb_{k+1}\|^2_{\Xi^{-1}}-\frac{2-\alpha}{2}\| \rb_{k+1}\|^2_{\Xi}\\
-&\frac{2-\alpha}{2\alpha}\|\y_{k+1}-\y_k\|^2_T+\frac{2-\alpha}{2\alpha}\|\y_{k}-\y_{k-1}\|_T^2- \frac{\left(2-\alpha\right)^2}{2\alpha}\|B\y_{k+1}-B\y_k \|_\Xi^2\\
+&\frac{2-\alpha}{2}\|B\y_{k}-B \y^*\|^2_\Xi - \frac{2-\alpha}{2}\|B\y_{k+1}-B \y^*\|^2_\Xi\\
+& \frac{1}{2}\|\x_k-\x^*\|_S^2- \frac{1}{2}\|\x_{k+1}-\x^*\|_S^2 - \frac{1}{2}\|\x_{k+1}-\x_{k}\|_S^2\\
+& \frac{1}{2}\|\y_k-\y^*\|_T^2 - \frac{1}{2}\|\y_{k+1}-\y^*\|_T^2-\frac{1}{2}\|\y_{k+1}-\y_k\|_T^2\geq 0    
\end{split}
\end{align}

Define
\begin{align*}
    V_{k}=\frac{1}{2\alpha}\| \vb^*-\vb_{k}\|^2_{\Xi^{-1}} + \frac{2-\alpha}{2\alpha}\|\y_k-\y_{k-1}\|^2_T + \frac{2-\alpha}{2}\|B\y_k-B\y^*\|_\Xi^2+ \frac{1}{2}\|\x_k-\x^*\|_S^2 + \frac{1}{2}\|\y_k-\y^*\|_T^2,
\end{align*}
and 
\begin{align*}
    D_{k+1}= \frac{2-\alpha}{2}\|\rb_{k+1}\|^2_{\Xi}+ \frac{(2-\alpha)^2}{2\alpha}\|B\y_{k+1}-B\y_k\|_\Xi^2+ \frac{1}{2}\|\x_{k+1}-\x_k\|^2_S+\frac{1}{2}\|\y_{k+1}-\y_k\|_T^2.
\end{align*}

\eqref{eqn:key_step5} turns into the form $V_k-V_{k+1}\geq D_{k+1}$.

\textbf{Step 7:} Summing \(V_k-V_{k+1}\geq D_{k+1}\) over \(k=1,\ldots,K\), we obtain
\begin{equation}
0
\leq
\sum_{k=1}^{K} D_{k+1}
\leq
\sum_{k=1}^{K}(V_k-V_{k+1})
=
V_1-V_{K+1}
\leq
V_1.
\end{equation}
Hence, \(\sum_{k=2}^{\infty}D_k<\infty\), which implies that \(D_k\to 0\). Moreover, since \(D_{k+1}\geq 0\), we have \(V_{k+1}\leq V_k\leq V_1\) for all \(k\geq 1\). This implies that (i) $\lim_{k\to\infty}\|\rb_{k+1}\|^2_{\Xi}=0$, (ii) $\|\y_k-\y^*\|_{T+B^\top \Xi B}$ is bounded, and (iii) $\|\vb^*-\vb_k\|_{\Xi^{-1}}^2$ is bounded. Because $T+B^\top \Xi B$ is positive definite by assumption, $\|\y_k-\y^*\|^2_2$ is also bounded. In addition, because $\vb_k=\ub_k-(1-\alpha)\Xi B\y_k$ and $\Xi$ is positive definite, $\|\ub_k-\ub^*\|^2_2$ is bounded as well. 

Moreover, because $\Xi$ is positive definite, we have
\begin{equation}
    \|A\x_{k}-A\x^*\|_\Xi\leq \|A\x_k-A\x^*+B\y_k-B\y^*\|_\Xi+ \|B\y_k-B\y^*\|_\Xi, 
\end{equation}
which implies that $\|A\x_k-A\x^*\|_\Xi^2$ is bounded. Because both $\|A\x_k-A\x^*\|_\Xi^2$ and $\|\x_k-\x^*\|_S^2$ are bounded and $S+A^\top \Xi A$ is positive definite by assumption, we have that $\|\x_k-\x^*\|$ is bounded as well. 

Hence the sequence $\{\left(\x_k,\y_k,\ub_k\right)\}_k$ is bounded and it has a convergent subsequence $\{\left(\x_{k_i},\y_{k_i},\ub_{k_i}\right)\}_i$. We denote its limit as $(\x^\infty,\y^\infty,\ub^\infty)$. 

\textbf{Step 8:} We know that $\lim_{k\to\infty}\|\y_{k+1}-\y_k\|_T^2=0$, $\lim_{k\to\infty}\|\x_{k+1}-\x_k\|_S^2=0$ and $\lim_{k\to\infty}\|B\y_{k+1}-B\y_k\|_\Xi^2=0$. Because $T$, $S$ and $\Xi$ are positive semidefinite, we have 
\begin{align*}
    T\left(\y_{k+1}-\y_k\right)\to 0,\quad S\left(\x_{k+1}-\x_k\right)\to 0,\quad \text{ and } \quad \Xi B\y_{k+1}-\Xi B\y_k \to 0.
\end{align*}

We also have $\rb_{k+1}\to 0$. Hence along the subsequence $\{\left(\x_{k_i},\y_{k_i},\ub_{k_i}\right)\}_i$ and by Proposition 20.38 of \cite{bauschke2017convex}, \eqref{eqn:foc_admm_2} leads to
\begin{align*}
    -A^\top\ub^\infty \in \partial f(\x^\infty)\text{ and } -B^\top\ub^{\infty}\in \partial g(\y^\infty).
\end{align*}
This implies that $(\x^\infty,\y^\infty)$ is an optimal solution to the primal problem and $\ub^\infty$ is an optimal dual solution. 

\textbf{Step 9:} Replace the $(\x^*,\y^*,\ub^*)$ in Steps 1--6 above with $(\x^\infty,\y^\infty,\ub^\infty)$. Because $V_1\geq ...\geq V_k\geq V_{k+1}\geq ...$, $\Xi$ is positive definite and $T+B^\top\Xi B$ is positive definite, we have, for the entire sequence, $\vb_k\to \vb^\infty= \ub^\infty-(1-\alpha)\Xi B\y^{\infty} $, $\y_k\to \y^\infty$ and hence $\ub_k\to\ub^\infty$. Moreover, because $\|\rb_{k+1}\|^2_\Xi\to 0$, we have $\|A\x_k-A\x^{\infty}\|_\Xi + \|\x_k-\x^\infty\|_S\to 0$. This implies that $\|\x_k-\x^\infty\|_2^2\to 0$ because $S+A^\top \Xi A$ is positive definite by assumption.

This concludes the proof.

%% file: paper_appendix/Z_proof_admm_certificate.tex
We prove the result for BM-ADMM; the proof for Flow-ADMM is analogous. Consider the equivalent scaled version of the BM
formulation in \eqref{eqn:bm_split}:
\begin{align*}
\min_{\mathbf p,\overline{\mathbf{p}},\mathbf s}\quad
& \|\pcalD-\widehat p\|_{\Omega}^2
+ \iota_{\mathbb R_+^{n2^{n-1}}}(\mathbf s)
+ \sum_{D\in 2^X}\iota_{\Delta^D}\left(\pD\right) \\
\text{s.t.}\quad
& \widetilde{K}_n\mathbf p-\mathbf s=0, \qquad \mathbf p-\overline{\mathbf{p}}=0.
\end{align*}
The same argument applies when $\widetilde K_n$
is replaced by $K_n$.
Algorithm~\ref{app:bm_admm_alg_full} is a special case of the general ADPMM template in Algorithm~\ref{algo:adpmm}. In particular, take the first block of variables to be $\mathbf p$ and the second block to be $(\mathbf s,\overline{\mathbf{p}})$. Define
$f(\mathbf p)=0$, so that
$\operatorname{dom}(f)=\mathbb R^{n2^{n-1}}$, and define
$g(\mathbf s,\overline{\mathbf{p}})=\iota_{\mathbb R_+^{n2^{n-1}}}(\mathbf s)+\|\pcalD-\widehat p\|_{\Omega}^2
+\sum_{D\in 2^X}\iota_{\Delta^D}(\pD)$, so that
$\operatorname{dom}(g)
=\mathbb R_+^{n2^{n-1}}
\times
\prod_{D\in2^X}\Delta^D$.

The linear constraint is represented by
\[
A=
\begin{bmatrix}
\widetilde{K}_n \\
I 
\end{bmatrix},
\qquad
B=
\begin{bmatrix}
-I_{n2^{n-1}} & 0\\
0 & -I_{n2^{n-1}}
\end{bmatrix},
\qquad
c=0.
\]
The augmented-Lagrangian weighting matrix is
$\Xi
=\operatorname{diag}(\beta_1 I_{n2^{n-1}},\beta_2 I_{n2^{n-1}})$,
and the relaxation parameter is $\alpha\in(0,2)$. We take $S=0$ and let $T$ be the zero-padded extension of
$2\Omega^{\mathrm{prox}}$ to the full $(\mathbf s,\overline{\mathbf{p}})$-space. Here, $\Omega^{\mathrm{prox}}\succeq0$ is defined only on the
coordinates of $\overline{\mathbf{p}}$ corresponding to the observed menus
$\mathcal D$; accordingly, $T$ is obtained by padding
$2\Omega^{\mathrm{prox}}$ with zeros on the $\overline{\mathbf{p}}$-coordinates
corresponding to unobserved menus and on all $\mathbf s$-coordinates.
The factor of two accounts for the coefficient $1/2$ multiplying
the proximal quadratic in the general ADPMM template.

It is clear that $A$ and $B$ have full column rank and that
$T+B^\top\Xi B$ and
$S+A^\top\Xi A$ are positive definite. The feasible set within the effective domain is nonempty
and compact, and the objective is continuous on this set,
so an optimal solution exists. The constraint qualification is satisfied by taking $\mathbf p=\overline{\mathbf{p}}$ to be the choice probabilities induced by the uniform distribution over all strict preference orderings and setting $\mathbf s=\widetilde{K}_n\mathbf p$; this point lies in the relative interior of the domain and satisfies the linear constraints. Hence, by Proposition~\ref{prop:admm_convergence}, $(\mathbf p_t,\mathbf s_t,\overline{\mathbf{p}}_t)$ converges to an optimal solution $(\mathbf p^*,\mathbf s^*,\overline{\mathbf{p}}^*)$.

Since $\mathbf p_t\to\mathbf p^*$, the associated flow $\mathbf f_t=K_n\mathbf p_t$ converges to $\mathbf f^*=K_n\mathbf p^*$. Because $(\mathbf p^*,\mathbf s^*,\overline{\mathbf p}^*)$
is feasible for the scaled formulation above, we have
$\mathbf p^*=\overline{\mathbf p}^*$ and
$\widetilde K_n\mathbf p^*=\mathbf s^*\geq0$.
Since positive diagonal row scaling preserves
nonnegativity, $\mathbf f^*=K_n\mathbf p^*\geq0$,
and $\mathbf f^*$ is therefore a feasible flow. It follows that $A\mathbf f_t-b\to 0$ and $\min\{\mathbf f_t,0\}\to 0$, where $A$ here denotes the flow incidence matrix. Proposition~\ref{prop:repair_flow} then implies
$\|\mathbf f_t^{\mathrm{feas}}-\mathbf f_t\|_1\to 0$.

Since $Z_n\mathbf f_t=\mathbf p_t$, we also have
$Z_n\mathbf f_t^{\mathrm{feas}}-\mathbf p_t
=Z_n(\mathbf f_t^{\mathrm{feas}}-\mathbf f_t)\to 0$.
Hence
$Z_{\mathcal D}\mathbf f_t^{\mathrm{feas}}-\mathbf p_{\mathcal D,t}\to 0$,
and therefore
$Z_{\mathcal D}\mathbf f_t^{\mathrm{feas}}\to\mathbf p^*_{\mathcal D}$.
Since $\mathbf p^*_{\mathcal D}$ is an optimizer of \eqref{eqn:DC-RUM_Proj},
$\operatorname{dist}(Z_{\mathcal D}\mathbf f_t^{\mathrm{feas}},\mathcal P^*)\to 0$.
Applying the bound in Proposition~\ref{prop:convergence_of_FW} gives
$c(Z_{\mathcal D}\mathbf f_t^{\mathrm{feas}})\to 0$.

Finally, $Z_{\mathcal D}\mathbf f_t^{\mathrm{feas}}$ is feasible for every $t$, so the bounds in \eqref{eqn:lb} apply at every iteration. Moreover,
$Z_{\mathcal D}\mathbf f_t^{\mathrm{feas}}\to\mathbf p^*_{\mathcal D}$ implies
$\|Z_{\mathcal D}\mathbf f_t^{\mathrm{feas}}-\widehat p\|_{\Omega}^2\to v^*$.
Together with
$c(Z_{\mathcal D}\mathbf f_t^{\mathrm{feas}})\to 0$,
this implies that both the lower and upper bounds converge to $v^*$, completing the proof.

%% file: paper_appendix/Z_Appendix_Delete_Rankings_Proof.tex
We first show that the restricted flow polytope satisfies $\mathcal{F}^{\mathrm r}=\operatorname{conv}\left(\mathcal{R}^{\mathrm r}\right)$. Using the notation in \eqref{eqn:flow_polytope}, the restricted polytope can be represented as
\begin{equation*}
\mathcal{F}^{\mathrm r}=\{f: Af=b,\ f\geq 0,\ f_e=0\ \text{for all }e\in E\setminus E^{\mathrm r}\}.
\end{equation*}
Note that by Corollary 4.8 of \cite{ip}, $A$ is totally unimodular and, by Theorem 4.5 of \cite{ip}, $\mathcal{F}^{\mathrm r}$ is an integral polytope. It is easy to see that every vertex of $\mathcal{F}^{\mathrm r}$ corresponds to the incidence vector of a path in $\mathcal{R}^{\mathrm r}$ and each path in $\mathcal{R}^{\mathrm r}$ is contained in $\mathcal{F}^{\mathrm r}$. Hence $\mathcal{F}^{\mathrm r}=\operatorname{conv}\left(\mathcal{R}^{\mathrm r}\right)$.

Rankings in $\Pi^{\mathrm r}$ are in one-to-one correspondence with paths in $\mathcal R^{\mathrm r}$. Recall that each ranking $\pi\in\Pi$ determines a unique
$\emptyset$--$X$ path by adding alternatives in increasing
order of $\pi(x)$, from least to most preferred. Conversely,
the order in which a path adds alternatives uniquely
determines the ranking. If $\pi\in\Pi^{\mathrm r}$, then
the restricted alternatives are added in the relative order
$x_k,\ldots,x_1$. Consequently, whenever a node on the path
contains $x_i$, it also contains every $x_j$ with $j>i$.
Thus every node satisfies \eqref{eqn:suffix_condition},
and the path belongs to $\mathcal R^{\mathrm r}$.
Conversely, if $r\in\mathcal R^{\mathrm r}$, then
\eqref{eqn:suffix_condition} ensures that, when $x_i$ is
added, every $x_j$ with $j>i$ has already been added.
The associated ranking therefore satisfies
$\pi(x_1)>\cdots>\pi(x_k)$ and belongs to
$\Pi^{\mathrm r}$. This establishes a one-to-one
correspondence between $\Pi^{\mathrm r}$ and
$\mathcal R^{\mathrm r}$.

For each corresponding ranking--path pair $(\pi,r)$,
we have $Z_n r=p^\pi$. Therefore, by the linearity of $Z_n$, $\operatorname{conv}\left(
\{p^\pi\}_{\pi\in\Pi^{\mathrm r}}
\right)=
\operatorname{conv}\left(Z_n\mathcal R^{\mathrm r}\right)=
Z_n\operatorname{conv}\left(\mathcal R^{\mathrm r}\right)
=
Z_n\mathcal F^{\mathrm r}$.

%% file: paper_appendix/Z_Proof_Simplication.tex
\begin{definition}[Projected ranking]\label{def:projected_ranking}
Let $X$ be a set of alternatives and let $I\subseteq X$ be a nonempty subset.
For a ranking $\pi:X\to\{1,\dots,|X|\}$, define its \emph{projection relative to $I$} as the mapping $\sigma:X\to\{1,\dots,|X|\}$ such that 
\begin{equation*}
\sigma(x)=
\begin{cases}
\pi(x), & \text{if } x\in X\setminus I,\\[6pt]
\max\left\{\pi(y)+1 : \begin{aligned}y\in X\setminus I,\\\ \pi(y)<\pi(x)\end{aligned}\right\},
& \text{if } x\in I \text{ and } \exists y \in X\setminus I\text{ s.t. } \pi(y)<\pi(x),\\[6pt]
1, & \text{if } x\in I \text{ and } \pi(y)>\pi(x)\text{ for all }y\in X\setminus I.
\end{cases}
\end{equation*}
For every nonempty $D\subseteq X$ satisfying $|D\cap I|\leq 1$
and every $a\in D$, define the induced choice probabilities by
\begin{equation}\label{eqn:aD}
p(a,D)
=
\mathbf{1}\!\left\{\sigma(a)\ge \sigma(d)\ \text{for all } d\in D\right\},
\quad \forall a\in D.
\end{equation}
We denote the resulting choice probability vector by \(p^\sigma\).
\end{definition}

The induced choices on $\mathcal D$ depend only on the ranking
on $X\setminus I$ and the insertion position of each $x\in I$
relative to that ranking. The ordering of alternatives assigned to the same insertion position is irrelevant. We formalize this point in the lemma below.

\begin{lemma}\label{lemma:lemma3}
Let \(X\) be a finite set of alternatives, let \(\mathcal{D}\) be a collection of menus, and let \(I \subseteq X\) be an independent set in the co-occurrence graph defined in Section~\ref{section:simplification}. Let $\pi:X\to\{1,\dots,|X|\}$ be a complete ranking and let $p^\pi$ denote the corresponding deterministic choice probabilities on $2^X$. Let $\sigma$ be the projected ranking of $\pi$ with respect to $I$ as defined in Definition~\ref{def:projected_ranking}, and let $p^\sigma$ denote the induced deterministic choice probabilities on $\mathcal{D}$.

Denote the set of all projected rankings by
\begin{equation}\label{def:Sigma}
\widetilde{\Sigma}=\{\sigma : \sigma \text{ is the projection of some complete ranking } \pi\}.
\end{equation}
Then,
\begin{equation*}
p^\sigma=\Proj_{\mathcal D}(p^\pi) \quad
\text{ and } \qquad
\mathcal{P}^{\mathcal D}_{\mathrm{RUM}}
=\conv\!\left(\{p^\sigma\}_{\sigma\in\widetilde{\Sigma}}\right),
\end{equation*}
where $\Proj_{\mathcal D}$ denotes the projection from choice probabilities indexed by $2^X$ onto those indexed by $\mathcal D$.
\end{lemma}
\begin{proof}
First note that, since $I$ is an independent set,
$|D\cap I|=1$ for every $D\in\mathcal D$ with
$D\cap I\neq\emptyset$. Let $\sigma$ be a projected ranking of $\pi$. We show that $\sigma$ preserves the ordering information of $\pi$ on $\mathcal{D}$. Fix a menu $D\in\mathcal{D}$ and first suppose there exists an $x\in D\cap I$. Compare the ordering of $x$ and $y\in D\cap (X\setminus I)$ under $\pi$ and its projection $\sigma$.

\begin{itemize}
\item If $\pi(x)>\pi(y)$ for all $y\in D\cap (X\setminus I)$, then by construction of the projection $\sigma(x)>\sigma(y)$ for all such $y$.

\item If there exists $y\in D\cap (X\setminus I)$ with $\pi(y)>\pi(x)$, then either $x$ is dominated by all $y\in X\setminus I$, in which case $\sigma(x)=1$ and hence $\sigma(y)>\sigma(x)$, or there exists $y'\in X\setminus I$ such that $\pi(y)>\pi(x)>\pi(y')$. In the latter case, choose $y'$ to be the highest-ranked
alternative in $X\setminus I$ below $x$. Then,
\[
\sigma(x)=\pi(y')+1\le \pi(x),
\]
which again implies $\sigma(y)>\sigma(x)$.
\item $\sigma$ preserves the ordering information among $y\in D\cap \left(X\setminus I\right)$.
\end{itemize}
When $D$ does not contain any element in $I$, $\sigma$ preserves all the ordering information of $\pi$. 

Hence $p^\sigma=\Proj_{\mathcal D}(p^\pi)$. The second statement follows because the projection of a polytope is the convex hull of the projections of its vertices.
\end{proof}

To prove Proposition~\ref{theorem:simplification}, we first note that the optimal value of the maximization problem on the left-hand side is attained at one of the vertices $\{p^\sigma\}_{\sigma\in\widetilde{\Sigma}}$ defined in Lemma \ref{lemma:lemma3}, where $\widetilde{\Sigma}$ is defined in \eqref{def:Sigma}.

Choose $\sigma^*\in\widetilde{\Sigma}$ such that
$p^{\sigma^*}$ is an optimizer of the left-hand side.
Then
\begin{equation*}
    \max_{\mathbf{p}\in \mathcal{P}^{\mathcal{D}}_{\textrm{RUM}}}\sum_{D\in \D}\sum_{a\in D}\mathbf{p}(a,D)w(a,D)= \sum_{D\in \D}\sum_{a\in D}p^{\sigma^*}(a,D)w(a,D).
\end{equation*}
Moreover, we have
\begin{align}\label{eqn:63_20260420}
\begin{split}
     &\sum_{D\in \D}\sum_{a\in D}p^{\sigma^*}(a,D)w(a,D)\\
     =&\sum_{D\in\mathcal{D}^{X\setminus I}}\sum_{\substack{a\in D}} p^{\sigma^*}\left(a,D\right)w(a,D)+\sum_{x\in I}\sum_{D\in \mathcal{D}^x}\sum_{\substack{a\in D}}p^{\sigma^*}(a,D)w(a,D)
\end{split}
\end{align}
To see why this equality holds, first note that
$|D\cap I|=1$ for every $D\in\mathcal D$ with
$D\cap I\neq\emptyset$.
Each such menu $D$ is uniquely associated with an
element $x\in I$, and the equality follows immediately.

Now let \(\sigma^*\) induce a ranking \(\widetilde{\sigma}^*\) on \(X\setminus I\). For each \(x\in I\), the set \(\mathcal{P}_x^{\widetilde{\sigma}^*}\) characterizes the possible relative positions of \(x\) with respect to the elements in \(X\setminus I\). Moreover, the restriction of \(p^{\sigma^*}\) to the coordinates $\mathcal{J}_x=\{(a,D) : a\in D, D\in \mathcal{D}^x\}$
corresponds to an element of \(\mathcal{P}_x^{\widetilde{\sigma}^*}\). It follows that, for each \(x\in I\),
\begin{equation*}
\sum_{D\in \mathcal{D}^x}
\sum_{\substack{a\in D\\ (a,D)\in\mathcal{J}}}
p^{\sigma^*}(a,D) w(a,D)
\;\le\;
\max_{\mathbf{p}\in \mathcal{P}_x^{\widetilde{\sigma}^*}}\sum_{D\in \mathcal{D}^x}
\sum_{\substack{a\in D\\ (a,D)\in\mathcal{J}}}
\mathbf{p}(a,D) w(a,D).
\end{equation*}
Summing the right-hand side over \(x\in I\), adding the term $\sum_{D\in\mathcal{D}^{X\setminus I}}\sum_{\substack{a\in D}} p^{\sigma^*}\left(a,D\right)w(a,D)$, and then maximizing over \(\sigma\in \Sigma\) (noting that \(\widetilde{\sigma}^*\in \Sigma\)) yields that the left-hand side of \eqref{eqn:23_04192026} is bounded above by the right-hand side.

Now we show the reverse direction. Fix a ranking \(\widetilde{\sigma}\) on \(X\setminus I\). For each \(x\in I\), let an optimizer over \(\mathcal{P}_x^{\widetilde{\sigma}}\) determine the relative position of \(x\) with respect to the elements in \(X\setminus I\). These positions can be combined to form a projected ranking \(\sigma\), which is induced by some complete ranking \(\pi\). Consequently, the value attained by the maximization problem on the right-hand side can also be achieved by the problem on the left-hand side, by an argument similar to the one above.

Therefore the two maximization problems attain the same optimal value, which proves the proposition. The construction of an optimizer for the left-hand side from the optimizers of the right-hand side is straightforward.

%% file: paper_appendix/Z_KS_Network.tex
\begin{proof}
As noted after Definition~\ref{def:ufp}, the vertices of
$\mathcal{F}_{\mathcal{G}}$ are the incidence vectors of directed
paths from $\varnothing$ to $\mathcal{B}$. We first show that each
such path $r$ is mapped by $L_{\mathcal{G}}$ to a rational type.

We first show that $L_{\mathcal{G}}r$ is a choice type. Along a path,
each budget $\B_j$ is added exactly once, namely, at the unique
transition
\[
    S\longrightarrow S\cup\{\B_j\}.
\]
The corresponding arc has exactly one patch label $x_{i\mid j}$ for some $i$.
Hence, exactly one patch is selected from each budget $\B_j$.
Therefore, $L_{\mathcal{G}}r$ is a choice type.

We next show that $L_{\mathcal{G}}r$ is a rational type. We verify that $L_{\mathcal{G}}r$ is rational by showing that the induced revealed-preference graph on budgets is acyclic according to Definition \ref{def:rp} and Definition \ref{def:rationaltype}.

To construct such a graph, consider
the patch $x_{i\mid j}$ selected when the path moves from $S$ to
$S\cup\{\B_j\}$. By the definition of the multi-arc power-set
digraph, $x_{i\mid j}$ lies strictly above every budget in
\[
    \mathcal{B}\setminus(S\cup\{\B_j\}).
\]
Hence, if the selected patch $x_{i\mid j}$ lies strictly below some
budget $\B_\ell$, then necessarily $\B_\ell\in S$. 

For each $\B_j$ and $\B_\ell\in S$ that appears before $\B_j$ along the path $r$,  we add an arc
\[
    \B_\ell\to\B_j.
\]
Therefore, all arcs point from budgets that appear earlier
on the path to budgets that appear later. The resulting
graph contains every arc induced by the choice type
$L_{\mathcal{G}}r$ under Definition~\ref{def:rp}.

The order in which budgets are added along the path thus provides a
topological ordering of the induced revealed-preference graph on budgets as in Definition \ref{def:rp}.
Consequently, the graph is acyclic, and $L_{\mathcal{G}}r$ is a
rational type by Definition \ref{def:rationaltype}.

We now prove the reverse direction. Let $a$ be any rational type. We show that there exists a path $r$ in $\mathcal{F}_{\mathcal{G}}$ such that $L_{\mathcal{G}}r=a$.

Let $x_{a_j|j}$ denote the patch selected by the rational type $a$ from budget $\B_j$, for $j=1,\ldots,J$. By definition, $a$ induces an acyclic revealed-preference graph on the budgets $\mathcal{B}=\{\B_1,\ldots,\B_J\}$. Hence, this graph admits a topological ordering. Without loss of generality, after relabeling the budgets, write this ordering as
\begin{equation}\label{eqn:KS_network_ordering}
    \B_1\succ\B_2\succ\cdots\succ\B_J,
\end{equation}
where every revealed-preference edge $\B_j\to\B_k$ satisfies $j<k$.

Let $\mathcal{G}$ be the multi-arc power-set digraph that is induced by $\mathcal{B}=\{\B_1,\ldots,\B_J\}$ according to Definition \ref{def:ma_digraph}.
We now show that there exists a path in $\mathcal{G}$ that follows the ordering in \eqref{eqn:KS_network_ordering} and selects the patches specified by $a$.

Let $S_1=\varnothing$ and, for $k=1,\ldots,J$, define
\[
    S_k:=\{\B_1,\ldots,\B_{k-1}\}.
\]
We construct the path recursively by moving from $S_k$ to $S_k\cup\{\B_k\}$. Let $x_{a_k|k}$ denote the patch selected by $a$ from budget $\B_k$. We claim that the arc
\[
    S_k \xrightarrow{\;x_{a_k|k}\;} S_k\cup\{\B_k\}
\]
is present in $\mathcal{G}$.

Indeed, such an arc could be absent only if $x_{a_k|k}$ were dominated by the remaining budgets in the sense of Definition~\ref{def:undominated}, that is, if it lay strictly below some remaining budget $\B_\ell$ with $\ell>k$. But then the selected patch from $\B_k$ would lie strictly below $\B_\ell$, implying a revealed-preference edge
\[
    \B_\ell\to\B_k.
\]
This contradicts the topological ordering in \eqref{eqn:KS_network_ordering}, according to which every revealed-preference edge points from an earlier budget to a later budget. Hence, the required arc is present at every step.

It follows that
\[
    \varnothing
    \to \{\B_1\}
    \to \{\B_1,\B_2\}
    \to \cdots
    \to \mathcal{B}
\]
is a path in $\mathcal{G}$ whose arc labels are precisely
$x_{a_1|1},\ldots,x_{a_J|J}$. Therefore, if $r$ denotes the incidence vector of this path, then
\[
    L_{\mathcal{G}}r=a.
\]

Combining the two directions, the vertices of $\mathcal{F}_{\mathcal{G}}$ map exactly onto the rational types. Since $\mathcal{F}_{\mathcal{G}}$ is the convex hull of its path vertices and linear maps commute with convex hulls, we obtain
\[
    \mathcal{P}^{\mathrm{KS}}
    =
    \left\{
        L_{\mathcal{G}}f:
        f\in\mathcal{F}_{\mathcal{G}}
    \right\}.
\]
\end{proof}